\documentclass[11pt]{article}
\usepackage[latin9]{inputenc}
\usepackage{geometry}
\usepackage{color}
\usepackage{verbatim}
\usepackage{mathrsfs}
\usepackage{bm}
\usepackage{amsmath}
\usepackage{amssymb}
\usepackage{graphicx}
\usepackage{esint}

\makeatletter

\title{Center of mass, spin supplementary conditions, and the momentum of spinning particles}

\author{L. Filipe O. Costa\footnote{Email: lfpocosta@math.ist.utl.pt, $^{\dagger}$Email: jnatar@math.ist.utl.pt}, Jos\'e Nat\'ario$^{\dagger}$ \\
        $^{*\dagger}$CAMGSD,  Instituto Superior T\'ecnico, Universidade de Lisboa\\
        Lisboa, Portugal\\
$^{*}$Centro de F\'isica do Porto -- CFP, Departamento de F\'isica e Astronomia,\\
        Universidade do Porto, Porto, Portugal}

\date{\today}

\makeatother

\begin{document}
\maketitle 
\begin{abstract}
We discuss the problem of defining the center of mass in general relativity
and the so-called spin supplementary condition. The different spin
conditions in the literature, their physical significance, and the
momentum-velocity relation for each of them are analyzed in depth.
The reason for the non-parallelism between the velocity and the momentum,
and the concept of ``hidden momentum'', are dissected. It is argued
that the different solutions allowed by the different spin conditions
are equally valid descriptions for the motion of a given test body,
and their equivalence is shown to dipole order in curved spacetime.
These different descriptions are compared in simple examples. 
\end{abstract}

\section{Introduction}

An old problem in the description of the dynamics of test particles
endowed with multipole structure is the fact that, even for a free
\emph{pole-dipole} particle (i.e., with a momentum vector $P^{\alpha}$,
and a spin 2-form $S_{\alpha\beta}$ as its only two relevant moments)
in flat spacetime, the equations of motion resulting from the conservation
laws $T_{\ \ ;\beta}^{\alpha\beta}=0$ do not yield a determinate
system, since there exist three more unknowns than equations. The
so-called ``spin supplementary condition'', $S^{\alpha\beta}u_{\beta}=0$,
for some unit timelike vector $u^{\alpha}$, first arose as a means
of closing the system, by killing \textcolor{black}{off three components
of} $S^{\alpha\beta}$. Its physical significance remained however
obscure, especially in the earlier treatments that dealt with point
particles \cite{FrenkelZphys:1926,FrenkelNature:1926,BhabhaCorben:1941,Corben:1961}
(see also in this respect \cite{Dixon:1967}). Later treatments, most
notably the works by Möller \cite{MollerDublin:1949,Moller:AIH1949},
dealing with extended bodies, shed some light on the interpretation
of the spin condition, as it being a choice of representative point
in the body; more precisely, choosing it as the center of mass (``centroid'')
as measured in the rest frame of an observer of 4-velocity $u^{\alpha}$
--- since in relativity, the center of mass of a spinning body is
an observer-dependent point. Different choices have been proposed;
the best known ones are the Frenkel-Mathisson-Pirani (FMP) condition
\cite{FrenkelZphys:1926,Mathisson:1937}, which chooses the centroid
as measured in a frame comoving with it; the Corinaldesi-Papapetrou
(CP) condition \cite{CorinaldesiPapapetrou:1951}, which chooses the
centroid measured by the observers of zero 3-velocity ($u^{i}=0$)
in a given coordinate system; and the Tulczyjew-Dixon (TD) condition
\cite{Tulczyjew:1959,Dixon:1964}, which chooses the centroid measured
in the zero 3-momentum frame ($u^{\alpha}\propto P^{\alpha}$). A
more recent condition, proposed in \cite{Ohashi:2003,KyrianSemerak:2007},
dubbed herein the ``Ohashi-Kyrian-Semerák (OKS) condition'' (which,
as we shall see, seems to be favored in many applications), chooses
the centroid measured with respect to some $u^{\alpha}$ parallel-transported
along its worldline. The spin condition generally remained, however,
a not well understood problem (this is true even today), not being
clear, namely, its status as a choice (the discussion is sometimes
put in terms of which are the ``correct'' and the ``wrong'' conditions
for each type of particle, see see introduction of \cite{Semerak:1999}
for a review), the differences arising from the different choices,
and what it means to consider different solutions corresponding to
\emph{the same physical motion}. Also, some aspects of each condition
have been poorly understood, especially the FMP condition and its
famous helical motions \cite{MathissonZitterbewegung:1937}. The rules
for transition between spin conditions, and the quantities that are
fixed (for different solutions corresponding to the same physical
body), were established in \cite{KyrianSemerak:2007}, where the numerical
solutions were compared in the Kerr spacetime, and it was shown that,
within the limit of validity of the pole-dipole approximation, the
different solutions are contained within a minimal worldtube, formed
by all the possible positions of the center of mass, which lies inside
the convex hull of the body's worldtube. These rules were further
discussed in \cite{Costaetal:2012}, and used to show that the helical
motions are fully consistent solutions, always contained within the
minimal worldtube (and to clarify the misunderstanding that led to
the contrary claims in the literature).

The non-parallelism between the momentum and the velocity of a multipole
particle subject to external fields, and its relation with the spin
supplementary condition, is another old problem. A significant step
towards its understanding was taken in \cite{Grallaetal:2010}, where
a generalized concept of ``hidden momentum'' (first discovered in
the context of classical electrodynamics \cite{ShockleyJames,Vaidman:1990,HnizdoFluid:1997,ColemanVanVleck:1968,GriffithsAMJPhys:2009})
was introduced in general relativity, and applied to the study of
the TD and CP conditions (the latter designated therein by a different
name, the ``laboratory frame centroid''). These ideas were further
worked out, with emphasis on the FMP condition, in recent works by
the authors \cite{Costaetal:2012,CostaNatarioZilhao:2012}.

In this paper, we discuss in detail the different spin conditions
in general relativity, the centroids that they determine, their uniqueness/non-uniqueness,
and the momentum-velocity relation arising from each of them. The
different solutions given by the different spin conditions corresponding
to the same physical motion are compared in simple examples, and their
differences dissected. Building on the works in \cite{KyrianSemerak:2007}
and in \cite{Costaetal:2012} (where the equivalence was shown for
free particles in flat spacetime), we prove the equivalence of the
solutions to dipole order in curved spacetime; in particular, we clarify
the dependence of the spin-curvature force on the spin condition,
as being precisely what ensures the equivalence, and the connection
of that with the geodesic deviation equation.

\subsection{Notation and conventions}
\begin{enumerate}
\item \begin{flushleft}
Signature $-+++$; $\epsilon_{\alpha\beta\sigma\gamma}\equiv\sqrt{-g}[\alpha\beta\gamma\delta]$
is the Levi-Civita tensor, and we follow the orientation $[1230]=1$
(i.e., in flat spacetime $\epsilon_{1230}=1$); $\epsilon_{ijk}\equiv\epsilon_{ijk0}$.
Riemann tensor: $R_{\ \beta\mu\nu}^{\alpha}=\Gamma_{\beta\nu,\mu}^{\alpha}-\Gamma_{\beta\mu,\nu}^{\alpha}+...$. 
\par\end{flushleft}
\item \begin{flushleft}
$(h^{u})_{\ \beta}^{\alpha}\equiv\delta_{\beta}^{\alpha}+u^{\alpha}u_{\beta}$
denotes the projector orthogonal to a unit time-like vector $u^{\alpha}$.
\par\end{flushleft}
\item \emph{The three basic vectors in the description of an extended body.}
$P^{\alpha}$ is the momentum; $U^{\alpha}\equiv dz^{\alpha}/d\tau$
is the tangent vector to the reference worldline $z^{\alpha}(\tau)$;
the vector field involved in the \emph{spin condition} $S^{\alpha\beta}u_{\beta}=0$
is generically denoted by $u^{\alpha}$. 
\item ``Centroid'', ``center of mass'', ``CM'': have all the \emph{same
meaning} herein. $x_{{\rm CM}}^{\alpha}(u)\equiv$ centroid as measured
by an observer of 4-velocity $u^{\alpha}$.
\end{enumerate}

\section{Center of mass in relativity and the significance of the spin supplementary
condition\label{sec:Center-of-mass}}

\begin{figure}
\includegraphics[width=0.9\textwidth]{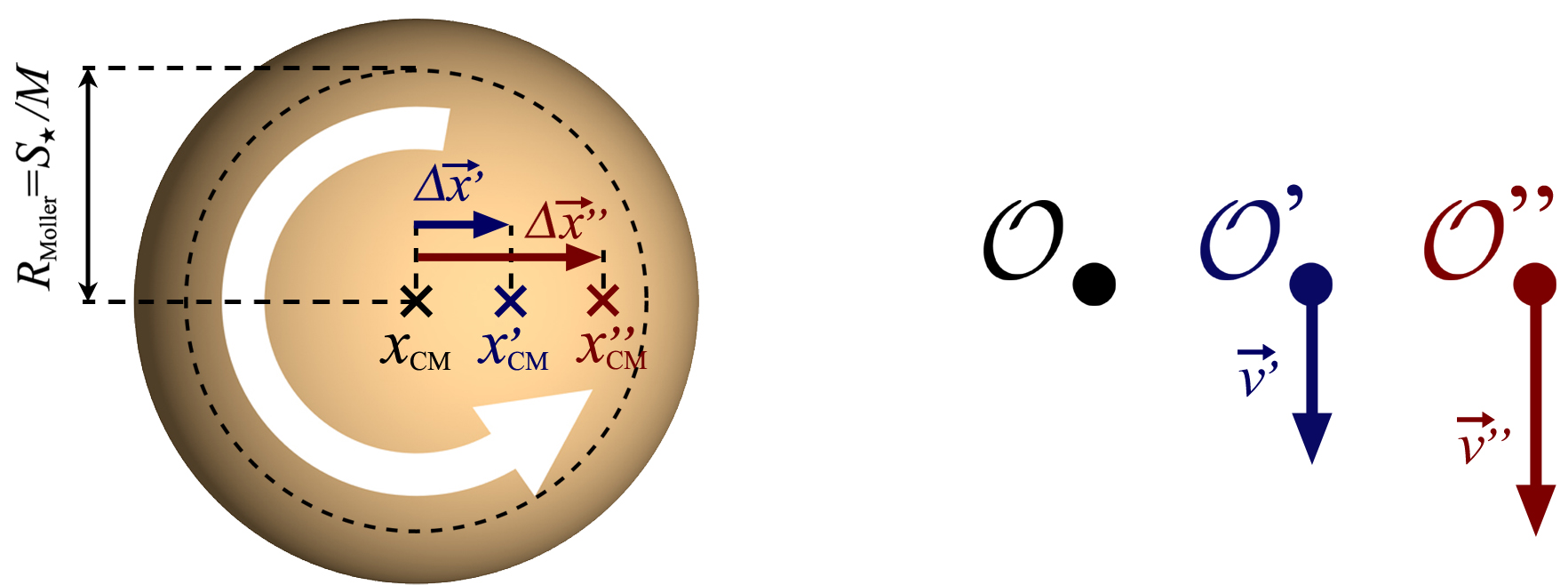}

\caption{\label{fig:CMShift}A free spinning spherical body in flat spacetime.
Observer $\mathcal{O}$, at rest relative to the axis of rotation,
measures the centroid $x_{{\rm CM}}^{\alpha}$ to coincide with the
sphere's geometrical center (and with the rotation axis). $x_{{\rm CM}}^{\alpha}\equiv x_{{\rm CM}}^{\alpha}(P)$
is the centroid as measured in the $P^{i}=0$ frame. Observers $\mathcal{O}'$
and $\mathcal{O}''$ moving (relative to $\mathcal{O}$) with velocities
$\vec{v}'$, $\vec{v}''$ \emph{opposite} to the rotation of the body,
see points on the right side of the body moving faster than those
on the left side; hence for these observers the right side of the
body is more massive, and the centroid they measure is shifted to
the right by $\Delta\vec{x}=\vec{v}\times\vec{S}_{\star}/M$. The
larger the speed $v$ the larger the shift; when $v$ equals the speed
of light, the shift takes its maximum value, with the centroid lying
in the circle of radius $R_{{\rm Moller}}=S_{\star}/M$.}
\end{figure}

In the multipole scheme an extended body is represented by a set of
moments of its current density 4-vector $j^{\alpha}$ (the ``electromagnetic
skeleton'') and a set of moments of the energy momentum tensor $T^{\alpha\beta}$,
called ``inertial'' or ``gravitational'' moments (forming the
so-called~\cite{Mathisson:1937} ``gravitational skeleton''), defined
with respect to a reference worldline $z^{\alpha}(\tau)$ which is
taken to be some representative point of the body, and whose motion
aims to represent the ``bulk'' motion of the body. The natural choice
for such point is the body's center of mass (CM); however, in relativity,
the CM of a spinning body is observer-dependent. This is illustrated
in Fig.~\ref{fig:CMShift}. In order to establish how the center
of mass changes with the observer, we need reasonable definitions
of momentum, angular momentum, mass and center of mass. In flat spacetime
these are all well defined notions; but it is not so in curved spacetime,
as they consist of integrals which amount to adding tensors defined
at different (albeit close, if the body is assumed small) points;
different generalizations of these notions have been proposed (see
e.g. \cite{Dixon:1964,Madore:1969,Dixon:1970}). The discussion herein
is aimed to be as general as possible; for that we use the following
definitions that hold reasonable (at least to lowest orders) regardless
of the particular multipole scheme followed.

Consider a system of Riemann normal coordinates $\{x^{\hat{\alpha}}\}$
(e.g. \cite{MTW,Madore:1969}) centered at the point $z^{\alpha}$
of the reference worldline, associated to the orthonormal frame $\mathbf{e}_{\hat{\alpha}}$
at that point, and take it to be momentarily comoving with some observer
$\mathcal{O}$ of 4-velocity $u^{\alpha}$ (\emph{not} necessarily
tangent to the curve $z^{\alpha}(\tau)$); that is, at $z^{\alpha}$,
$\mathbf{e}_{\hat{0}}=u^{\alpha}$ and the triad $\mathbf{e}_{\hat{i}}$
spans the instantaneous local rest space of $\mathcal{O}$. We define
the momentum $P^{\alpha}$, angular momentum $S^{\alpha\beta}$, mass
$m(u)$ and centroid $x_{{\rm CM}}^{\alpha}(u)$ of the particle with
respect to $\mathcal{O}$ as the tensors at $z^{\alpha}(\tau)$ (respectively
point) whose components (respectively coordinates) in this chart are
\begin{eqnarray}
P^{\hat{\alpha}} & \equiv & \int_{\Sigma(z,u)}T^{\hat{\alpha}\hat{\beta}}d\Sigma_{\hat{\beta}}\ ,\label{eq:Pgeneral}\\
S^{\hat{\alpha}\hat{\beta}} & \equiv & 2\int_{\Sigma(z,u)}x^{[\hat{\alpha}}T^{\hat{\beta}]\hat{\gamma}}d\Sigma_{\hat{\gamma}}\ ,\label{eq:Sab}\\
m(u) & \equiv & -P^{\alpha}u_{\alpha}\ =\ \int_{\Sigma(z,u)}T^{\hat{0}\hat{\gamma}}d\Sigma_{\hat{\gamma}}\ ,\label{eq:m(u)}\\
x_{{\rm CM}}^{\hat{\alpha}}(u) & \equiv & \frac{\int_{\Sigma(z,u)}x^{\hat{\alpha}}T^{\hat{0}\hat{\gamma}}d\Sigma_{\hat{\gamma}}}{m(u)}\ .\label{eq:XcmDef}
\end{eqnarray}
Here $\Sigma(z,u)\equiv\Sigma(z(\tau),u)$ is the spacelike hypersurface
generated by all geodesics orthogonal to the timelike vector $u^{\alpha}$
at the point $z^{\alpha}$ (in normal coordinates it coincides with
the spatial hypersurface $x^{\hat{0}}=0$), $d\Sigma$ is the 3-volume
element on $\Sigma(z,u)$, and $d\Sigma_{\gamma}\equiv-n_{\gamma}d\Sigma$,
where $n^{\alpha}$ is the (future-pointing) unit vector normal to
$\Sigma(z,u)$ (at $z^{\alpha}$, $n^{\alpha}=u^{\alpha}$). These
definitions correspond to the ones given in \cite{Madore:1969}, and
have a well defined mathematical meaning, which can be written in
the manifestly covariant form (\ref{eq:PCov})-(\ref{SCov}) below.
They also correspond, to a good approximation, to the ones given in
Dixon's schemes \cite{Dixon:1964,Dixon:1970}. This is discussed in
detail in Appendix \ref{sec:Momentum-and-angular in curved spacetime}.
Note that although we used normal coordinates to perform the integrations
above, the end results $P^{\alpha}$ and $S^{\alpha\beta}$ are tensors,
which can now be expressed in any frame%
\footnote{One could say the same about the point $x_{{\rm CM}}^{\alpha}(u)$,
although one must bear in mind when transforming its coordinates to
the new frame that it will still be the CM as measured by the specific
observer $u^{\alpha}$, and \emph{not} the CM as measured in the new
frame.%
}.

The vector 
\begin{equation}
(d_{{\rm G}}^{u})^{\alpha}\equiv-S^{\alpha\beta}u_{\beta}\label{eq:MassDipCov}
\end{equation}
yields the ``mass dipole moment'' as measured by the observer $\mathcal{O}$
(of 4-velocity $u^{\alpha}$), and 
\begin{equation}
\Delta x^{\alpha}=-\frac{S^{\alpha\beta}u_{\beta}}{m(u)}\label{eq:DeltaxCov}
\end{equation}
can be interpreted as the shift, or the ``displacement'', of the
centroid $x_{{\rm CM}}^{\alpha}(u)$ relative to the reference worldline
$z^{\alpha}(\tau)$. This is readily seen in the coordinate system
$\{x^{\hat{\alpha}}\}$, where $u^{\hat{i}}=0$ and $S^{\hat{i}\hat{\beta}}u_{\hat{\beta}}=-S^{\hat{i}\hat{0}}$,
and so from Eq.~(\ref{eq:Sab}) we have 
\begin{equation}
S^{\hat{i}\hat{0}}=2\int_{\Sigma(z,u)}x^{[\hat{i}}T^{\hat{0}]\hat{\gamma}}d\Sigma_{\hat{\gamma}}=\int_{\Sigma(z,u)}x^{\hat{i}}T^{\hat{0}\hat{\gamma}}d\Sigma_{\hat{\gamma}}\equiv m(u)x_{{\rm CM}}^{\hat{i}}(u);\label{eq:Massdipole}
\end{equation}
note that $x^{\hat{0}}=0$, since the integration is performed in
the geodesic hypersurface $\Sigma(z,u)$ orthogonal to $u^{\alpha}$
at $z^{\alpha}(\tau)$. Hence $\Delta x^{\hat{i}}=S^{\hat{i}\hat{0}}/m(u)$
yields the coordinates $x_{{\rm CM}}^{\hat{i}}(u)$ of the center
of mass measured by $\mathcal{O}$, in the normal system $\{x^{\hat{\alpha}}\}$.
Since the latter is constructed from geodesics radiating out of $z^{\alpha}$,
$\Delta\mathbf{x}$ is the vector at $z^{\alpha}$ tangent to the
geodesic connecting $z^{\alpha}$ and $x_{{\rm CM}}^{\alpha}(u)$,
and whose length equals that of the geodesic; that is, $x_{{\rm CM}}^{\alpha}(u)$
is the image by the geodesic exponential map of $\Delta\mathbf{x}$:
$x_{{\rm CM}}^{\alpha}(u)=\exp_{z}(\Delta\mathbf{x})$. In flat spacetime
(where vectors are arrows connecting two points), $\Delta\mathbf{x}$
reduces to the displacement vector from $z^{\alpha}$ to $x_{{\rm CM}}^{\alpha}(u)$;
in curved spacetime it is still a reasonable notion of center of mass
shift, and so (\ref{eq:MassDipCov}) is a sensible definition of mass
dipole moment. In particular, its vanishing for some observer means
that one is choosing $z^{\alpha}$ as the center of mass $x_{{\rm CM}}^{\alpha}(u)$
as measured by that observer. That is, the condition 
\begin{equation}
S^{\alpha\beta}u_{\beta}=0,\label{eq:GenSpinCondition}
\end{equation}
implying, in the system $\{x^{\hat{\alpha}}\}$, $S^{\hat{i}\hat{0}}=0\Rightarrow x_{{\rm CM}}^{\hat{i}}(u)=0$,
states that the reference worldline is the center of mass as measured
by the observer $\mathcal{O}(u)$ (or, equivalently, that the mass
dipole vanishes for $\mathcal{O}(u)$). Eq.~(\ref{eq:GenSpinCondition}),
for some timelike vector field $u^{\alpha}$ defined (at least) along
$z^{\alpha}(\tau)$, is known as the ``spin supplementary condition'',
which one needs to impose in order to have a determined system of
equations of motion, as we shall see in the next section. As we have
just seen, one can generically interpret it \emph{as a choice of center
of mass}.

In order to see how the center of mass changes with the observer,
let us for simplicity consider the case with no electromagnetic field,
$F^{\alpha\beta}=0$; in this case, as explained in detail in Appendix
\ref{sec:AppendixHypersurfaces}, under the assumption that the size
of the body is small compared with the scale of the curvature, the
moments (\ref{eq:Pgeneral})-(\ref{eq:Sab}) do not depend on the
argument $u^{\alpha}$ of $\Sigma(z,u)$; that is, they depend on
the point along the reference worldline $z^{\alpha}(\tau)$, but not
on the particular \emph{geodesic} hypersurface $\Sigma$ through it.
We may thus regard $P^{\alpha}(\tau)$ and $S^{\alpha\beta}(\tau)$
as well defined functions on $z^{\alpha}(\tau)$. We shall also introduce
the following relations which will be useful throughout this paper.
Let $u^{\alpha}$ and $u'^{\alpha}$ be the 4-velocities of two different
observers. We can write\textcolor{black}{{} (e.g. \cite{JantzenCariniBini:1992}){}}
\begin{equation}
u'^{\alpha}=\gamma(u,u')(u^{\alpha}+v^{\alpha}(u',u));\quad\gamma(u,u')\equiv-u^{\alpha}u'_{\alpha}=\frac{1}{\sqrt{1-v^{\alpha}v_{\alpha}}}\ ,\label{eq:U_u-1}
\end{equation}
where $v^{\alpha}(u',u)$ is a vector orthogonal to $u^{\alpha}$,
whose space components $v^{i}$ yield the ordinary 3-velocity of the
observer $u'^{\alpha}$ in the frame $u^{i}=0$ (i.e., the velocity
of the observer $u'^{\alpha}$ relative to the observer $u^{\alpha}$).
Choose $z^{\alpha}$ to be the CM as measured by $u^{\alpha}$: $z^{\alpha}=x_{{\rm CM}}^{\alpha}(u)$;
that is, choose $S^{\alpha\beta}u_{\beta}=0$. In order to obtain
the mass dipole measured by $u'^{\alpha}$, one just has to contract
$S^{\alpha\beta}$ with $u'_{\beta}$: $(d_{G}^{u'})^{\alpha}\equiv-S^{\alpha\beta}u'_{\beta}$;
this is because, under the assumptions above, $S^{\alpha\beta}$ does
not depend on the normal to the hypersurface $\Sigma(z)$, and thus,
in the $u'^{i}=0$ frame, we may write $-S^{\alpha\beta}u'_{\beta}$
in the form (\ref{eq:Massdipole}), only with $u'$ in the place of
$u$. The shift of the centroid $x_{{\rm CM}}^{\alpha}(u')$ measured
by $u'^{\alpha}$ relative to $x_{{\rm CM}}^{\alpha}(u)$ is thus
\begin{equation}
\Delta x^{\alpha}=-\frac{S^{\alpha\beta}u'_{\beta}}{m(u')}=-\gamma(u,u')\frac{S^{\alpha\beta}v_{\beta}(u',u)}{m(u')}\ ,\label{eq:Shiftuu'Gen}
\end{equation}
cf. Eq.~(\ref{eq:DeltaxCov}). Especially interesting is the case
$u^{\alpha}=P^{\alpha}/M$, where we denoted $M\equiv\sqrt{-P^{\alpha}P_{\alpha}}$;
this amounts to choosing $z^{\alpha}$ as the CM as measured in the
$P^{i}=0$ frame, $z^{\alpha}=x_{{\rm CM}}^{\alpha}(P)$. In this
case 
\begin{equation}
\Delta x^{\alpha}=-\frac{S_{\star}^{\alpha\beta}v_{\beta}}{M}\ ,\label{eq:CMShiftPCov}
\end{equation}
where $v^{\alpha}\equiv v^{\alpha}(u',P)$ is the velocity of the
observer $u'^{\alpha}$ relative to the $P^{i}=0$ frame, and we denoted
by $S_{\star}^{\alpha\beta}$ the angular momentum taken with respect
to $z^{\alpha}=x_{{\rm CM}}^{\alpha}(P)$ (note that the tensor $S^{\alpha\beta}$
depends on the choice of $z^{\alpha}$, cf. Eq.~(\ref{eq:Sab});
for the same body, $S^{\alpha\beta}$ is in general different for
different $z^{\alpha}$'s). Let us denote also the corresponding spin
vector by $S_{\star}^{\alpha}$, so that $S_{\star}^{\alpha\beta}=\epsilon_{\ \ \gamma\delta}^{\alpha\beta}S_{\star}^{\gamma}P^{\delta}/M$.
The space part (both in the $u'^{i}=0$ and in the $P^{i}=0$ frames,
as $\Delta x^{\alpha}$ is orthogonal to both $u'^{\alpha}$ and $P^{\alpha}$)
reads 
\begin{equation}
\Delta x^{i}=\frac{(\vec{S}_{\star}\times\vec{v})^{i}}{M}\ .\label{eq:CMShift}
\end{equation}
Thus the set of all shift vectors corresponding to all possible observers
spans a disk of radius $R_{{\rm Moller}}=S_{\star}/M$, centered at
$x_{{\rm CM}}^{\alpha}(P)$ and orthogonal to $S_{\star}^{\alpha}$
and $P^{\alpha}$, in the tangent space at $x_{{\rm CM}}^{\alpha}(P)$.
This statement can roughly be rephrased as saying that the set of
all possible positions of the center of mass as measured by the different
observers is contained (and fills) such disk (in flat spacetime this
is an exact statement, originally by Möller \cite{Moller:AIH1949}).
Let us dub such disk the ``disk of centroids'', and its radius $R_{{\rm Moller}}$
the \emph{Möller radius}.

In order to illustrate how this works, consider for simplicity the
setup in Fig.~\ref{fig:CMShift}: a free spinning spherical body
in flat spacetime. Observer $\mathcal{O}$, at rest relative to the
axis of rotation, clearly must (by symmetry) measure the CM to coincide
with the body's geometrical center (and with the rotation axis). The
rest frame of such an observer corresponds in this case to the $P^{i}=0$
frame (this statement will be made obvious in Sec.~\ref{sub:The-Mathisson-Pirani-condition.}
by Eq.~(\ref{eq:UperpInertial})). Consider now other observers,
$\mathcal{O}'$ and $\mathcal{O}''$, moving (relative to $\mathcal{O}$)
with velocities $\vec{v}'$ and $\vec{v}''$, opposite to the rotation
of the body; for these observers the center of mass is shifted to
the right, as they measure the right side of the body to be more massive.
The larger the speed $v$ the larger the shift; when $v$ equals the
speed of light, the shift takes its maximum value, with the centroid
lying in the circle of radius $R_{{\rm Moller}}$.

\textcolor{black}{In spacetime, the set of all possible centroid worldlines
forms a worldtube --- the ``minimal worldtube'' \cite{KyrianSemerak:2007},
see Fig. \ref{fig:Figure-TWO} --- typically very narrow}%
\footnote{For the fastest spinning celestial body known to date, the pulsar
PSR J1748-2446ad (rotation frequency $716$~Hz, estimated radius
$a=16$~km), whose equatorial velocity is 0.23$c$, $R_{{\rm Moller}}\simeq0.1a$,
see also the contribution by D. Giulini in this volume.%
}\textcolor{black}{}%
\textcolor{black}{, and always }contained within the \emph{convex
hull} of the body's worldtube (see \cite{Madore:1969} for its precise
definition). This can be shown in different ways. In flat spacetime,
it is not difficult to show (see e.g. \cite{Synge:1956} p. 313),
that if the mass density-energy density $\rho(u)=T^{\alpha\beta}u_{\beta}u_{\alpha}$
is positive everywhere within the body and with respect to all observers
$u^{\alpha}$ (i.e., if the weak energy condition holds everywhere
within the body), then the center of mass with respect to any $u^{\alpha}$
must be within the body's convex hull. The flat spacetime arguments
apply just as well in a local Lorentz frame $\{x^{\hat{\alpha}}\}$
(under the assumption above that the body is small enough so that
we can take it to be nearly orthonormal throughout it). In the same
framework one can show that $R_{{\rm Moller}}$ is the minimum size
that a \emph{classical} particle can have in order to have finite
spin without containing mass-energy flowing faster than light, that
is, without violating the dominant energy condition. The dominant
energy condition implies $\rho\ge|\vec{J}|$, where $J^{\hat{i}}\equiv-T^{\hat{\alpha}\hat{i}}u_{\hat{\alpha}}$.
Let $a$ be the largest dimension of the body; in the local Lorentz
frame centered at $x_{{\rm CM}}^{\alpha}(P)$ and such that $P^{\hat{i}}=0$,
we may write, 
\begin{equation}
S_{\star}=\left|\int\vec{x}\times\vec{J}d\Sigma\right|\le\int x|\vec{J}|d\Sigma\le\int\rho xd\Sigma\le Ma\ \Leftrightarrow\ a\ge\frac{S_{\star}}{M}\ .\label{eq:MinSize}
\end{equation}

\section{The momentum-velocity relation\label{sec:The-momentum-velocity-relation}}

The force and the spin evolution equations for a multipole particle
in an external electromagnetic and gravitational field are \cite{Dixon:1964}
\begin{eqnarray}
\frac{DP^{\alpha}}{d\tau} & = & qF_{\ \beta}^{\alpha}U^{\beta}+\frac{1}{2}F^{\mu\nu;\alpha}\mu_{\mu\nu}+F_{\ \gamma;\beta}^{\alpha}U^{\gamma}d^{\beta}+F_{\ \beta}^{\alpha}\frac{Dd^{\beta}}{d\tau}\ ,\nonumber \\
 &  & -\frac{1}{2}R_{\ \beta\mu\nu}^{\alpha}S^{\mu\nu}U^{\beta}+F^{\alpha}(2^{{\rm N}>1})\label{eq:ForceDS0}
\end{eqnarray}
\begin{equation}
\frac{DS^{\alpha\beta}}{d\tau}=2P^{[\alpha}U^{\beta]}+\tau^{\alpha\beta}\label{eq:DSabdtHidden}
\end{equation}
where $q$, $d^{\alpha}$ and $\mu_{\alpha\beta}$ are, respectively,
the particle's charge, electric dipole vector, and magnetic dipole
2-form (for their precise definitions, see \cite{CostaNatarioZilhao:2012}).
$F^{\alpha}(2^{{\rm N}>1})$ denotes the force (gravitational and
electromagnetic) due to the quadrupole and higher moments, and $\tau^{\alpha\beta}$
is sometimes called the ``torque'' tensor. $U^{\alpha}\equiv dz^{\alpha}/d\tau$
is the tangent to \textcolor{black}{the reference worldline $z^{\alpha}(\tau)$.}
These equations form an undetermined system even in the case $DP^{\alpha}/d\tau=0$
and $\tau^{\alpha\beta}=0$ \textcolor{black}{(for there would be
13 unknowns:} \textcolor{black}{$P^{\alpha}$, 3 independent components
of $U^{\alpha}$, and 6 independent components of $S^{\alpha\beta}$,
for only 10 equations), manifesting the need for a supplementary condition,
which amounts to specify the worldline $z^{\alpha}(\tau)$, relative
to which the moments are taken. The condition} $S^{\alpha\beta}u_{\beta}=0$,
\textcolor{black}{for some unit timelike vector field $u^{\alpha}$
defined along $z^{\alpha}$, kills off 3 components of the angular
momentum and makes that choice, requiring, as explained in the previous
section, $z^{\alpha}(\tau)$ to be the centroid as }measured by an
observer of 4-velocity $u^{\alpha}$. Contracting (\ref{eq:DSabdtHidden})
with $u_{\beta}$ one obtains an expression for the momentum of the
particle, 
\begin{equation}
P^{\alpha}=\frac{1}{\gamma(u,U)}\left(m(u)U^{\alpha}+S^{\alpha\beta}\frac{Du_{\beta}}{d\tau}+\tau^{\alpha\beta}u_{\beta}\right)\ ,\label{eq:Momentum}
\end{equation}
where \textcolor{black}{$\gamma(U,u)\equiv-U^{\alpha}u_{\alpha}$,
$m(u)\equiv-P^{\alpha}u_{\alpha}$, and }in the second term we used
\textcolor{black}{$S^{\alpha\beta}u_{\beta}=0$}. Eq.~(\ref{eq:Momentum})
tells us that, in general, $P^{\alpha}$ is not parallel to the CM
4-velocity $U^{\alpha}$; in this section we will discuss the reason
for that.

The vector $P^{\alpha}$ can be split in its projections parallel
and orthogonal to the CM 4-velocity $U^{\alpha}$: 
\begin{equation}
P^{\alpha}=P_{{\rm kin}}^{\alpha}+P_{{\rm hid}}^{\alpha};\quad P_{{\rm kin}}^{\alpha}\equiv mU^{\alpha},\; P_{{\rm hid}}^{\alpha}\equiv(h^{U})_{\ \beta}^{\alpha}P^{\beta}\ ,\label{eq:HiddenMomentum}
\end{equation}
where $m\equiv-P^{\alpha}U_{\alpha}$ is the the ``proper mass'',
i.e., the energy of the particle as measured in the CM frame, and
\[
(h^{U})_{\ \beta}^{\alpha}\equiv U^{\alpha}U_{\beta}+\delta_{\ \beta}^{\alpha}
\]
is the projector orthogonal to $U^{\alpha}$. We dub the parallel
projection $P_{{\rm kin}}^{\alpha}=mU^{\alpha}$ ``kinetic momentum''
associated with the motion of the center of mass; it is the most familiar
part of $P^{\alpha}$, formally similar to the momentum of a monopole
particle. The component $P_{{\rm hid}}^{\alpha}$ orthogonal to $U^{\alpha}$
is the so-called ``hidden momentum'' (e.g. \cite{Grallaetal:2010}).
The reason for the latter denomination is seen taking the perspective
of an observer $\mathcal{O}(U)$ comoving with the particle: in the
frame of $\mathcal{O}(U)$ (i.e., the $U^{i}=0$ frame) the 3-momentum
is in general not zero: $\vec{P}=\vec{P}_{{\rm hid}}\ne0$; however,
by definition, the particle's CM is at rest in that frame; hence this
momentum must be somehow hidden in the particle. $P_{{\rm hid}}^{\alpha}$
consists of two parts of distinct origin: $P_{{\rm hid}}^{\alpha}=P_{{\rm hidI}}^{\alpha}+P_{{\rm hid\tau}}^{\alpha}$,
\begin{align}
P_{{\rm hidI}}^{\alpha} & \equiv\frac{1}{\gamma(u,U)}(h^{U})_{\ \sigma}^{\alpha}S^{\sigma\beta}\frac{Du_{\beta}}{d\tau}\ ;\label{eq:HiddenInertial}\\
P_{{\rm hid\tau}}^{\alpha} & \equiv\frac{1}{\gamma(u,U)}(h^{U})_{\ \sigma}^{\alpha}\tau^{\sigma\beta}u_{\beta},\label{eq:Phidtau}
\end{align}
which we shall explain. \textcolor{black}{$P_{{\rm hidI}}^{\alpha}$
is a} term that depends only on the spin supplementary condition,
i.e., on the choice of the field $u^{\alpha}$ relative to which the
centroid is computed. In this sense we say it is \emph{gauge}. This
type of \textcolor{black}{hidden momentum was first discussed in~\cite{Grallaetal:2010}
(dubbed ``kinematical''} therein). The vector field $u^{\alpha}$
needs only to be defined \emph{along} $z^{\alpha}(\tau)$; but if
one takes it as belonging to some observer congruence in spacetime
(one can always do such an extension), and decomposing 
\begin{equation}
u_{\alpha;\beta}=-(a^{u})_{\alpha}u_{\beta}-\epsilon_{\alpha\beta\gamma\delta}\omega^{\gamma}u^{\delta}+\theta_{\alpha\beta}\label{eq:Kinematics}
\end{equation}
where $(a^{u})^{\alpha}\equiv u_{\ ;\beta}^{\alpha}u^{\beta}$ is
the acceleration \emph{of the observers} $u^{\alpha}$, $\omega^{\alpha}=\frac{1}{2}\epsilon^{\alpha\lambda\sigma\tau}u_{\tau}u_{[\sigma;\lambda]}$
their vorticity, and $\theta_{\alpha\beta}\equiv(h^{u})_{\alpha}^{\lambda}(h^{u})_{\beta}^{\nu}u_{(\lambda;\nu)}$
the shear/expansion, we may write 
\begin{equation}
P_{{\rm hidI}}^{\alpha}=\frac{1}{\gamma(u,U)}(h^{U})_{\ \sigma}^{\alpha}S_{\ \beta}^{\sigma}\left(\gamma(u,U)(a^{u})^{\beta}-\epsilon_{\ \mu\gamma\delta}^{\beta}u^{\delta}U^{\mu}\omega^{\gamma}+\theta^{\beta\gamma}U_{\gamma}\right)\ .\label{eq:PhidIGEM}
\end{equation}
The kinematical quantities in Eq.~(\ref{eq:Kinematics}) are connected
to ``inertial forces'', namely $G^{\alpha}=-(a^{u})^{\alpha}$ and
$H^{\alpha}=\omega^{\alpha}$ are, respectively, the ``gravitoelectric
field'' and the ``Fermi-Walker gravitomagnetic field'' as measured
by the congruence of observers $u^{\alpha}$, see \cite{CostaNatario:2012,JantzenCariniBini:1992}.
For this reason we dub $P_{{\rm hidI}}^{\alpha}$ ``inertial'' hidden
momentum.

$P_{{\rm hid\tau}}$ is associated to the ``torque'' tensor $\tau^{\alpha\beta}$
and (in general) consists of two parts: one which is again gauge and
arises for certain choices of reference worldline $z^{\alpha}$ (i.e.,
of the field $u^{\alpha}$) \emph{when} a physical torque acts on
the particle, plus another part which is not gauge, and cannot be
made to vanish by any center of mass choice. Following \cite{Grallaetal:2010}
we dub the latter ``dynamical hidden momentum''. To dipole order,
this dynamical part consists of a form of mechanical momentum that
arises in electromagnetic systems, first discovered in \cite{ShockleyJames},
and since discussed in number of papers, e.g.~\cite{Vaidman:1990,HnizdoFluid:1997,ColemanVanVleck:1968},
including recent works~\cite{GriffithsAMJPhys:2009,Grallaetal:2010,GrallaHerrmann:2013}.
To quadrupole and higher orders, there are both electromagnetic and
gravitational contributions to $\tau^{\alpha\beta}$, and thus to
$P_{{\rm hid\tau}}$.

\subsection{``Inertial hidden momentum'': center of mass shift and the decoupling
of $U^{\alpha}$ from $P^{\alpha}$\label{sub:Inertial-hidden-momentum:DecouplingUP}}

\begin{figure}
\includegraphics[width=0.85\textwidth]{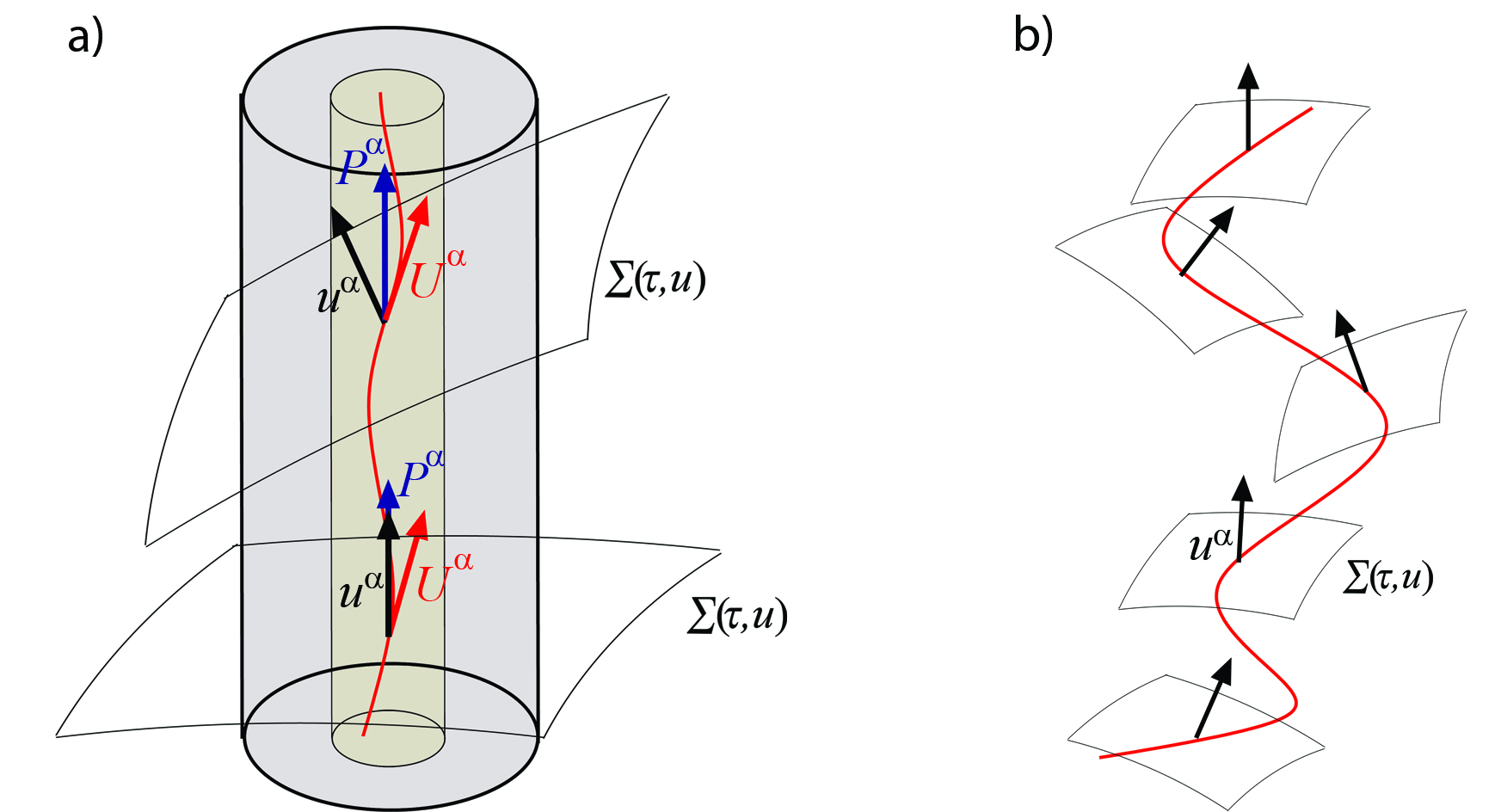}

\caption{\textcolor{black}{\label{fig:Figure-TWO}a) the body's worldtube (larger
cylinder), the worldtube of centroids (narrow inner cylinder), and
the }\textcolor{black}{\emph{three}}\textcolor{black}{{} basic vectors
involved in the description of the motion: the momentum $P^{\alpha}$,
the 4-velocity $U^{\alpha}=dz^{\alpha}/d\tau$, and the vector field
$u^{\alpha}$ involved in the spin supplementary condition $S^{\alpha\beta}u_{\beta}=0$.
The vector $u^{\alpha}$ is orthogonal to the hypersurfaces $\Sigma(\tau,u)$
at $z^{\alpha}$, and has the interpretation of the 4-velocity of
the observer measuring the centroid. These three vectors are }\textcolor{black}{\emph{not}}\textcolor{black}{{}
parallel in general. b) A curve with a varying $u^{\alpha}$ along
it; that leads to a varying shift, see Fig.~\ref{fig:CMShift}, leading
to a non-zero velocity of the centroid in the $P^{i}=0$ frame, cf.
Eq.~(\ref{eq:UperpInertial}), and possibly to an acceleration without
any force involved.}}
\end{figure}

Eq.~(\ref{eq:HiddenInertial}) tells us that when $u^{\alpha}$ varies
along $z^{\alpha}(\tau)$ (i.e., $Du^{\alpha}/d\tau\ne0$), in general
$P_{{\rm hidI}}^{\alpha}\ne0$, thus $U^{\alpha}$ is not parallel
to $P^{\alpha}$. This comes as a natural consequence of what we discussed
in Sec.~\ref{sec:Center-of-mass} about the observer dependence of
the center of mass. Recall the situation in Fig.~\ref{fig:CMShift},
a \emph{free} spinning particle in \emph{flat spacetime}: the centroid
measured by observers moving relative to $\mathcal{O}$ are shifted
relative to $x_{{\rm CM}}(\mathcal{O})$. If the velocity of these
observers changes along $z^{\alpha}(\tau)$, e.g., if at an instant
$\tau'$ we have $u^{\alpha}(\tau')=u'^{\alpha}$, and at $\tau''$
$u^{\alpha}(\tau'')=u''^{\alpha}$, the shift changes accordingly,
giving rise to a non-trivial velocity of the centroid. That is, superfluous
centroid motions can be generated just by changing $u^{\alpha}$ along
$z^{\alpha}(\tau)$. The momentum, however, remains the same, $DP^{\alpha}/d\tau=0$,
cf. Eq.~(\ref{eq:ForceDS0}); thus the situation may be cast as the
centroid acquiring a non-zero velocity in the $P^{i}=0$ frame (which
is in this case the rest frame of the observer $\mathcal{O}$). This
amounts to saying that $U^{\alpha}$ gains a component orthogonal
to $P^{\alpha}$ (denote it by $U_{\perp}^{\alpha}$); conversely,
there is a component of $P^{\alpha}$ orthogonal to $U^{\alpha}$,
which is the hidden momentum. Let us see this in detail. Denote by
$U_{\parallel}^{\alpha}$ and $U_{\perp}^{\alpha}$, respectively,
the components of $U^{\alpha}$ parallel and orthogonal to $P^{\alpha}$,
\begin{equation}
U^{\alpha}=U_{\parallel}^{\alpha}+U_{\perp}^{\alpha}\ ;\qquad U_{\parallel}^{\alpha}\equiv\frac{m}{M^{2}}P^{\alpha}\ ;\qquad U_{\perp}^{\alpha}\equiv(h^{P})_{\ \beta}^{\alpha}U^{\beta}\ ,\label{eq:Udecomp}
\end{equation}
where $m\equiv-U^{\alpha}P_{\alpha}$, and $(h^{P})_{\ \beta}^{\alpha}\equiv P^{\alpha}P_{\beta}/M^{2}+\delta_{\ \beta}^{\alpha}$
denotes the projector in the direction orthogonal to $P^{\alpha}$.
$U_{\perp}^{\alpha}$ is, up to a $\gamma$ factor, \emph{the 3-velocity
of the centroid} in the $P^{i}=0$ frame, cf. Eq.~(\ref{eq:U_u-1})
above (substitute therein $u'^{\alpha}=U^{\alpha}$, $u^{\alpha}=P^{\alpha}/M$).
In the\emph{ }special case $\tau^{\alpha\beta}=0$, we have from Eq.
(\ref{eq:Momentum}) 
\begin{equation}
U_{\perp}^{\alpha}=-\frac{1}{m(u)}(h^{P})_{\ \sigma}^{\alpha}S^{\sigma\beta}\frac{Du_{\beta}}{d\tau}\ ,\label{eq:UperpInertial}
\end{equation}
showing that indeed the variation of $u^{\alpha}$ along $z^{\alpha}(\tau)$
leads to a centroid moving in the zero 3-momentum frame, and to a
non-parallelism between $U^{\alpha}$ and $P^{\alpha}$ (it is actually
the sole reason for that in the special case $\tau^{\alpha\beta}=0$).
\emph{If we further specialize} to the case of a free particle (depicted
in Figs.~\ref{fig:CMShift}-\ref{fig:Figure-TWO}), $DP^{\alpha}/d\tau=0$,
and noting that the centroid shift can be written as $\Delta x^{\alpha}=-(x_{{\rm CM}}^{\alpha}(P)-x_{{\rm CM}}^{\alpha}(u))=S^{\alpha\beta}P_{\beta}/M^{2}$,
cf. Eq.~(\ref{eq:Shiftuu'Gen}), the shift variation along $z^{\alpha}$
becomes 
\begin{equation}
\frac{D\Delta x^{\alpha}}{d\tau}=\frac{P_{\beta}}{M^{2}}\frac{DS^{\alpha\beta}}{d\tau}=U^{\alpha}-\frac{m}{M^{2}}P^{\alpha}=U_{\perp}^{\alpha}\ .\label{eq:ShifUperp}
\end{equation}
In the second equality we used Eq.~(\ref{eq:DSabdtHidden}), in the
third we used Eqs.~(\ref{eq:Udecomp}). That is, the variation of
the shift equals the component of $U^{\alpha}$ orthogonal to $P^{\alpha}$,
mathematically formalizing the heuristic arguments in Figs.~\ref{fig:CMShift}
and \ref{fig:Figure-TWO}b). One should note however that, although
this reasoning is useful to gain intuition, in the general case ($DP^{\alpha}/d\tau\ne0$)
Eq.~(\ref{eq:ShifUperp}) does not hold, and $U_{\perp}^{\alpha}$
is not just the variation of $\Delta x^{\alpha}$; this is because
the centroid $x_{{\rm CM}}^{\alpha}(P)$ is in general no longer at
rest in the $P^{i}=0$ frame. (When one employs the TD condition,
$u^{\alpha}=P^{\alpha}/M$, if $DP^{\alpha}/d\tau\ne0$, then the
centroid 4-velocity is not in general parallel to $P^{\alpha}$, cf.
Eq. (\ref{eq:UperpInertial}) or, explicitly, Eq.~(\ref{eq:PTulczyjewDixon0})).
For the general case the argument can be given as follows: \emph{the
centroid position depends on the field $u^{\alpha}$ relative to which
it is measured, and its velocity on the variation of $u^{\alpha}$;
$P^{\alpha}$, however, is unaffected by that, which means that in
general $P^{\alpha}\nparallel U^{\alpha}$}. This is precisely what
Eq.~(\ref{eq:UperpInertial}) says.

When $U_{\perp}^{\alpha}\ne0$, then (obviously) $P^{\alpha}$ has
a component $P_{{\rm hid}}^{\alpha}$ orthogonal to $U^{\alpha}$.
Noting from (\ref{eq:Udecomp}) that $P^{\alpha}=M^{2}(U^{\alpha}-U_{\perp}^{\alpha})/m$,
and from Eqs.~(\ref{eq:HiddenMomentum}) that $U^{\alpha}=(P^{\alpha}-P_{{\rm hid}}^{\alpha})/m$,
we obtain the following relations between the hidden momentum and
$U_{\perp}^{\alpha}$: 
\begin{equation}
U_{\perp}^{\alpha}=-\frac{1}{m}(h^{P})_{\ \beta}^{\alpha}P_{{\rm hid}}^{\beta};\qquad P_{{\rm hid}}^{\alpha}=-\frac{M^{2}}{m}(h^{U})_{\ \beta}^{\alpha}U_{\perp}^{\beta}\label{eq:UperpPhid}
\end{equation}
(these are fully general expressions, valid when $\tau^{\alpha\beta}\ne0$).

\textcolor{black}{Differentiating (\ref{eq:UperpInertial}) with respect
to $\tau$, we see that when $D^{2}u^{\alpha}/d\tau^{2}\ne0$, in
general the centroid acceleration $a^{\alpha}=DU^{\alpha}/d\tau$
will be non-zero, i.e., it will accelerate without the action of a
force. That can lead to exotic motions; an example of that are the
famous Mathisson helical motions, as shown in \cite{Costaetal:2012};
the same principle also leads to the bobbings in the ``tetherballs''
studied in \cite{Grallaetal:2010} (in this case a force is involved,
but it is not parallel to the acceleration), or the ones studied in
Sec. \ref{sub:Comparison-of-the}. Of course, such effects can always
be made to vanish by a choosing some $u^{\alpha}$ parallel transported
along $z^{\alpha}(\tau)$; hence one can say that they are a complicated
description for the same physics that, in principle, could be described
in a simpler manner. In Fig. \ref{fig:Figure-TWO} we illustrate the
situation for a free particle in flat spacetime: the worldline $z^{\alpha}(\tau)$
of a centroid measured by a field} of observers $u^{\text{\ensuremath{\alpha}}}$
that varies along it has, in general, superfluous motions. These \textcolor{black}{are
confined to the worldtube of centroids, which is a straight tube (always
within the convex hull of the body's worldtube, see Sec.~\ref{sec:Center-of-mass})
parallel to the constant momentum $P^{\alpha}$, and whose cross section
orthogonal to $P^{\alpha}$ is the disk of centroids, orthogonal to
$S_{\star}^{\alpha}$, illustrated in Fig. \ref{fig:CMShift}. Choosing
$Du^{\alpha}/d\tau=0$ (e.g., inertial frames), the centroid worldlines
obtained are straight lines parallel to $P^{\alpha}$, yielding the
simplest description possible for this problem.}

\subsection{Center of mass and momentum-velocity relation of the different spin
conditions\label{sub:Center-of-mass-Shift_and_P_U}}

In this section we shall consider, for simplicity, the case $\tau^{\alpha\beta}=0$,
so that the only hidden momentum present is the inertial hidden momentum
$P_{{\rm hidI}}^{\alpha}$. Although all forms of hidden momentum
have some sort of dependence on the spin condition, by the circumstance
that $U^{\alpha}\equiv dz^{\alpha}/d\tau$ depends on the reference
worldline $z^{\alpha}(\tau)$ chosen, $P_{{\rm hidI}}^{\alpha}$ is
the part that arises \emph{solely} from it. Note that $\tau^{\alpha\beta}=0$
corresponds for instance to the case of pole-dipole particles in purely
gravitational systems.

\subsubsection{The Corinaldesi-Papapetrou (CP) condition\label{sub:The-Corinaldesi-Papapetrou-(CP)}}

\begin{figure}
\includegraphics[width=0.9\textwidth]{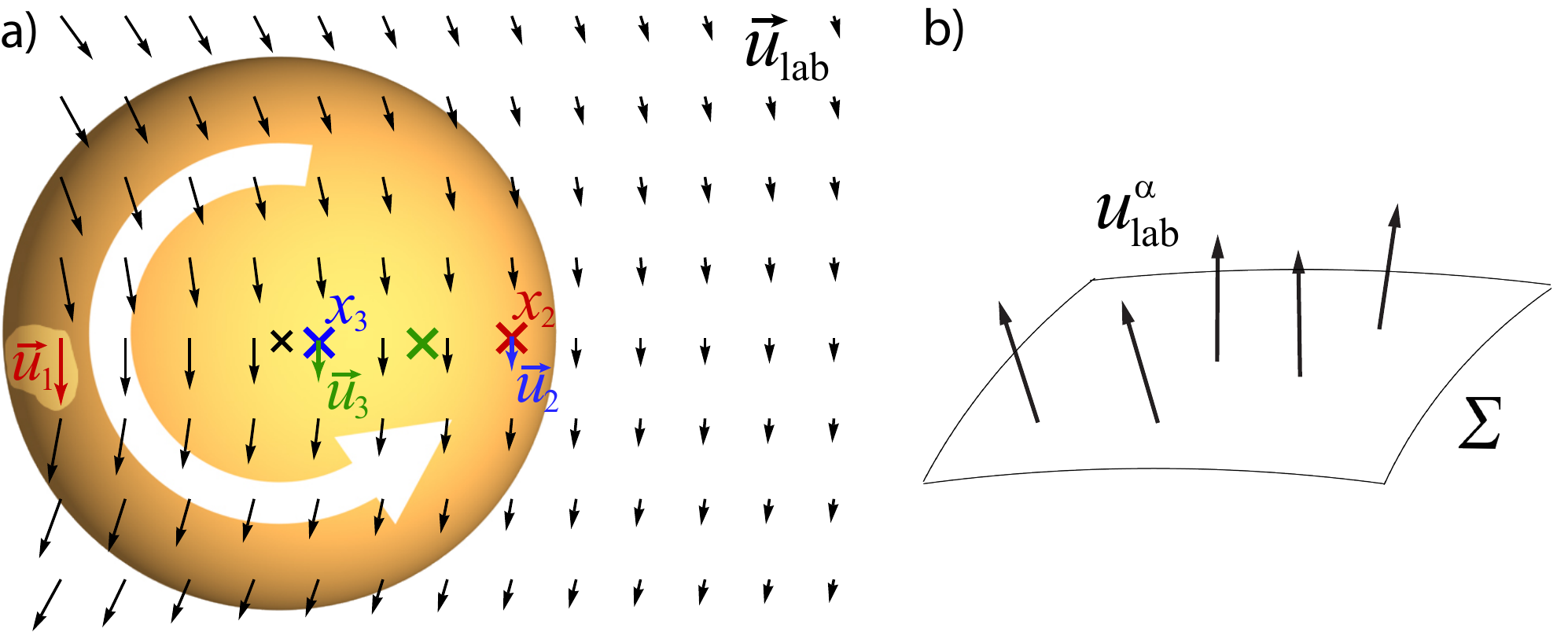}

\caption{\label{fig:CPMont}a) Centroid as measured by different observers
of the congruence $u_{{\rm lab}}^{\alpha}$. Colors specify an observer
and the corresponding centroid. Observer $u_{{\rm lab}}^{\alpha}(x_{1})\equiv u_{1}^{\alpha}$
measures the centroid to be at $x_{2}^{\alpha}\equiv x_{{\rm CM}}^{\alpha}(u_{1})$.
The observer $u_{{\rm lab}}^{\alpha}(x_{2})\equiv u_{2}^{\alpha}$
at $x_{2}^{\alpha}$ is a different one, thus its centroid will in
general be at a different point $x_{3}^{\alpha}$. Observer $u_{{\rm lab}}^{\alpha}(x_{3})\equiv u_{3}^{\alpha}$
at $x_{3}^{\alpha}$ measures the centroid to be at yet another different
point, and so on. b) For the observers $u_{{\rm lab}}^{\alpha}$ to
agree on the centroid position, they must be orthogonal to the \emph{same}
totally geodesic hypersurface $\Sigma$. In this case the condition
$S_{\ \beta}^{\alpha}u_{{\rm lab}}^{\beta}=0$ fixes an \emph{unique}
worldline.}
\end{figure}

This spin condition was introduced in \cite{CorinaldesiPapapetrou:1951}
for the Schwarzschild spacetime, where it was cast, in Schwarzschild
coordinates, as $S^{i0}=0$. One can write it covariantly as $S_{\ \beta}^{\alpha}u_{{\rm lab}}^{\beta}=0$,
with $u_{{\rm lab}}^{\alpha}$ corresponding to observers that have
zero 3-velocity in such coordinates, $u_{{\rm lab}}^{i}=0$. These
are the so-called \emph{``static observers'',} whose 4-velocity
is parallel to the time Killing vector: $u_{{\rm lab}}^{\alpha}=u_{{\rm static}}^{\alpha}\propto\partial/\partial t$.
Hence, this condition chooses as reference worldline the centroid
measured by the static observers. It can be generalized by taking
the static observers of other stationary spacetimes, or, as done in
\cite{KyrianSemerak:2007}, to arbitrary metrics taking the congruence
of observers with zero 3-velocity in the coordinate system chosen
(let us dub it the ``laboratory'' frame). This effectively amounts
to considering an arbitrary congruence of observers, which will be
the problem discussed below: take a matter distribution described
by the energy-momentum tensor $T^{\alpha\beta}(x)$, and a congruence
of observers $u_{{\rm lab}}^{\alpha}$ one may arbitrarily fix; then
find the worldlines $z^{\alpha}$ obeying the condition $S_{\ \beta}^{\alpha}(z)u_{{\rm lab}}^{\beta}(z)=0$
--- which demands $z^{\alpha}$ to be the center of mass as measured
by the observer $u_{{\rm lab}}^{\alpha}(z)$ \emph{located} \emph{at
that precise point}. At first sight, it does not even seem obvious
that such solutions exist. For when one considers an observer $u_{{\rm lab}}^{\alpha}(x_{1})$
at a given point $x_{1}^{\alpha}$, the centroid with respect to $u_{{\rm lab}}^{\alpha}(x_{1})$
will be at some point $x_{2}^{\alpha}$, in general not coinciding
with $x_{1}^{\alpha}$; and then at the site $x_{2}^{\alpha}$, the
observer $u_{{\rm lab}}^{\alpha}(x_{2})$ that lies there is a different
one, and measures its centroid to be in yet another different point
$x_{3}^{\alpha}$, and so on. This is illustrated in Fig. \ref{fig:CPMont}a).

\begin{figure}
\includegraphics[width=0.9\textwidth]{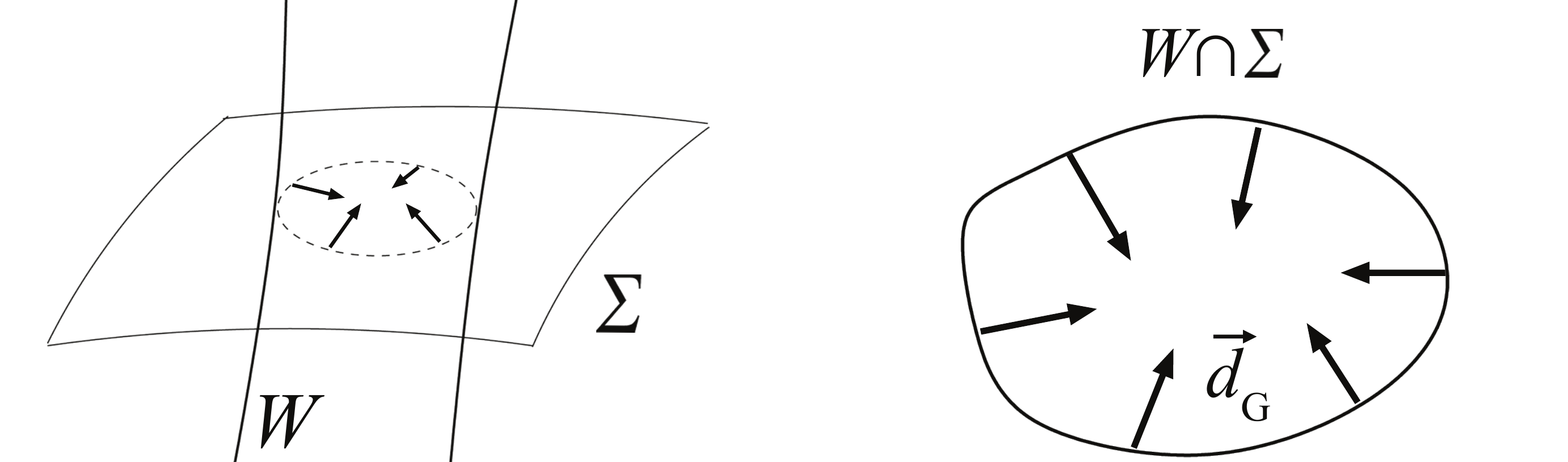}

\caption{\label{fig:Brouwer}The vector field $d_{{\rm G}}^{\alpha}(z)$ (i.e.,
the mass dipole as measured by the observers $u_{{\rm lab}}^{\alpha}$
at $z^{\alpha}$), at the boundary of the region formed by the intersection
of the convex hull $W$ of the body's worldtube with some space-like
hypersurface $\Sigma$, always points inwards. Since the field $d_{{\rm G}}^{\alpha}(z)$
is spacelike and continuous, the Brouwer fixed point theorem ensures
that $d_{{\rm G}}^{\alpha}(z)=0$ at \emph{at least one} point $z^{\alpha}\in W\cap\Sigma$.
In other words, there is at least one point which is the center of
mass as measured by the observer $u_{{\rm lab}}^{\alpha}$ \emph{at
that point}.}
\end{figure}

We shall now show that the solution indeed always exists, but in general
it is not unique. Consider the vector field (the mass dipole with
respect to the observer $u_{{\rm lab}}^{\beta}(z)$) 
\[
d_{{\rm G}}^{\alpha}(z)=-S_{\ \beta}^{\alpha}(z,u_{{\rm lab}})u_{{\rm lab}}^{\beta}(z)\ ,
\]
which is a function of $z^{\alpha}$, where $S_{\ \beta}^{\alpha}(z,u_{{\rm lab}})$
is the angular momentum taken about $z^{\alpha}$ and in the geodesic
hypersurface orthogonal to $u_{{\rm lab}}^{\alpha}$ at $z^{\alpha}$.
Consider moreover the intersection of the convex hull of the body's
worldtube $W$ with some arbitrary spacelike hypersurface $\Sigma$,
see Fig.~\ref{fig:Brouwer}; and let $\vec{d}_{{\rm G}}(z)$ be the
projection of $d_{{\rm G}}^{\alpha}(z)$ on $\Sigma$. At the boundary
of the region $W\cap\Sigma$ it is clear from the definition of $S^{\alpha\beta}(z,u)$
in Eq.~(\ref{eq:Sab}) that $\vec{d}_{{\rm G}}(z)$ points inwards
(since, by virtue of the weak energy condition, $T_{\alpha\beta}u^{\alpha}u^{\beta}>0$
for any time-like vector $u^{\alpha}$). Given that $d_{{\rm G}}^{\alpha}(z)$
is a \emph{continuous} vector field (since $u_{{\rm lab}}^{\beta}$
is an observer \emph{congruence}), the Brouwer fixed point theorem
implies that the flow of $\vec{d}_{{\rm G}}$ must have a fixed point;
i.e., $\vec{d}_{{\rm G}}=0$ at at least one point within $W\cap\Sigma$.
Since $d_{{\rm G}}^{\alpha}(z)$ is a space-like vector, this effectively
means that $d_{{\rm G}}^{\alpha}(z)=0$ at that point. 

The argument above is analogous to the one followed by Madore \cite{Madore:1969}
to produce a similar proof for the vector field $S_{\ \beta}^{\alpha}(z,P)P^{\beta}(z)$.

Hence, at least one worldline $z^{\alpha}$ will exist such that $S_{\ \beta}^{\alpha}(z)u_{{\rm lab}}^{\beta}(z)=0$;
but in general it is not unique. The analysis in Sec.~\ref{sub:The-Mathisson-Pirani-condition.}
below provides an example. Take, in flat spacetime, the congruence
$u_{{\rm lab}}^{\alpha}$ to be observers rotating rigidly with the
angular velocity of Mathisson's helical motions, $\Omega=M/S_{\star}$
(i.e., take the laboratory frame to be the observers at rest in a
frame rotating with angular velocity $\Omega$), \emph{opposite} to
the sense of rotation of the body, and around the centroid measured
in the $P^{i}=0$ frame, $x_{{\rm CM}}^{\alpha}(P)$. In this case,
every point $z^{\alpha}$ within the worldtube of centroids is a center
of mass with respect to the observers at rest in this frame, i.e.,
every such point is a solution of $S_{\ \beta}^{\alpha}(z)u_{{\rm lab}}^{\beta}(z)=0$.

In some cases the solution is unique; it is clearly so when the observers
of the congruence agree on the centroid position (note however that
this is a sufficient, but \emph{not} necessary condition for uniqueness).
The moments (thus the centroid, Eq. (\ref{eq:XcmDef})) are defined
integrating on geodesic hypersurfaces $\Sigma(z,u)$ orthogonal to
$u^{\alpha}(z)$ at some $z^{\alpha}$; in order for different observers,
e.g. $u_{1}^{\alpha}$ and $u_{2}^{\alpha}$, to agree on the centroid
position, their hypersurfaces $\Sigma(z_{1},u_{1})$ and $\Sigma(z_{2},u_{2})$
should be the same. Since $\Sigma(z_{1},u_{1})$ is geodesic \emph{at}
$z_{1}^{\alpha}$ (i.e., it is constructed from geodesics orthogonal
to $u_{1}^{\alpha}$ radiating out of the point $z_{1}^{\alpha}$),
and $\Sigma(z_{2},u_{2})$ is geodesic \emph{at} $z_{2}^{\alpha}$,
and that must be true for every other point, then this means that
the congruence $u_{{\rm lab}}^{\alpha}(x)$ must be orthogonal to
a \emph{totally geodesic hypersurface} $\Sigma$. This implies $u_{{\rm lab}}^{\alpha}(x)$
to be \emph{vorticity-free}, $\omega^{\alpha}=0$ (so that it is hypersurface
orthogonal), and \emph{rigid}, $\theta_{\alpha\beta}=0$ (so that
the second fundamental form of $\Sigma$ vanishes). Since, from the
general decomposition (\ref{eq:Kinematics}), we have, for any spatial
vector $X^{\alpha}$ tangent to $\Sigma(z,u_{{\rm lab}})$, 
\begin{equation}
\nabla_{\mathbf{X}}u_{{\rm lab}}^{\alpha}=-\epsilon_{\alpha\beta\gamma\delta}X^{\beta}\omega^{\gamma}u^{\delta}+\theta_{\alpha\beta}X^{\beta}\ ,\label{eq:DerAlongSigma}
\end{equation}
then this implies that the congruence $u_{{\rm lab}}^{\alpha}$ is
parallel%
\footnote{Such observers are said to be ``kinematically comoving'' (see \cite{Bolos:2007}
Sec.~6.1). %
} along $\Sigma(z,u_{{\rm lab}})$, $\nabla_{\mathbf{X}}u_{{\rm lab}}^{\alpha}=0$.
This is the case of static observers in a static spacetime. When these
conditions hold, and within the regime where the normal coordinates
can be taken as nearly rectangular throughout the body%
\footnote{To see the reason for this assumption, consider two observers $u_{1}^{\alpha}=u_{{\rm lab}}^{\alpha}(x_{1})$
and $u_{2}^{\alpha}=u_{{\rm lab}}^{\alpha}(x_{2})$, orthogonal to
the same geodesic hypersurface $\Sigma$. Let $\{x^{\hat{\alpha}}\}$
and $\{x^{\tilde{\alpha}}\}$, respectively, denote their normal coordinate
systems, related by $x^{\tilde{\alpha}}=\Lambda_{\ \hat{\beta}}^{\tilde{\alpha}}(x^{\hat{\beta}}-x_{2}^{\hat{\beta}})$
(where $\Lambda_{\ \hat{\beta}}^{\tilde{\alpha}}$ is a function of
$x^{\hat{\beta}}$). They will agree on the centroid position if $x_{{\rm CM}}^{\tilde{\alpha}}(u_{2})=\Lambda_{\ \hat{\beta}}^{\tilde{\alpha}}(x_{{\rm CM}}^{\hat{\beta}}(u_{1})-x_{2}^{\hat{\beta}})$.
From Eq. (\ref{eq:XcmDef}) we see that it is the case when $\mathbf{d}x^{\hat{0}}=\mathbf{d}x^{\tilde{0}},$
$x^{\tilde{i}}=\Lambda_{\ \hat{j}}^{\tilde{i}}(x^{\hat{j}}-x_{2}^{\hat{j}})$,
with $\Lambda_{\ j}^{\tilde{i}}$ a \emph{constant} matrix. Due to
the curvature, however, this cannot be exactly so; choosing $\partial_{\tilde{\alpha}}|_{x_{2}}\simeq\partial_{\hat{\alpha}}|_{x_{2}}$,
we have $\Lambda_{\ \hat{\beta}}^{\tilde{\alpha}}=\delta_{\ \hat{\beta}}^{\hat{\alpha}}+\mathcal{O}(\|\mathbf{R}\|\hat{x}\hat{x}_{2})+\mathcal{O}(\|\mathbf{R}\|\hat{x}_{2}^{2})$,
e.g. Eq. (11.12) of \cite{Brewin:2009}. It follows that, for all
observers \emph{within} the body's convex hull, $\|x_{{\rm CM}}^{\tilde{\alpha}}(u_{2})-x_{{\rm CM}}^{\tilde{\alpha}}(u_{1})\|/a\lesssim\lambda$,
$\lambda=\|\mathbf{R}\|a^{2}$; hence $x_{{\rm CM}}^{\tilde{\alpha}}(u_{2})\simeq x_{{\rm CM}}^{\tilde{\alpha}}(u_{1})$
if $\lambda\ll1$. This is, as expected, the condition that the metric
$g_{\hat{\alpha}\hat{\beta}}=\eta_{\hat{\alpha}\hat{\beta}}+\mathcal{O}(\|\mathbf{R}\|\hat{x}^{2})$
can be taken as nearly flat throughout the body.%
} (that amounts to taking $\lambda\ll1$ in Eq. (\ref{eq:Lambda}),
which is reasonable in this context, see Appendix \ref{sec:Momentum-and-angular in curved spacetime}
and footnote \ref{fn:Assumption}), the observers $u_{{\rm lab}}^{\alpha}$
will agree on the centroid position. If one starts with an observer
$u_{1}^{\alpha}\equiv u_{{\rm lab}}^{\alpha}(x_{1})$ at a point $x_{1}^{\alpha}$,
and computes the centroid it measures from Eq. (\ref{eq:XcmDef}),
the worldline $z^{\alpha}=x_{{\rm CM}}^{\alpha}(u_{{\rm 1}})$ obtained
will therefore obey the CP condition $S_{\ \beta}^{\alpha}(z)u_{{\rm lab}}^{\beta}(z)=0$,
since repeating the computation in the normal coordinates of the observer
$u_{{\rm lab}}^{\alpha}(z)$ at $z^{\alpha}$ yields the same result.
An example is when $u_{{\rm lab}}^{\alpha}$ are the observers associated
to a global inertial frame in flat spacetime. In this case $u_{{\rm lab}}^{\alpha}$
not only is the same vector everywhere, as one can set the Lorentz
frames of each observer $u_{{\rm lab}}^{\alpha}(x)$ in an hyperplane
$\Sigma$ to be the same up to spatial translations; so all observers
of this frame will measure the centroid at the same point (i.e. there
is a well defined, unique centroid associated to such frame). This
is an \emph{exact} statement in this case.

\emph{Momentum-velocity relation.}--- Since $u_{{\rm lab}}^{\alpha}$
is a well defined vector field in the region of interest, we may write
$Du_{{\rm lab}}^{\alpha}/d\tau=u_{{\rm lab}}^{\alpha;\beta}U_{\beta}$,
and therefore, from Eq.~(\ref{eq:PhidIGEM}), the momentum reads
\begin{equation}
P^{\alpha}=mU^{\alpha}+\frac{1}{\gamma}(h^{U})_{\ \sigma}^{\alpha}S_{\ \beta}^{\sigma}\left(-\gamma G^{\beta}-\epsilon_{\ \mu\gamma\delta}^{\beta}u_{{\rm lab}}^{\delta}U^{\mu}\omega^{\gamma}+\theta^{\beta\gamma}U_{\gamma}\right)\ ;\label{eq:MomentumCP}
\end{equation}
here $\gamma\equiv-u_{{\rm lab}}^{\alpha}U_{\alpha}$, $G^{\alpha}=-\nabla_{\mathbf{u}_{{\rm lab}}}u_{{\rm lab}}^{\alpha}$
is minus the acceleration of the laboratory observers (i.e., the gravitoelectric
field), $\omega^{\gamma}$ is their vorticity (or the Fermi-Walker
gravitomagnetic field \cite{CostaNatario:2012,JantzenCariniBini:1992}),
and $\theta_{\alpha\beta}$ their shear/expansion tensor. Hence we
have a well defined expression for $P^{\alpha}$ in terms of $U^{\alpha}$,
$S^{\alpha\beta}$ and the kinematics of the congruence $u_{{\rm lab}}^{\alpha}$,
telling us that $P^{\alpha}$ differs from $mU^{\alpha}$ only if
the laboratory observers measure inertial forces (i.e., if they are
accelerated, rotating, or shearing/expanding).

\subsubsection{The Tulczyjew-Dixon (TD) condition}

The condition $S^{\alpha\beta}P_{\beta}=0$ amounts to choosing $u^{\alpha}=P^{\alpha}/M$,
i.e., the centroid is the one as measured in the zero 3-momentum frame.
As shown in \cite{Beiglbock:1967,Schattner:1979}, for a given matter
distribution, described by the energy-momentum tensor $T^{\alpha\beta}(x)$,
there is only one worldline $z^{\alpha}(\tau)$ such that $S^{\alpha\beta}P_{\beta}=0$
($S^{\alpha\beta}$ and $P^{\alpha}$ being both evaluated at $z^{\alpha}$,
and using the hypersurface $\Sigma(z,P)$ orthogonal to $P^{\alpha}$
at $z^{\alpha}$). In other words, this spin condition \emph{specifies
an unique worldline}. It is the central worldline of the worldtube
of centroids, as can be seen from Eq. (\ref{eq:CMShift}). From (\ref{eq:Momentum})-(\ref{eq:HiddenMomentum}),
we have the expressions for the momentum 
\begin{equation}
P^{\alpha}=\frac{1}{m}\left(M^{2}U^{\alpha}+S^{\alpha\beta}\frac{DP_{\beta}}{d\tau}\right)=mU^{\alpha}+\frac{1}{m}(h^{U})_{\ \sigma}^{\alpha}S^{\sigma\beta}\frac{DP_{\beta}}{d\tau}\ .\label{eq:PTulczyjewDixon0}
\end{equation}
Here $DP^{\alpha}/d\tau$ is the force, given by Eq.~(\ref{eq:ForceDS0});
in the absence of electromagnetic field, and to pole-dipole order,
this expression can be manipulated into the well known expression
(e.g. \cite{Semerak:1999}) 
\begin{equation}
U^{\alpha}=\frac{m}{M^{2}}\left(P^{\alpha}+\frac{2S^{\alpha\nu}R_{\nu\tau\kappa\lambda}S^{\kappa\lambda}P^{\tau}}{4M^{2}+R_{\alpha\beta\gamma\delta}S^{\alpha\beta}S^{\gamma\delta}}\right)\ ,\label{eq:P_U_Dixon}
\end{equation}
determining $U^{\alpha}$ \emph{uniquely} in terms%
\footnote{The factor $m/M^{2}$ (involving $U^{\alpha}$ via $m$) can be determined
by the normalization condition $U^{\alpha}U_{\alpha}=-1$.%
} of $P^{\alpha}$, $S^{\alpha\beta}$ and $R_{\alpha\beta\gamma\delta}$.
A more general expression for the case when $F_{\alpha\beta}\ne0$,
and to arbitrary multipole order, is given in Eq.~(35) of \cite{Grallaetal:2010}.

\subsubsection{The Frenkel-Mathisson-Pirani (FMP) condition. Helical motions.\label{sub:The-Mathisson-Pirani-condition.}}

The condition $S^{\alpha\beta}U_{\beta}=0$, i.e., $u^{\alpha}=U^{\alpha}$,
states that the centroid is measured in its own rest frame; in other
words, it chooses the center of mass as measured by an observer comoving
with it. This condition does not yield an unique worldline though:
it is infinitely degenerate. For a given matter distribution, described
by $T^{\alpha\beta}(x)$, there are infinitely many worldlines $z^{\alpha}$
through it such that $S^{\alpha\beta}U_{\beta}=0$. Indeed, any point
within the disk of centroids can be a solution (i.e., can be a center
of mass as measured in its proper frame) provided that it moves with
the appropriate velocity. In order to see that, we start by an heuristic
argument, originally due to Möller \cite{Moller:AIH1949}: consider,
in Fig.~\ref{fig:CMShift}, a point in circular motion \emph{opposite}
to the rotation of the body, with a radius $R=v'S_{\star}/M$ such
that it passes through the centroid $x_{{\rm CM}}^{\alpha}(u')$ measured
by the observer $\mathcal{O}'$, and having \emph{therein} the same
velocity as $\mathcal{O}'$. Such point instantaneously coincides
with $x_{{\rm CM}}^{\alpha}(u')$, and at the same time is at rest
with respect to $\mathcal{O}'$; it is thus a center of mass computed
in its own rest frame, and will be so at every instant as the motion
is circular. The angular velocity of such points is constant, $\Omega=v'/R=M/S_{\star}$
(i.e., does not depend on $R$). That is, consider a disk of the same
size of the disk of centroids, \emph{rigidly} rotating about the centroid
$x_{{\rm CM}}^{\alpha}(P)$ measured by $\mathcal{O}$ ; any point
of such disk is a centroid computed in its rest frame, and is thus
a solution of $S^{\alpha\beta}U_{\beta}=0$. This is the origin of
the helical motions (in a frame moving with respect to $\mathcal{O}$,
the circular motions become helices).

These facts can be explicitly checked from the equations of motion.
First we note that, with this spin condition, the momentum becomes,
cf. Eq.~(\ref{eq:Momentum}), 
\begin{equation}
P^{\alpha}=mU^{\alpha}+S^{\alpha\beta}a_{\beta}=mU^{\alpha}+\epsilon_{\ \ \gamma\delta}^{\alpha\beta}a_{\beta}S^{\gamma}U^{\delta}\ ,\label{eq:momentumhelical}
\end{equation}
where $S^{\alpha}$ is the spin vector defined by 
\begin{equation}
S^{\alpha}=\frac{1}{2}\epsilon_{\ \beta\mu\nu}^{\alpha}S^{\mu\nu}U^{\beta};\qquad S^{\alpha\beta}=\epsilon^{\alpha\beta\mu\nu}S_{\mu}U_{\nu}\ .\label{eq:SpinVecMP}
\end{equation}
Noting from (\ref{eq:momentumhelical}) that $P^{\alpha}a_{\alpha}=P^{\alpha}S_{\alpha}=0$,
and using $S^{\alpha\beta}U_{\beta}=0$, the component of the 4-velocity
orthogonal to $P^{\alpha}$ is, from Eq.~(\ref{eq:UperpInertial}),
\begin{equation}
U_{\perp}^{\alpha}=-\frac{1}{M^{2}}\epsilon^{\alpha\beta\mu\nu}S_{\mu}P_{\nu}\frac{DU_{\beta}}{d\tau}\label{eq:UperpMathisson}
\end{equation}
which in the $P^{i}=0$ frame reads 
\begin{equation}
\vec{U}+\frac{1}{M}\frac{D\vec{U}}{d\tau}\times\vec{S}=0\ .\label{eq:MOmVelMPnonCov}
\end{equation}
This is a differential equation for the space components $\vec{U}$;
as discussed above, $\vec{v}=\vec{U}/\gamma(P,U)$ has the interpretation
of 3-velocity of the centroid in the $P^{i}=0$ frame. Take now for
simplicity the case of a free particle in flat spacetime; in this
case, from Eq.~(\ref{eq:ForceDS0}) we have $DP^{\alpha}/d\tau=0$;
also, from Eq.~(\ref{eq:Fermi-Walker}) below, it follows that $DS^{\alpha}/d\tau=0$
(since $S^{\alpha}a_{\alpha}=0$, which can be seen substituting (\ref{eq:momentumhelical})
in $DP^{\alpha}/d\tau=0$); thus $M$ and $\vec{S}$ in (\ref{eq:MOmVelMPnonCov})
are constants, and the solution for the reference worldline $z^{\alpha}$
($\vec{U}=d\vec{z}/d\tau$) is, in rectangular coordinates (taking
$\vec{S}$ along $\vec{e}_{z}$), 
\begin{equation}
z^{\alpha}(\tau)=\left(\gamma\tau,-R\cos\left(\frac{v\gamma}{R}\tau\right),R\sin\left(\frac{v\gamma}{R}\tau\right),0\right)\label{eq:PositionMathisson}
\end{equation}
where $R=v\gamma S/M$, and $\gamma\equiv\gamma(P,U)=-P^{\alpha}U_{\alpha}/M=\sqrt{1-v^{2}}$;
$v$ can take any value between 0 and 1. These are circular motions
of radius $R$ and frequency $\Omega=M/\gamma S$, centered about
the centroid $x_{{\rm CM}}^{\alpha}(P)$ measured in the $P^{i}=0$
frame. They may not seem at first the same motions we deduced from
the heuristic argument above; in particular, the fact $\gamma$ can
be arbitrarily large has led some authors to believe that the radius
of these motions, for a given body, can be arbitrary, and for this
reason deemed them unphysical \cite{Weyssenhoff:1946,WeyssenhoffRaabe:1947,Dixon:1965}.
That is not the case; the reason for that is that $S$ is different
for all the helical representations corresponding to the same body.
Let $z^{\alpha}$ and $z'^{\alpha}$ denote two different helical
solutions. The scalar $S=\sqrt{S^{\alpha}S_{\alpha}}=\sqrt{S^{\alpha\beta}S_{\alpha\beta}}/2$,
for a spin tensor obeying $S^{\alpha\beta}U_{\beta}=0$, is the magnitude
of the angular momentum taken about $z^{\alpha}=x_{{\rm CM}}^{\alpha}(U)$.
It should in be different, \emph{for the same matter distribution}
$T_{\alpha\beta}(x)$, from $S'=\sqrt{S'^{\alpha\beta}S'_{\alpha\beta}}/2$,
since $S'^{\alpha\beta}$, obeying $S'^{\alpha\beta}U'_{\beta}=0$,
is the angular momentum about a different point, $z'^{\alpha}=x_{{\rm CM}}^{\alpha}(U')$.
It is shown in Sec.~IV of \cite{Costaetal:2012} that, for all helical
motions, $S=S_{\star}/\gamma$, where $S_{\star}=\sqrt{S_{\star}^{\alpha\beta}S_{\star\alpha\beta}}/2$
is the magnitude of the angular momentum taken about $x_{{\rm CM}}^{\alpha}(P)$
(i.e., $S_{\star}^{\alpha\beta}P_{\beta}=0$). So indeed these motions
have a \emph{finite} radius and constant frequency, as deduced above:
\[
\Omega=\frac{M}{S_{\star}};\qquad R=\frac{vS_{\star}}{M}\ .
\]

Hence we see that the famous helical motions are just another exotic
effect generated by the variation, along $z^{\alpha}(\tau)$, of the
field of observers $u^{\alpha}$ ($=U^{\alpha}$, in this case) with
respect to which the centroid is computed; what is special about them
is that in this case the non-trivial motion induced on the centroid
is such that the latter is always at rest with respect to the observer
measuring it. Thus they are \emph{not} unphysical, contrary to some
claims in the literature; but they do not contain new physics either,
they are just alternative, unnecessarily complicated descriptions
for physical motions that can be described through simpler representations:
for example, the non-helical solution that this spin condition also
allows, which in the case of a free particle in flat spacetime is
uniform straight line motion (corresponding to $v=0$, $R=0$, in
Eq.~(\ref{eq:PositionMathisson}) above).

It is also worth noting that, from a dynamical perspective, the consistency
of the helical motions (namely, the fact the centroid accelerates
\emph{without any force}) is explained through an interchange between
kinetic momentum $P_{{\rm kin}}^{\alpha}=mU^{\alpha}$ and hidden
momentum $P_{{\rm hidI}}^{\alpha}=S^{\alpha\beta}a_{\beta}$, which
occurs in a way that their variations cancel out at every instant,
such that $P^{\alpha}=mU^{\alpha}+P_{{\rm hidI}}^{\alpha}$ remains
constant; see Fig.~3 of \cite{Costaetal:2012}. This is exactly the
same principle behind the bobbings due to $P_{{\rm hidI}}^{\alpha}$
discussed in Sec.~\ref{sub:Comparison-of-the} below.

\subsubsection*{Features of the FMP condition: Fermi-Walker transport and gravito-electromagnetic
analogies}

If one employs the Frenkel-Mathisson-Pirani condition, the spin vector
of a gyroscope (if $\tau^{\alpha\beta}=0$) is Fermi-Walker transported
along the worldlines of \emph{any of the centroids }obeying this condition.
This can easily be seen substituting Eq. (\ref{eq:SpinVecMP}) in
(\ref{eq:DSabdtHidden}) to obtain 
\begin{equation}
\frac{DS^{\alpha}}{d\tau}=S_{\nu}a^{\nu}U^{\alpha}\ .\label{eq:Fermi-Walker}
\end{equation}

This is the most natural description for the spin evolution, where
the mathematical definition of a locally non-rotating frame meets
the physical one: gyroscopes ``oppose'' to changes in direction
of their rotation axes; the axis of torque-free gyroscopes define
\emph{physically} the non-rotating frames. On the other hand, Fermi-Walker
transport is the \emph{mathematical} definition of a non-rotating
frame $\mathbf{e}_{\hat{\alpha}}$ adapted to an arbitrarily accelerated
observer: $\nabla_{\mathbf{U}}\mathbf{e}_{\hat{\beta}}=\Omega_{\,\,\hat{\beta}}^{\hat{\alpha}}\mathbf{e}_{\hat{\alpha}}$,
$\Omega^{\alpha\beta}=2U^{[\alpha}a^{\beta]}$; that is, it admits
``rotation'' (actually boost) in the time-space plane formed by
$U^{\alpha}$ and $a^{\alpha}$, unavoidable to keep the time axis
of the tetrad parallel to the 4-velocity ($\mathbf{U}=\mathbf{e}_{\hat{0}}$),
so that the triad $\mathbf{e}_{\hat{i}}$ spans the observer's local
rest space; but no additional spatial rotation (i.e., the axes $\mathbf{e}_{\hat{i}}$
orthogonal to $a^{\alpha}$ are parallel transported).

Another interesting feature of this spin condition is that is gives
rise to three \emph{exact} gravito-electromagnetic analogies \cite{CostaNatarioZilhao:2012,CostaNatario:2012}:
i) the spin-curvature force (penultimate term of Eq.~(\ref{eq:ForceDS0}))
becomes $F_{{\rm G}}^{\alpha}=-\mathbb{H}^{\beta\alpha}S_{\beta}$,
where $\mathbb{H}_{\alpha\beta}\equiv\star R_{\alpha\mu\beta\nu}U^{\mu}U^{\nu}$,
analogous to the force on a magnetic dipole (second term of Eq.~(\ref{eq:ForceDS0})),
$F_{{\rm EM}}^{\alpha}=B^{\beta\alpha}\mu_{\beta}$, where $B_{\alpha\beta}\equiv\star F_{\alpha\mu;\beta}U^{\mu}$;
ii) Eq.~(\ref{eq:Fermi-Walker}) becomes, in an orthonormal frame
``adapted'' to a congruence of observers, $dS^{\hat{i}}/d\tau=(\vec{S}\times\vec{H})^{\hat{i}}/2$,
where $\vec{H}$ is the ``gravitomagnetic field'', analogous to
the precession of a magnetic dipole, $D\vec{S}/d\tau=\vec{\mu}\times\vec{B}$
(first term of Eq.~(\ref{eq:tauDEM})); iii) the inertial hidden
momentum, cf. Eq.~(\ref{eq:momentumhelical}), is $P_{{\rm hidI}}^{\alpha}=\epsilon_{\ \beta\gamma\delta}^{\alpha}U^{\delta}S^{\beta}G^{\gamma}$,
with $G^{\alpha}=-a^{\alpha}$ the ``gravitoelectric'' field as
measured in the centroid frame, formally analogous to the electromagnetic
hidden momentum, Eq.~(\ref{eq:PhidEMMP}) below. These analogies
(apart from their theoretical interest) provide useful insight to
study some problems; they are discussed in detail in \cite{CostaNatarioZilhao:2012}.

The downside of this condition is the fact that it is not always easy
to set up the non-helical solution. It is done through suitable ansatzs
in Sec. \ref{sub:Comparison-of-the} below (at an approximate level),
or at an exact level, in \emph{very special} systems, in \cite{CostaNatarioZilhao:2012}
(therein it is seen to be a good choice, as it takes advantage of
the symmetries of the problems to yield the simplest equations). Prescriptions
in the case of Schwarzschild and Kerr spacetimes are also proposed
in \cite{Plyatsko:2008,Plyatsko:2011}; however no general rule is
known.

\subsubsection{The Ohashi-Kyrian-Semerák (OKS) spin condition\label{sub:The-Ohashi-Kyrian-Semerak}}

This condition, introduced in \cite{Ohashi:2003}, and first discussed
in depth in \cite{KyrianSemerak:2007}, amounts to choosing a vector
field $u^{\alpha}$ parallel transported along $z^{\alpha}(\tau)$,
$Du^{\alpha}/d\tau=0$, which causes the inertial hidden momentum
$P_{{\rm hidI}}^{\alpha}$ and its associated gauge motions to vanish,
cf. Eq.~(\ref{eq:HiddenInertial}). In the general case where the
torque tensor $\tau^{\alpha\beta}$ is non-zero, as we shall see in
Sec. \ref{sub:gauge-hidden-torque} below, some superfluous motions
may still be present though, due to the pure gauge part of the hidden
momentum $P_{{\rm hid\tau}}^{\alpha}$ related to $\tau^{\alpha\beta}$
(in gravitational systems, $P_{{\rm hid\tau}}^{\alpha}$ is usually
less important, as it involves the particle's quadrupole moment).
When $\tau^{\alpha\beta}=0$ (the problem at hand herein), it yields
the simplest momentum velocity relation possible, $P^{\alpha}=mU^{\alpha}$,
and a centroid that accelerates only when there is a force, $ma^{\alpha}=(h^{U})_{\ \beta}^{\alpha}F^{\beta}$
(this becomes $ma^{\alpha}=F^{\alpha}$ for pole-dipole particles
in gravitational fields, since $m=M\equiv\sqrt{-P^{\alpha}P_{\alpha}}$
is constant, as readily seen contracting Eqs. (\ref{eq:ForceDS0})
or (\ref{eq:SpinCurvatureForce}) with $P^{\alpha}$). Eq.~(\ref{eq:DSabdtHidden})
also takes a simple form, yielding a spin tensor $S^{\alpha\beta}$
parallel transported along $z^{\alpha}(\tau)$, $DS^{\alpha\beta}/d\tau=0$.

This condition obviously does not specify an unique worldline through
the body; it is infinitely degenerate, because there are infinite
possible choices of $u^{\alpha}$ (the only restriction imposed is
$Du^{\alpha}/d\tau=0$); but another of its advantages \cite{KyrianSemerak:2007}
is that one does not need%
\footnote{We thank O. Semerák and A. Harte for discussions on these issues.%
} to explicitly determine $u^{\alpha}$ to solve the equations of motion
(for pole-dipole particles), only its value at the initial point is
needed. These properties together make this condition the most suitable
(at least in that case) for numerical implementation.

\subsubsection{Uniqueness of the centroid vs determinacy of the equations}

There are some \emph{apparent} contradictions in the literature regarding
the uniqueness of the worldline specified by the different spin conditions,
and what that means in terms of the determinacy of the equations of
motion. On the one hand most authors (e.g. \cite{Moller:AIH1949,Tulczyjew:1959,Dixon:1970,KyrianSemerak:2007,Grallaetal:2010})
argue, in agreement with the discussion above, that the FMP condition
does not uniquely specify a worldline through the body; on the other
hand, it has recently been argued \cite{KudryashovaObukhov:2010,ObukhovPuetzfeld:2011}
that it uniquely specifies the motion, given certain initial conditions.
Also, in \cite{Tulczyjew:1959,KyrianSemerak:2007}, it is said that
the CP condition yields an unique solution, whereas in the analysis
above we have seen that, depending on the coordinate system chosen,
it may or may not yield an unique center of mass. Our considerations
above are based on starting with a test body whose matter distribution
is described by an energy-momentum tensor $T^{\alpha\beta}(x)$, and
asking the following question: given $T^{\alpha\beta}(x)$, does the
condition $S^{\alpha\beta}u_{\beta}=0$ yield an unique worldline?
As we have seen, from the four conditions studied above, the answer
is affirmative, as a general statement, only for the TD condition.

But if one takes the perspective of the the initial value problem
for the equations of motion (\ref{eq:ForceDS0})-(\ref{eq:DSabdtHidden}),
the impact of the uniqueness/non-uniqueness of the center of mass
definition is not straightforward. First of all one should notice
that, without further assumptions, the system (\ref{eq:ForceDS0})-(\ref{eq:DSabdtHidden}),
supplemented by (\ref{eq:GenSpinCondition}), can be determined only
to dipole order and if $F^{\alpha\beta}=0$ (otherwise one needs evolution
laws for $\mu^{\alpha\beta}$, $d^{\alpha}$, and the higher order
electromagnetic and gravitational moments). In this case, all the
conditions yield a well defined solution if sufficient initial conditions
are provided; and it is \emph{the type of initial data} needed to
determine the equations that depends on the nature of center of mass
definition given by each of the conditions.

On general grounds one can say that if the equations of motion can
be written as the \emph{explicit} functions (dot denotes ordinary
derivative along $\mathbf{U}$) 
\[
\dot{z}^{\alpha}(\tau)\equiv U^{\alpha}=U^{\alpha}(\mathbf{z},\mathbf{P},S^{\mu\nu});\qquad\dot{P}^{\alpha}=f^{\alpha}(\mathbf{z},\mathbf{P},S^{\mu\nu});\qquad\dot{S}^{\alpha\beta}=g^{\alpha\beta}(\mathbf{z},\mathbf{P},S^{\mu\nu})
\]
then, given the initial values $\{z^{\alpha},P^{\alpha},S^{\alpha\beta}\}|_{{\rm in}}$,
the system is determined. The first equation is the \emph{explicit}
velocity-momentum relation; but the other two also require such a
relation, as can be seen by writing explicitly 
\[
\dot{P}^{\alpha}=\Gamma_{\nu\mu}^{\alpha}P^{\mu}U^{\nu}-\frac{1}{2}R_{\ \beta\gamma\delta}^{\alpha}U^{\beta}S^{\gamma\delta};\qquad\dot{S}^{\alpha\beta}=2\Gamma_{\nu\mu}^{[\alpha}S^{\beta]\mu}U^{\nu}+2P^{[\alpha}U^{\beta]}.
\]
Thus, in order to have $\dot{P}^{\alpha}$ and $\dot{S}^{\alpha\beta}$
as explicit functions of $(\mathbf{z},\mathbf{P},S^{\mu\nu})$, we
need to have an explicit relation $U^{\alpha}(\mathbf{z},\mathbf{P},S^{\mu\nu})$.

In the case of the OKS condition, since one has simply $U^{\alpha}=P^{\alpha}/m=P^{\alpha}/M$,
cf.~Sec.~\ref{sub:The-Ohashi-Kyrian-Semerak}, the statements above
obviously hold, and the solution is determined given $\{z^{\alpha},P^{\alpha},S^{\alpha\beta}\}|_{{\rm in}}$,
or, equivalently, $\{z^{\alpha},S^{\alpha\beta},U^{\alpha},m\}|_{{\rm in}}$.

The situation is similar for the TD condition (only with a more complicated
velocity-momentum relation). Eq.~(\ref{eq:P_U_Dixon}) is an explicit
relation $U^{\alpha}(\mathbf{z},\mathbf{P},S^{\mu\nu})$; thus, given
$\{z^{\alpha},P^{\alpha},S^{\alpha\beta}\}|_{{\rm in}}$, the solution
is determined. The initial data $\{z^{\alpha},S^{\alpha\beta},U^{\alpha},m\}|_{{\rm in}}$
is equally sufficient because one can extract $P^{\alpha}|_{{\rm in}}$
from Eq.~(\ref{eq:PTulczyjewDixon0}) (substituting therein $DP^{\alpha}/d\tau$
by the explicit expression $-R_{\ \beta\gamma\delta}^{\alpha}U^{\beta}S^{\gamma\delta}/2$;
an equation for $M^{2}$ is obtained by squaring (\ref{eq:PTulczyjewDixon0})).

The case of the CP condition is also essentially similar. One obtains
an explicit relation $U^{\alpha}(\mathbf{z},\mathbf{P},S^{\mu\nu})$
as follows%
\footnote{We thank O. Semerák for his input on this issue.%
}. Substitute decomposition (\ref{eq:Kinematics}) into Eq. (\ref{eq:Momentum}),
with $u^{\alpha}=u_{{\rm lab}}^{\alpha}$, to obtain 
\begin{equation}
P^{\alpha}=\frac{-P_{\alpha}u_{{\rm lab}}^{\alpha}}{\gamma}U^{\alpha}+\frac{1}{\gamma}S_{\ \beta}^{\alpha}\left(-\gamma G^{\beta}-\epsilon_{\ \mu\gamma\delta}^{\beta}u_{{\rm lab}}^{\delta}U^{\mu}\omega^{\gamma}+\theta^{\beta\gamma}U_{\gamma}\right)\ ,\label{eq:MomentumCP2}
\end{equation}
where $\gamma\equiv-U_{\alpha}u_{{\rm lab}}^{\alpha}$. This is an
equation for $P^{\alpha}$ in terms of $U^{\alpha}$, $S^{\alpha\beta}$,
and the quantities $G^{\alpha}$, $\omega^{\alpha}$, and $\theta_{\alpha\beta}$
which are given in advance (see Sec. \ref{sub:The-Corinaldesi-Papapetrou-(CP)}
and the equivalent Eq. (\ref{eq:MomentumCP})). We need now to solve
for $U^{\alpha}$. Expressing (\ref{eq:MomentumCP2}) in its components
in a frame where $u_{{\rm lab}}^{i}=0$, we obtain 
\[
A_{\ k}^{i}v^{k}=P^{i}+S_{\ j}^{i}G^{j}\ ,\qquad A_{\ k}^{i}\equiv\left[P^{0}\delta_{\ k}^{i}-S_{\ j}^{i}\left(\epsilon_{\ kl}^{j}\omega^{l}-\theta_{\ k}^{j}\right)\right]\ ,
\]
where $v^{i}=U^{i}/U^{0}$ is the centroid velocity in the $u_{{\rm lab}}^{i}=0$
frame. This is a system of linear equations for the three components
$v^{k}$, with solution $v^{i}=[A^{-1}]_{\ k}^{i}[P^{k}+S_{\ j}^{k}G^{j}]$.
The component $U^{0}$ (and subsequently, $U^{i}$) is then obtained
from the normalization condition $-1=U^{\alpha}U_{\alpha}=-(U^{0})^{2}(1-v^{2})$.
We thus end up with an explicit relation $U^{\alpha}(\mathbf{z},\mathbf{P},S^{\mu\nu})$,
meaning that, given the initial values $\{z^{\alpha},S^{\alpha\beta},P^{\alpha}\}|_{{\rm in}}$,
the solution is determined, as asserted in \cite{KyrianSemerak:2007}.
The set $\{z^{\alpha},S^{\alpha\beta},U^{\alpha},m\}|_{{\rm in}}$
is also sufficient, in agreement with the claim in \cite{Tulczyjew:1959},
because one immediately obtains $P^{\alpha}|_{{\rm in}}$ from (\ref{eq:MomentumCP}).
Finally, note that this is a distinct problem from the one addressed
in Sec.~\ref{sub:The-Corinaldesi-Papapetrou-(CP)} (where we started
just with a matter distribution $T^{\alpha\beta}(x)$ and imposed
$S_{\ \beta}^{\alpha}u_{{\rm lab}}^{\beta}=0$, in which case, as
we have seen, the solution always exists but in general is \emph{not}
unique). Herein one \emph{assumes} the existence of some $T^{\alpha\beta}(x)$
that is compatible with the initial conditions prescribed (conversely,
in the prescription of Sec. \ref{sub:The-Corinaldesi-Papapetrou-(CP)}
there is no longer freedom to choose an arbitrary initial position
$z^{\alpha}$).

The case of the FMP condition has some important differences. The
momentum-velocity relation is (\ref{eq:momentumhelical}); the acceleration
can be written as (cf.~Eq.~(24) of \cite{KyrianSemerak:2007}) 
\begin{equation}
a^{\alpha}(\mathbf{z},\mathbf{U},S^{\mu\nu})=\frac{1}{S^{2}}\left(\frac{1}{m}F^{\mu}S_{\mu}S^{\alpha}-P_{\gamma}S^{\alpha\gamma}\right)\label{eq:ExplicitaFMP}
\end{equation}
with $S^{\alpha}$ defined by (\ref{eq:SpinVecMP}), $S^{2}\equiv\sqrt{S^{\alpha}S_{\alpha}}$,
and $F^{\alpha}\equiv DP^{\alpha}/d\tau$. Substituting in (\ref{eq:momentumhelical})
one obtains an explicit relation $P^{\alpha}(\mathbf{z},\mathbf{U},S^{\mu\nu})$.
However, one cannot a priori guarantee that such relation can be inverted
into a relation $U^{\alpha}(\mathbf{z},\mathbf{P},S^{\mu\nu})$ (such
problem, in the general case, has not yet been tackled in the literature,
to the authors' knowledge). In the \emph{special case of a free particle
in flat spacetime}, we have, from (\ref{eq:momentumhelical}), 
\begin{equation}
a^{\alpha}(\mathbf{P},S^{\mu\nu})=-\frac{S^{\alpha\beta}P_{\beta}}{S^{2}};\qquad U^{\alpha}(\mathbf{P},S^{\mu\nu})=\frac{1}{m}\left(P^{\alpha}+\frac{1}{S^{2}}S^{\alpha\mu}S_{\mu\beta}P^{\beta}\right).\label{eq:aUMPflat}
\end{equation}
This is an explicit relation $U^{\alpha}(\mathbf{P},S^{\mu\nu})$
($m$ can be determined through the condition $U^{\alpha}U_{\alpha}=-1$);
therefore, in agreement with the claims in \cite{KudryashovaObukhov:2010,ObukhovPuetzfeld:2011},
the motion is indeed determined given the initial data $\{z^{\alpha},S^{\alpha\beta},P^{\alpha}\}|_{{\rm in}}$
(i.e., this set of data specifies one particular solution of the \emph{degenerate}
condition $S^{\alpha\beta}U_{\beta}=0$). On the other hand (unlike
the situation for the other three spin conditions), the set of data
$\{z^{\alpha},S^{\alpha\beta},U^{\alpha},m\}|_{{\rm in}}$ is \emph{not}
enough; one needs, additionally, the initial acceleration $a^{\alpha}$,
in agreement with the claims in e.g. \cite{Dixon:1964,Grallaetal:2010}.
This is clear from Eqs.~(\ref{eq:momentumhelical}), (\ref{eq:aUMPflat}):
the set of initial data $\{S^{\alpha\beta},P^{\alpha}\}|_{{\rm in}}$
\emph{is equivalent} to $\{S^{\alpha\beta},U^{\alpha},m,a^{\alpha}\}|_{{\rm in}}$.
These features are readily understood in the framework of the discussion
in Sec.~\ref{sub:The-Mathisson-Pirani-condition.}: as we have seen,
the motion of an helical solution $z^{\alpha}=x_{{\rm CM}}^{\alpha}(U)$
is a superposition of a circular motion centered at the centroid measured
in the $P^{i}=0$ frame, $x_{{\rm CM}}^{\alpha}(P)$, of radius $R=\|\Delta\mathbf{x}\|$
and angular velocity $\vec{\omega}=-M\vec{S}_{\star}/S_{\star}^{2}$,
combined with a boost of 4-velocity $P^{\alpha}/M$. $\Delta x^{\alpha}=z^{\alpha}-x_{{\rm CM}}^{\alpha}(P)$
is the shift of $z^{\alpha}$ relative to the center of the helix.
Given $z_{{\rm in}}^{\alpha}$, $S_{{\rm in}}^{\alpha\beta}$, and
$P^{\alpha}$, one obtains $x_{{\rm CM}}^{\alpha}(P)|_{{\rm in}}=z_{{\rm in}}^{\alpha}-\Delta x_{{\rm in}}^{\alpha}$
from the expression $\Delta x^{\alpha}=S^{\alpha\beta}P_{\beta}/M^{2}$,
cf. Eq. (\ref{eq:Shiftuu'Gen}); $S_{\star}^{\alpha\beta}$ follows
using $S^{\alpha\beta}=S_{\star}^{\alpha\beta}+2P^{[\alpha}\Delta x^{\beta]}$,
and therefore the motion is completely determined. On the other hand,
if instead of $P^{\alpha}|_{{\rm in}}$ one is given $\{U^{\alpha},m\}|_{{\rm in}}$,
one cannot determine $\Delta x^{\alpha}$; that is the reason why
one needs the acceleration, as it contains precisely the same information:
$a^{\alpha}=-\Delta x^{\alpha}M^{2}/S^{2}$, cf. Eq. (\ref{eq:aUMPflat}a).

\subsection{The dependence of the spin-curvature force on the spin condition;
equivalence of the spin conditions.\label{sub:The-dependence-of-the-force}}

\begin{figure}
\includegraphics[width=1\textwidth]{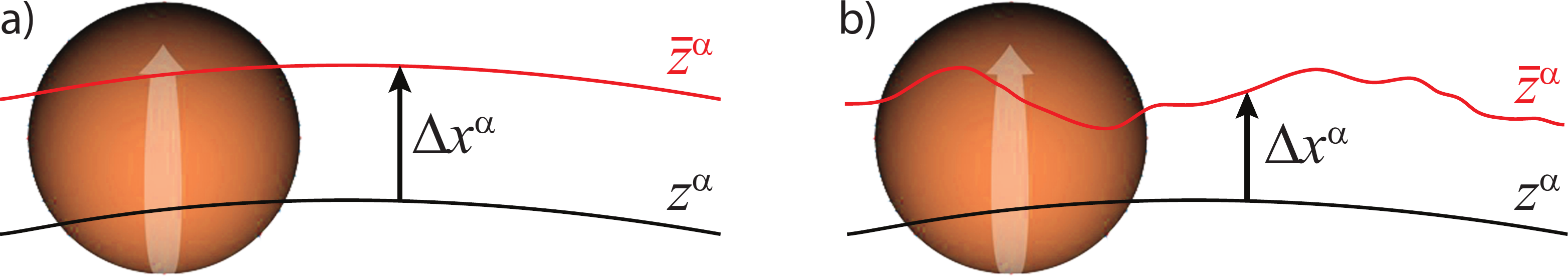}

\caption{\label{fig:Equivalence}a) Two different centroids of the OKS condition
move nearly parallel ($D\Delta x^{\alpha}/d\tau=0$). In a curved
spacetime, that implies that the force $DP^{\alpha}/d\tau$ along
the two worldlines must be different (e.g., if $z^{\alpha}(\tau)$
is a geodesic, $\bar{z}^{\alpha}(\bar{\tau})$ cannot be). b) Centroids
$\bar{z}^{\alpha}(\bar{\tau})$ of other spin conditions accelerate
relative to OKS centroids due to the gauge motions induced by the
variation of $u^{\alpha}$ along $\bar{z}^{\alpha}$ (inertial hidden
momentum). }
\end{figure}

We have seen that the significance of the spin condition $S^{\alpha\beta}u_{\beta}=0$
is that of a choice of representative worldline $z^{\alpha}$ in the
body, more precisely requiring such worldline to be, at each event,
the center of mass as measured by an observer of 4-velocity $u^{\alpha}$.
We have thereby implied that the different spin conditions yield different,
but \emph{equivalent} descriptions of the motion of a given body,
all contained within the worldtube of centroids, which in turn lies
within the convex hull of the body's worldtube. That is easy to see
for a free particle in flat spacetime, where indeed the different
solutions stay close forever and within the straight worldtube depicted
in Fig. \ref{fig:Figure-TWO}. However, when external non-homogenous
fields are present, changing $z^{\alpha}$ means not only changing
the point where the fields (i.e., $F^{\alpha\beta}$ and $R_{\alpha\beta\gamma\delta}$)
are evaluated, but also changing the moments ($S^{\alpha\beta}$,
$\mu_{\alpha\beta}$, $d^{\alpha}$, and the $2^{N>1}$ moments) themselves,
on which the forces and torques also depend. These two changes would
in principle compensate each other; the larger part of the compensation
comes from the lower order terms, and a smaller part (negligible to
some extent) from the higher order terms. Hence, in an approximation
where only moments up to $2^{N}$th order are kept, the different
worldlines will eventually diverge. However, this does not mean that
the spin condition is not a gauge choice after all; in fact, it just
marks the limit of validity of the given approximation \cite{KyrianSemerak:2007}.
The subtlety involved in this compensation is that, except for the
case of flat spacetime, it does \emph{not} mean that the force is
the same for different choices of $z^{\alpha}$.

In order to see this, let us consider first, in Newtonian mechanics,
the problem of describing an extended body through different reference
points; for more details on this problem, we refer to Sec. 3 of \cite{Dixon:1973}.
Consider a spherical body in a gravitational field $\vec{G}(\mathbf{x})$.
If one takes $z^{i}$ to be the body's center of mass $z^{i}=x_{{\rm CM}}^{i}\equiv\int\rho x^{i}d^{3}x/m$
--- a unique point, in Newtonian mechanics --- then, with respect
to $z^{i}$, the body is effectively a monopole, and the only force
present is the usual (monopole) gravitational force $\vec{F}=\vec{F}_{{\rm g}}=m\vec{G}(\mathbf{z})$.
Now take a different reference point $\bar{z}^{i}=z^{i}+\Delta x^{i}$
(not a centroid) say, at the boundary of the sphere. The monopole
force changes to $\vec{\bar{F}}_{{\rm g}}=m\vec{G}(\mathbf{z}+\Delta\mathbf{x})$;
but, on the other hand, the particle has a mass dipole moment $\vec{d}_{{\rm G}}=\int\rho\vec{x}d^{3}x=-m\Delta\vec{x}$
about $\bar{z}^{i}$ (as well as quadrupole and higher order moments).
The dipole force is $\bar{F}_{{\rm dip}}^{i}=\nabla_{j}G^{i}d_{{\rm G}}^{j}$;
hence to dipole order, we have the \emph{same net} Newtonian force:
\begin{equation}
\bar{F}^{i}=\bar{F}_{{\rm g}}^{i}+\bar{F}_{{\rm dip}}^{i}=mG^{i}(\mathbf{z}+\Delta\mathbf{x})-m\nabla_{j}G^{i}(\mathbf{z}+\Delta\mathbf{x})\Delta x^{j}\simeq mG^{i}(\mathbf{z})=F^{i}\ .\label{eq:F_bar_Newt}
\end{equation}
In General Relativity the situation is different because the lowest
order gravitational force is the (dipole order) spin-curvature force
\begin{equation}
F^{\alpha}=-\frac{1}{2}R_{\ \beta\mu\nu}^{\alpha}S^{\mu\nu}U^{\beta}={\displaystyle \star R_{\beta\nu\mu}^{\ \ \ \ \alpha}U^{\mu}S^{\beta}u^{\nu}\ ,}\label{eq:SpinCurvatureForce}
\end{equation}
cf. Eq. (\ref{eq:ForceDS0}), which depends explicitly on the spin
condition $S^{\alpha\beta}u_{\beta}=0$, i.e., on the choice of the
centroid $z^{\alpha}$. Such dependence is not compensated by a change
in the monopole force (which does not exist), nor by the higher order
terms (if that was the case, the pole-dipole approximation would not
even make sense, for, as we shall see in Sec. \ref{sub:Gravitational-system},
the differences in the force under different spin conditions are of
the same order of magnitude as the force itself). Hence, the \emph{net
force} $F^{\alpha}=\nabla_{\mathbf{U}}P^{\alpha}$ is different for
different $z^{\alpha}$'s, which is natural in a curved spacetime,
since the differentiation is along different curves. On the other
hand, although the monopole force $\vec{F}_{{\rm g}}$ (or $\vec{G}$)
has no \emph{physical} existence in the relativistic theory, there
is a counterpart to the \emph{tidal forces }arising from the variation
of these fields from point to point, $\nabla_{j}G_{i}$, which comes
from the curvature tensor (see below). And the crucial point here
is that the change in the force $F^{\alpha}$ when one changes $z^{\alpha}$
is precisely the one needed to compensate for the tidal forces which
``try'' to make the worldlines diverge.

This can be formalized as follows. Take two different centroids with
worldlines $z^{\alpha}$ and $\bar{z}^{\alpha}$, defined by $S^{\alpha\beta}u_{\beta}=0$
and $\bar{S}^{\alpha\beta}\bar{u}_{\beta}=0$, respectively. $S^{\alpha\beta}$
is the angular momentum about $z^{\alpha}$ and $\bar{S}^{\alpha\beta}$
the angular momentum about $\bar{z}^{\alpha}$. Extend (in a region
small enough so that they do not intersect) these worldlines to a
congruence of curves encompassing both $z^{\alpha}$ and $\bar{z}^{\alpha}$;
take them to be infinitesimally close, so that one can employ the
usual first order deviation equations (Eq. (\ref{eq:Deviation}) below),
and write a connecting vector as $\Delta x^{\alpha}=\bar{z}^{\alpha}(\tau)-z^{\alpha}(\tau)$.
Take moreover $u^{\alpha}$ to be parallel transported along $z^{\alpha}$
(i.e., it obeys Ohashi-Kyrian-Semerák spin condition), so that $P_{{\rm hidI}}^{\alpha}=0\Rightarrow P^{\alpha}=MU^{\alpha}$,
and let the field $\bar{u}^{\alpha}$ be arbitrary. Noting that $P^{\alpha}$
can be taken as the same for $z^{\alpha}$ and $\bar{z}^{\alpha}$
(see Appendix \ref{sec:AppendixHypersurfaces}), it follows from Eqs.
(\ref{eq:HiddenMomentum}) that $P^{\alpha}=\bar{m}\bar{U}^{\alpha}+\bar{P}_{{\rm hid}}^{\alpha}$,
where $\bar{m}\equiv-P^{\alpha}\bar{U}_{\alpha}$; contracting with
$P_{\alpha}$ to obtain an expression for $\bar{m}/M$, and using
Eqs. (\ref{eq:Udecomp}), one obtains 
\begin{equation}
\bar{U}^{\alpha}=\frac{P^{\alpha}}{M}+\left(\sqrt{1+\frac{\bar{P}_{{\rm hid}}^{\alpha}P_{\alpha}}{M^{2}}}-1\right)\frac{P^{\alpha}}{M}+\bar{U}_{\perp}^{\alpha}=\frac{P^{\alpha}}{M}+\left[\mbox{terms in }\bar{U}_{\perp}^{\alpha}\right].\label{eq:barU_P}
\end{equation}
$U_{\perp}^{\alpha}$ and $P_{{\rm hid}}^{\alpha}=P_{{\rm hidI}}^{\alpha}$
are gauge and reciprocal quantities; one can write one in terms of
the other using Eqs. (\ref{eq:UperpPhid}). $\bar{U}^{\alpha}\ne P^{\alpha}/M$
only if $\bar{U}_{\perp}^{\alpha}\ne0$ (or equivalently if $P_{{\rm hid}}^{\alpha}\ne0$).
From the deviation equation for accelerated worldlines \cite{BiniStrains:2006},
we have 
\begin{align}
\frac{D^{2}\Delta x^{\alpha}}{d\tau^{2}}= & -\mathbb{E}^{\alpha\beta}\Delta x_{\beta}+\nabla_{\Delta\mathbf{x}}a^{\alpha}=-\mathbb{E}^{\alpha\beta}\Delta x_{\beta}+\left(\nabla_{\bar{\mathbf{U}}}\bar{U}^{\alpha}-\nabla_{\mathbf{U}}U^{\alpha}\right)\nonumber \\
= & -\mathbb{E}^{\alpha\beta}\Delta x_{\beta}+\frac{1}{M}\left(\bar{F}^{\alpha}-F^{\alpha}\right)+\left[\mbox{terms in }\bar{U}_{\perp}^{\alpha}\right]\ ,\label{eq:Deviation}
\end{align}
where $\mathbb{E}_{\alpha\beta}\equiv R_{\alpha\mu\beta\nu}U^{\mu}U^{\nu}$
is the ``gravitoelectric'' tidal tensor, which is the relativistic
counterpart of the Newtonian tidal tensor $\nabla_{j}G_{i}$. In the
third equality we used Eq. (\ref{eq:barU_P}) and the following: $M$
is a conserved quantity for the OKS spin condition ($\nabla_{\mathbf{U}}M=0$
along $z^{\alpha}$ if $P^{\alpha}\parallel U^{\alpha}$, as readily
seen contracting (\ref{eq:SpinCurvatureForce}) with $P^{\alpha}$),
so that $F^{\alpha}=DP^{\alpha}/d\tau=Ma^{\alpha}$; and that along
$\bar{z}^{\alpha}$ one has $\nabla_{\bar{\mathbf{U}}}M=0+\left[\mbox{terms in }\bar{U}_{\perp}^{\alpha}\right]$.

Since $\Delta x^{\alpha}$ is infinitesimal, we can write (as in flat
spacetime), 
\begin{equation}
\bar{S}^{\alpha\beta}=S^{\alpha\beta}+2P^{[\alpha}\Delta x^{\beta]}\label{eq:SbarS}
\end{equation}
and therefore the difference between the forces is 
\begin{align}
\bar{F}^{\alpha}-F^{\alpha} & =-\frac{1}{2M}R_{\ \beta\gamma\delta}^{\alpha}P^{\beta}\left(\bar{S}^{\gamma\delta}-S^{\gamma\delta}\right)+\left[\mbox{terms in }\bar{U}_{\perp}^{\alpha}\right]\nonumber \\
 & =M\mathbb{E}^{\alpha\beta}\Delta x_{\beta}+\left[\mbox{terms in }\bar{U}_{\perp}^{\alpha}\right]\label{eq:DF}
\end{align}
where the terms in $\bar{U}_{\perp}^{\alpha}$ are of order $\mathcal{O}(S^{2})$.
Substituting in (\ref{eq:Deviation}), we obtain 
\[
\frac{D^{2}\Delta x^{\alpha}}{d\tau^{2}}=0+\left[\mbox{terms in }\bar{U}_{\perp}^{\alpha}\right]\ .
\]

That is, the worldline deviation of the two solutions reduces to terms
involving $U_{\perp}^{\alpha}$ (i.e., $P_{{\rm hidI}}^{\alpha}$),
that we have seen in Sec. \ref{sub:Inertial-hidden-momentum:DecouplingUP}
to be gauge (arising just from the choice of observers relative to
which the centroids are computed). This is illustrated in Fig. \ref{fig:Equivalence}.
In particular, if one takes two different solutions of the OKS condition
(so that no superfluous motions come into play%
\footnote{Only when $\nabla_{\mathbf{U}}u^{\alpha}=\nabla_{\bar{\mathbf{U}}}\bar{u}^{\alpha}=0$
should one expect two different centroids of the same body to move
parallel, even in flat spacetime, as explained in Sec.~\ref{sub:Inertial-hidden-momentum:DecouplingUP}
(see also Fig.~\ref{fig:Figure-TWO}b)). Otherwise (i.e., when $P_{{\rm hidI}}^{\alpha}\ne0$)
they can have an arbitrary relative motion, cf. Fig. \ref{fig:Equivalence}.%
}) we have simply $D^{2}\Delta x^{\alpha}/d\tau^{2}=0$, i.e., there
is no relative acceleration between the worldlines, which is \emph{guaranteed}
by the difference between the forces $F^{\alpha}=\nabla_{\mathbf{U}}P^{\alpha}$
and $\bar{F}^{\alpha}=\nabla_{\bar{\mathbf{U}}}P^{\alpha}$.

The situation becomes especially enlightening (and the correspondence
with the Newtonian theory closer) in the limit of weak \emph{static}
fields and slow motion of Sec.~\ref{sub:Comparison-of-the}. In this
case the \emph{coordinate} acceleration (for a stationary field) of
the centroid $z^{\alpha}$ is 
\[
m\frac{d^{2}z^{i}}{d\tau^{2}}=mG^{i}(\mathbf{z})+F^{i}-\frac{DP_{{\rm hid}}^{i}}{d\tau}
\]
where $\vec{G}$ is the Newtonian field (more precisely, a fictitious,
or ``inertial'' field, that mimics Newton's $\vec{G}$ in the coordinate
acceleration. Is is also known as the ``gravitoelectric'' field,
e.g. \cite{CostaNatario:2012}). The coordinate acceleration of the
centroid $\bar{z}^{\alpha}$ is 
\[
m\frac{d^{2}\bar{z}^{i}}{d\bar{\tau}^{2}}=mG^{i}(\bar{\mathbf{z}})+\bar{F}^{i}-\frac{D\bar{P}_{{\rm hid}}^{i}}{d\bar{\tau}}=mG^{i}(\bar{\mathbf{z}})+F^{i}+m\mathbb{E}^{ij}\Delta x_{j}-\frac{D\bar{P}_{{\rm hid}}^{i}}{d\bar{\tau}};
\]
in the second equality we used (\ref{eq:DF}) neglecting the $U_{\perp}^{\alpha}$
terms therein (as they are of order $\mathcal{O}(S^{2})$). To first
order in $\Delta\mathbf{x}$, $G^{i}(\bar{\mathbf{z}})\simeq G^{i}(\mathbf{z})+\nabla^{j}G^{i}\Delta x_{j}$;
and since, for a stationary field, to linear order (see \cite{CostaNatario:2012}),
$\mathbb{E}_{ij}=-\nabla_{j}G_{i}$, then $mG^{i}(\bar{\mathbf{z}})+\bar{F}^{i}=mG^{i}(\mathbf{z})+F^{i}$,
i.e., the sum of the spin curvature and the Newtonian forces is the
same for both worldlines, the change in one compensating for the other,
just like the case with the monopole and dipole forces in the Newtonian
problem above, cf. Eq. (\ref{eq:F_bar_Newt}). We have thus 
\begin{equation}
m\frac{d^{2}\bar{z}^{i}}{d\bar{\tau}^{2}}=mG^{i}(\mathbf{z})+F^{i}-\frac{D\bar{P}_{{\rm hid}}^{i}}{d\bar{\tau}}=m\frac{d^{2}z^{i}}{d\tau^{2}}+\frac{DP_{{\rm hid}}^{i}}{d\tau}-\frac{D\bar{P}_{{\rm hid}}^{i}}{d\bar{\tau}}\ .\label{eq:CoordAccelzz'}
\end{equation}
Hence, barring hidden momentum terms, the coordinate acceleration
for $\bar{z}^{\alpha}(\bar{\tau})$ is the same as for $z^{\alpha}(\tau)$.
This means, in particular, that the different solutions of the OKS
condition are trajectories that run parallel (as both the coordinate
acceleration and the velocity are the same for all of them).

\subsection{Comparison of the spin conditions in simple examples\label{sub:Comparison-of-the}}

\begin{figure}
\includegraphics[width=1\textwidth]{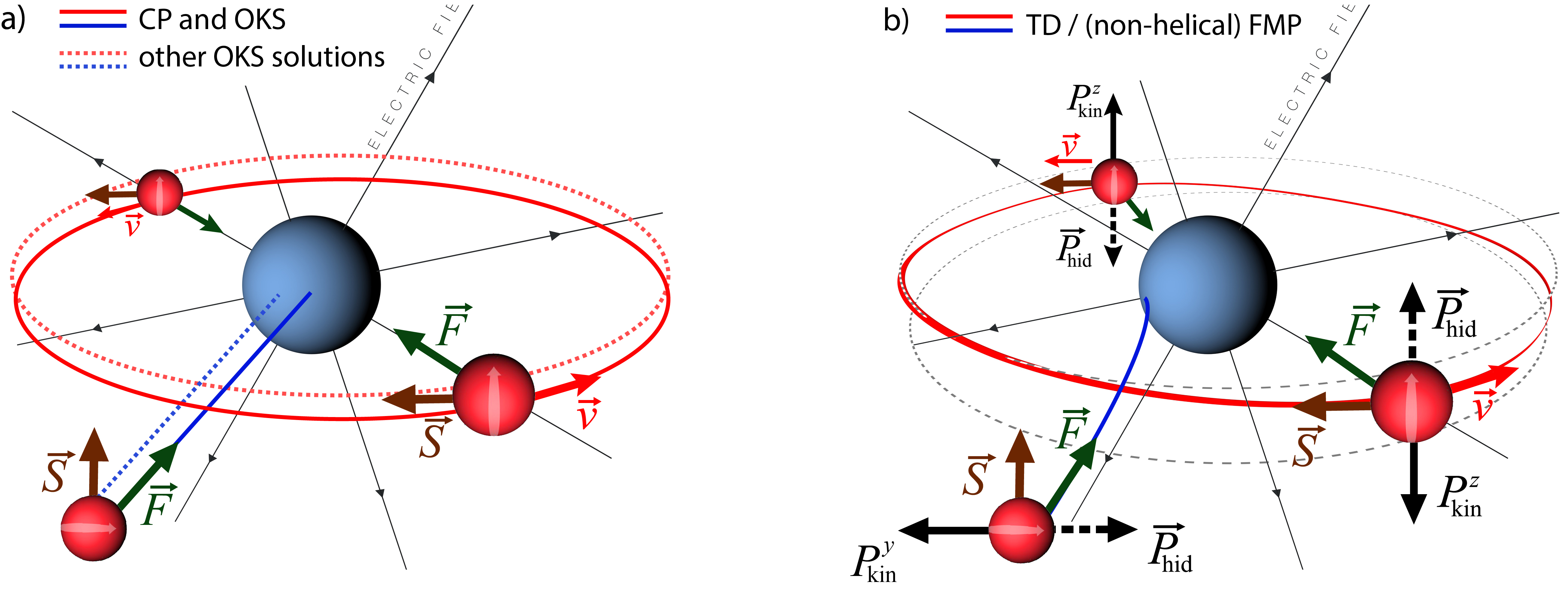}

\includegraphics[width=1\textwidth]{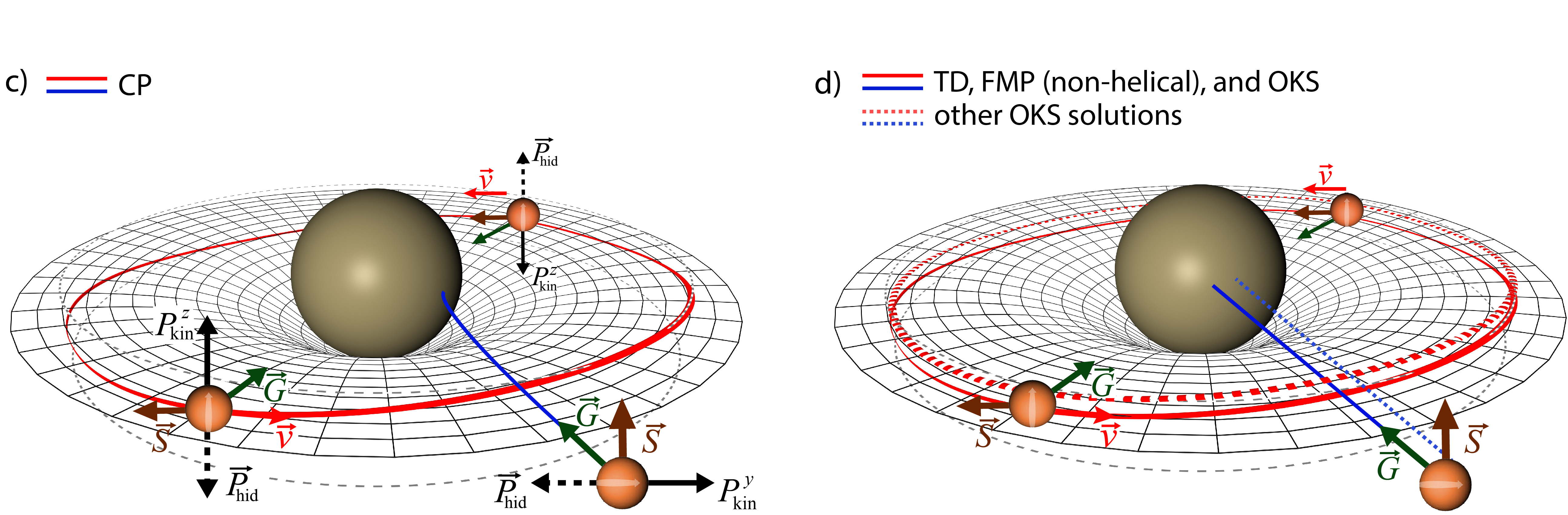}

\caption{\label{fig:Examples}{\small{{{Comparison of different spin conditions
($S^{\alpha\beta}u_{\beta}=0$) in two analogous physical systems:
a)-b) A spinning charged particle (but with $\vec{\mu}=0$) orbiting
a Coulomb charge in }}}}\emph{\small{{{flat spacetime}}}}{\small{{{;
c)-d) a spinning particle in the Schwarzschild spacetime. The CP condition,
$u^{\alpha}=u_{{\rm lab}}^{\alpha}$, chooses the centroid as measured
by the observers at rest in the background (the ``laboratory'' frame);
the FMP condition, $u^{\alpha}=U^{\alpha}$, and the TD condition,
$u^{\alpha}=P^{\alpha}/M$, choose the centroid as measured in the
frame comoving, or nearly comoving (respectively) with it. We consider
only the }}}}\emph{\small{{non-helical}}}{\small{{ FMP solution.
In the electromagnetic system, the CP condition yields $Du^{\alpha}/d\tau=0\Rightarrow P^{\alpha}=mU^{\alpha}$
(since the laboratory frame is inertial), thus there is no hidden
momentum nor exotic motions; trajectories are ellipses and, for particles
with initial radial velocity, straight lines, cf. Fig.~\ref{fig:Examples}a).
For the TD/FMP conditions, Fig.~\ref{fig:Examples}b), $Du^{\alpha}/d\tau\ne0$
since a force $\vec{F}$ acts on the particle, leading to a hidden
momentum $\vec{P}_{{\rm hid}}\simeq-\vec{S}\times\vec{F}/m$ that
modifies the trajectories. A bobbing is added to the elliptical trajectories
due to the oscillation of $\vec{P}_{{\rm hid}}=P^{z}\vec{e}_{z}$
along the orbit; and instead of a radial motion, the centroid deflects.
In the gravitational system the situation is }}}\emph{\small{{reversed}}}{\small{{:
$Du^{\alpha}/d\tau\approx0\Rightarrow P_{{\rm hid}}^{\alpha}\approx0$
for TD/FMP conditions, and it is for the CP condition (since the laboratory
observers are accelerated) that there is a hidden momentum $\vec{P}_{{\rm hid}}\simeq\vec{S}\times\vec{G}\ne0$.
The TD/FMP centroids with initial radial velocity move radially, whereas
the corresponding CP centroid deflects. $\vec{P}_{{\rm hid}}$ also
induces a bobbing in nearly elliptical orbits (adding to the existing
bobbing caused by the spin-curvature force, which is }}}\emph{\small{{not}}}{\small{{
gauge, but has the same form up to a factor of three).} In both systems
$P_{{\rm hid}}^{\alpha}$ and its induced motions are gauged away
by the Ohashi-Kyrian-Semerák (OKS) condition; different OKS centroids
move nearly parallel to each other.}}}
\end{figure}

In this section we consider the two simple setups illustrated in Fig.
\ref{fig:Examples} --- a spinning charged body (with $\vec{\mu}=0$,
and whose only non-vanishing electromagnetic moment is $q$, so that
$P_{{\rm hid\tau}}^{\alpha}=0$) orbiting a Coulomb charge in flat
spacetime, and a spinning body orbiting a Schwarzschild black hole,
both particles having spin $\vec{S}$ lying in the orbital plane ---
and compare the description of the motion given by the different spin
conditions. Such comparison will be done ensuring that one is dealing
with the worldlines of different centroids corresponding to the same
physical body (i.e., the same matter distribution $T^{\alpha\beta}(x)$).
We will be using the weak field slow motion approximation, for two
reasons: first, because it is sufficient to illustrate the effects
of interest; second, and more importantly, to make clear that the
choice of spin condition (and the resulting hidden momentum) impacts
the equations of motion at leading order, and thus these effects must
be taken into account in any linearized theory or Post-Newtonian approximation.

\subsubsection{Electromagnetic system\label{sub:Electromagnetic-system}}

The Corinaldesi-Papapetrou (CP) condition, which sets the reference
worldline $z^{\alpha}$ as being center of mass $x_{{\rm CM}}^{\alpha}(u_{{\rm lab}})$
measured in the ``laboratory'' frame (chosen as the congruence of
static observers $u_{{\rm lab}}^{\alpha}$, at rest with respect to
the source), coincides in this case with one of the solutions of the
Ohashi-Kyrian-Semerák (OKS) condition, because such frame is inertial,
and therefore $\nabla_{\mathbf{U}}u_{{\rm lab}}^{\alpha}=0\Rightarrow P_{{\rm hidI}}^{\alpha}=0$,
cf. Eq.~(\ref{eq:HiddenInertial}). The momentum is thus parallel
to the 4-velocity $P^{\alpha}=mU^{\alpha}$, and the equation of motion
for the centroid reduces to ($Q\equiv$ charge of the source) 
\begin{equation}
ma^{\alpha}=F^{\alpha}=qF^{\alpha\beta}U_{\beta}\ ;\qquad\vec{F}=q\vec{E}(U)=qQ\frac{\vec{r}}{r^{3}}+\mathcal{O}(v^{2})\ ,\label{eq:aEMCP}
\end{equation}
whose well known solution for a Coulomb field is an ellipse. In particular,
a particle with an initial velocity in the $x_{O}y$ plane, and equaling
that of a circular orbit, will follow a circular orbit in that plane
(regardless of its spin); and a particle with initial radial velocity
will move radially, cf. Fig.~\ref{fig:Examples}a). To compare with
the description given by other centroids (corresponding to other OKS
solutions, or to other spin conditions), we note that i) these have
worldlines $\bar{z}^{\alpha}(\bar{\tau})$ related to $z^{\alpha}(\tau)$
by (see Eq.~(\ref{eq:Shiftuu'Gen})) 
\begin{equation}
\bar{z}^{\alpha}=z^{\alpha}+\Delta x^{\alpha}\ ;\qquad\Delta\vec{x}\simeq\frac{\vec{S}\times\vec{v}}{m}\ ,\label{eq:zz'}
\end{equation}
where $\vec{v}$ is the particle's velocity with respect to the laboratory
observers (i.e., $U^{\alpha}=\gamma(U,u_{{\rm lab}})(u_{{\rm lab}}^{\alpha}+v^{\alpha})$,
cf. Eq.~(\ref{eq:U_u-1})); ii) the particle's momentum $P^{\alpha}$
is, to a good approximation, the same for all spin conditions, cf.
Appendix \ref{sub:The-case-withEM}; and iii) regardless of the reference
worldline chosen, to the accuracy at hand, the net force on the body
is the same, $\vec{F}=\vec{\bar{F}}$ (unlike one might expect, since
the fields are evaluated at different points). Point iii) is explained
by arguments analogous to the ones given in Sec.~\ref{sub:The-dependence-of-the-force}
for the Newtonian problem: when one changes $z^{\alpha}$, the particle's
moments change as well; if the particle was a monopole with respect
to $z^{\alpha}$, then about $\bar{z}^{\alpha}$ it will have an electric
dipole moment $\vec{d}=-q\Delta\vec{x}$, as well as higher order
moments. Whereas $\vec{F}$ is just the Coulomb force $\vec{F}=q\vec{E}(\mathbf{z})$,
to dipole order $\vec{\bar{F}}=q\vec{E}(\bar{\mathbf{z}})+\bar{F}_{{\rm dip}}^{i}$,
where $\bar{F}_{{\rm dip}}^{i}=\nabla_{j}E^{i}d^{j}+\mathcal{O}(v^{2})$
is the force due to the electric dipole, cf. third term of Eq.~(\ref{eq:ForceDS0}).
Hence 
\begin{equation}
\bar{F}^{i}=qE^{i}(\bar{\mathbf{z}})+\bar{F}_{{\rm dip}}^{i}=qE^{i}(\mathbf{z}+\Delta\mathbf{x})-q\nabla_{j}E^{i}(\mathbf{z}+\Delta\mathbf{x})\Delta x^{j}\simeq qE^{i}(\mathbf{z})=F^{i}\ .\label{eq:F_bar_FEM}
\end{equation}
Therefore, any difference in the acceleration of the two centroids
is due solely to the hidden momentum 
\begin{equation}
m\vec{\bar{a}}=\vec{\bar{F}}-\frac{D\vec{\bar{P}}_{{\rm hid}}}{d\tau}=\vec{F}-\frac{D\vec{\bar{P}}_{{\rm hid}}}{d\tau}=m\vec{a}-\frac{D\vec{\bar{P}}_{{\rm hid}}}{d\tau}\label{eq:mabar_ma_EM}
\end{equation}

This tells us that different solutions of the OKS condition, for which
$\bar{P}_{{\rm hidI}}^{\alpha}=0$ (corresponding to the centroids
as measured by observers moving with constant velocity with respect
to $u_{{\rm lab}}^{\alpha}$),\textcolor{black}{{} yield different
worldlines all (nearly) parallel to the CP solution, since they have
the same acceleration, and the same 4-velocity $U^{\alpha}\parallel P^{\alpha}$.
In particular, other OKS centroids $\bar{z}^{\alpha}$ corresponding
the same physical motion }whose description through $z^{\alpha}$
is radial motion, are non-radial straight lines parallel to $z^{\alpha}$;
and the \textcolor{black}{$\bar{z}^{\alpha}$'s} for which $z^{\alpha}$
is a circular motion are (non-concentric, in general non-coplanar)
circles, obtained from the latter by a constant spatial displacement
$\Delta\vec{x}$; see Fig. \ref{fig:Examples}a).

The situation is different if one chooses the Frenkel-Mathisson-Pirani
(FMP), $u^{\alpha}=\bar{U}^{\alpha}$, or the Tulczyjew-Dixon (TD)
condition, $u^{\alpha}=P^{\alpha}/M$, which pick as representative
point the centroid as measured in the frame comoving, or nearly comoving,
respectively, with it. Since a force acts on the particle (implying
$DP^{\alpha}/d\tau\ne0$ and $\bar{a}^{\alpha}\ne0$), it follows
that $Du^{\alpha}/d\tau\ne0$ and $P_{{\rm hidI}}^{\alpha}\ne0$,
and therefore $P^{\alpha}$ is not is not parallel to $\bar{U}^{\alpha}$,
cf. Eqs.~(\ref{eq:PTulczyjewDixon0}) and (\ref{eq:momentumhelical}).
From Eq. (\ref{eq:PTulczyjewDixon0}) (and noting from (\ref{eq:P_U_Dixon})
that $M=m+\mathcal{O}(S^{2})$), we have for the TD condition 
\begin{equation}
P^{\alpha}=m\bar{U}^{\alpha}+\frac{S^{\alpha\beta}}{m}F_{\beta}+\mathcal{O}(S^{2})=m\bar{U}^{\alpha}+\frac{1}{m}\epsilon_{\ \ \gamma\delta}^{\alpha\beta}F_{\beta}S^{\gamma}\bar{U}^{\delta}+\mathcal{O}(S^{2})\ .\label{eq:momentum2}
\end{equation}
This corresponds also to the momentum of the \emph{non-helical} FMP
solution. In order to see that, first take (\ref{eq:momentum2}) as
an ansatz, and observe it obeys, to the accuracy at hand, the FMP
equations of motion. Namely, substituting (\ref{eq:momentum2}) in
the explicit equation for the acceleration%
\footnote{In \cite{KyrianSemerak:2007}, where Eq. (\ref{eq:ExplicitaFMP})
was originally derived, $F^{\alpha}$ was taken to be the spin-curvature
force (\ref{eq:SpinCurvatureForce}); it is however easy to check,
following the derivation therein, that it holds for an arbitrary force,
as long as Eq. (\ref{eq:momentumhelical}) holds.%
} (\ref{eq:ExplicitaFMP}), one gets 
\begin{equation}
\bar{a}^{\alpha}=\frac{1}{S^{2}}\left(\frac{1}{m}F^{\mu}S_{\mu}S^{\alpha}-\frac{S_{\gamma}^{\ \beta}}{m}F_{\beta}S^{\alpha\gamma}\right)+\mathcal{O}(S)=\frac{F^{\alpha}}{m}+\mathcal{O}(S)\label{eq:aAprox}
\end{equation}
(in the second equality we noted that since $S_{\gamma}^{\ \beta}S^{\alpha\gamma}=S^{\alpha}S^{\beta}-(h^{U})^{\alpha\beta}S^{2}$
and $F^{\alpha}U_{\alpha}=0$, it follows that $-S_{\gamma}^{\ \beta}S^{\alpha\gamma}F_{\beta}/S^{2}=F_{\perp}^{\alpha}\equiv$
projection of $F^{\alpha}$ orthogonal to $S^{\alpha}$); then, substituting
(\ref{eq:aAprox}) in (\ref{eq:momentumhelical}), leads consistently
to (\ref{eq:momentum2}). Eq. (\ref{eq:aAprox}) states that the acceleration
comes, in first approximation (i.e., to zeroth order in $S$), from
the force $F^{\alpha}$, which is what one expects for a non-helical
solution (and the rationale for taking (\ref{eq:momentum2}) as an
ansatz for (\ref{eq:momentumhelical})). Note from Eq. (\ref{eq:PositionMathisson})
that, for a free particle in flat spacetime, the acceleration of the
helical motions is $a=\gamma^{2}vM/S=\mathcal{O}(S^{-1})\ne\mathcal{O}(S)$.
The helices are effectively precluded from the moment we impose $P^{\alpha}\simeq m\bar{U}^{\alpha}+S^{\alpha\beta}F_{\beta}/m$.

We can therefore write for the inertial hidden momentum of both the
TD or (non-helical) FMP conditions 
\begin{equation}
P_{{\rm hidI}}\simeq\frac{1}{m}\epsilon_{\ \ \gamma\delta}^{\alpha\beta}F_{\beta}S^{\gamma}\bar{U}^{\delta}\ .\label{eq:PhidI2}
\end{equation}
This hidden momentum leads to exotic motions of the centroid. From
Eq. (\ref{eq:mabar_ma_EM}), 
\begin{align}
m\vec{\bar{a}}= & m\vec{a}-\frac{D\vec{\bar{P}}_{{\rm hid}}}{d\tau}=m\vec{a}+\frac{1}{m}\vec{S}\times\vec{F}_{,l}\bar{v}^{l}+\mathcal{O}(v^{2})\nonumber \\
= & m\vec{a}+\frac{qQ}{m}\frac{1}{r^{3}}\left(\vec{\bar{v}}\times\vec{S}-3\frac{(\vec{\bar{v}}\cdot\vec{r})\vec{r}\times\vec{S}}{r^{2}}\right)\ .\label{eq:maEM_MP}
\end{align}
Here $\vec{r}=x\vec{e}_{x}+y\vec{e}_{y}$, since $\vec{F}(=\vec{\bar{F}})$
is the Coulomb force in the $x_{O}y$ plane.

\emph{Nearly circular motion. ---} Let us start by the motion above
whose description through $z^{\alpha}$ (i.e.~through the laboratory
frame centroid, given by the CP/OKS conditions) was a circular motion
in the $x_{O}y$ plane. Since we assume that $\vec{S}$ lies on the
$x_{O}y$ plane, it follows that $\vec{P}_{{\rm hidI}}=P_{{\rm hidI}}\vec{e}_{{\rm z}}$,
\[
P^{x}=m\bar{v}^{x}=mv^{x};\qquad P^{y}=m\bar{v}^{y}=mv^{y};\qquad P^{z}=m\bar{v}^{z}+P_{{\rm hidI}}
\]
(where we noted that since $P^{\alpha}$ is the same regardless of
the spin condition, the components of the centroid velocity in the
$x_{O}y$ plane are the same as for the CP condition: $\bar{v}^{x}=v^{x}$,
$\bar{v}^{y}=v^{y}$). Therefore, $\vec{\bar{v}}\cdot\vec{r}=0$,
$m\bar{a}^{x}=ma^{x}=F^{x}$, $m\bar{a}^{y}=ma^{y}=F^{y}$, and 
\begin{equation}
m\bar{a}^{z}=\frac{qQ}{m}\frac{1}{r^{3}}(\vec{v}\times\vec{S})^{z}\ ,\label{eq:aEMBobbings}
\end{equation}
cf. Eq. (\ref{eq:maEM_MP}). Thus, the projection of the motion in
the $x_{O}y$ plane is circular, identical to $z^{\alpha}$; and since
$\vec{S}$ is constant to this accuracy%
\footnote{Since $\tau^{\alpha\beta}=0$, $S^{\alpha}$ is Fermi-Walker transported
for the MP condition, Eq. (\ref{eq:Fermi-Walker}), and approximately
so for the TD condition (cf. Eq. (7.11) of \cite{Dixon:1970}); since
Eq. (\ref{eq:aEMBobbings}) is of first order in $v$, $\vec{S}$
can be taken constant therein.%
}, $\bar{a}^{z}$ oscillates between positive and negative values along
the orbit, leading to a bobbing motion, depicted in Fig.~\ref{fig:Examples}b).
This bobbing can easily be understood as follows. The particle's total
momentum along $z$ is constant (since there is no force along $z$,
$DP^{z}/d\tau=0$), and equal to zero, as one can see (since $P^{\alpha}$
is the same) from the results above of the CP/OKS conditions. On the
other hand, from Eq.~(\ref{eq:PhidI2}), there is a hidden momentum
along $z$, $P_{{\rm hid}}^{z}=-(\vec{S}\times\vec{F})^{z}/m$, oscillating
along the orbit (from $P_{{\rm min}}^{z}=-SF/m$ to $P_{{\rm max}}^{z}=+SF/m$);
this means that the centroid bobs up and down in order for the kinetic
momentum $m\bar{v}^{z}$ to cancel out $P_{{\rm hid}}^{z}$, keeping
$P^{z}=0$. Without loss of generality, we may take $\vec{S}=S\vec{e}_{x}$,
$v^{y}=v\cos\omega\tau$, and thus $(\vec{v}\times\vec{S})^{z}=-vS\cos\omega\tau$;
integrating $\ddot{\bar{z}}=\bar{a}^{z}$, Eq. (\ref{eq:aEMBobbings}),
and noticing that $m\omega^{2}r=Qq/r^{2}$ (as the motion is circular
in the $x_{O}y$ plane), we obtain $z=(vS/m)\cos\omega\tau$, describing
oscillations of half amplitude $vS/m$.

\emph{Nearly radial motion. --- }As we have seen above, for the CP
or the OKS conditions, it follows from Eqs. (\ref{eq:aEMCP}) that
a particle with radial initial velocity will move radially, regardless
of its spin. Take the case that the particle is dropped from rest
at some point on the $x$ axis; it will move in straight line along
$x$ towards the source, and we thus have that $P^{y}=P^{z}=0$, $P^{x}=P$.
Take $\vec{S}$ to be along $z$. For the FMP or the TD conditions
the situation is different; there is a hidden momentum, given by Eq.~(\ref{eq:PhidI2}),
\[
\vec{P}_{{\rm hid}}=-\frac{1}{m}\vec{S}\times\vec{F}=\frac{1}{m}SF\vec{e}_{y}
\]
(increasing as the particle approaches the source, since $F$ increases),
which causes the centroid to deflect in the negative $y$ direction,
in order to keep $P^{y}=P_{{\rm hid}}^{y}+m\bar{U}^{y}=0$. This is
depicted in Fig.~\ref{fig:Examples}b).

\subsubsection{Gravitational system\label{sub:Gravitational-system}}

In the gravitational case, the situation is reversed in comparison
to the electromagnetic system: now it is the ``laboratory frame''
(i.e., the observers at rest in the background) which is accelerated,
therefore $\nabla_{\mathbf{U}}u_{{\rm lab}}^{\alpha}\ne0$ and $P^{\alpha}\nparallel U^{\alpha}$
when the centroid is computed in such frame, cf. Eq.~(\ref{eq:MomentumCP});
and it is when the centroid is computed in the comoving (FMP condition)
or nearly comoving frame (TD condition) that we have $P^{\alpha}\simeq mU^{\alpha}$,
since the only force present is the spin curvature force (\ref{eq:SpinCurvatureForce}),
which yields a $\mathcal{O}(S^{2})$ contribution to the hidden momentum,
cf. Eq.~(\ref{eq:momentum2}). This is what we are now going to see
in detail.

As we have seen in Sec.~\ref{sub:The-dependence-of-the-force}, the
force (\ref{eq:SpinCurvatureForce}) depends explicitly on the spin
condition. For the FMP and the TD conditions, it can be written to
lowest order as \cite{CostaNatarioZilhao:2012} 
\begin{equation}
\bar{F}^{i}=-2\epsilon_{jkl}v^{k}G^{l,i}S^{j}+\epsilon_{\ lj}^{i}G_{\ ,k}^{l}v^{k}S^{j}\simeq m\bar{a}^{i}\ ,\label{eq:aGEM}
\end{equation}
where $\vec{G}=-m_{{\rm S}}\vec{r}/r^{3}$ is the Newtonian (or gravitoelectric)
field evaluated at $\bar{z}^{\alpha}$, $m_{{\rm S}}$ is the mass
of the Schwarzschild black hole, and in the second equality we used
$\vec{P}=m\vec{\bar{U}}+\mathcal{O}(S^{2})\Rightarrow m\vec{\bar{a}}\simeq\vec{\bar{F}}$,
as follows from (\ref{eq:momentum2}). Explicitly: 
\begin{equation}
m\vec{\bar{a}}\simeq\vec{\bar{F}}=-\frac{3m_{{\rm S}}}{r^{3}}\left[\vec{v}\times\vec{S}+\frac{2\vec{r}[(\vec{v}\times\vec{r})\cdot\vec{S}]}{r^{2}}+\frac{(\vec{v}\cdot\vec{r})\vec{S}\times\vec{r}}{r^{2}}\right]\ .\label{eq:maGravMP}
\end{equation}
Notice the first term, formally analogous to the first term of (\ref{eq:maEM_MP}),
which caused the bobbing in the electromagnetic system; but note as
well that despite the similarity, they have very different origins:
the latter comes from the inertial hidden momentum, whereas the former
comes from the spin-curvature force.

The coordinate acceleration is given by the sum of $m\vec{a}$ with
the (radial) Newtonian ``force'' $m\vec{G}$, 
\[
m\frac{d^{2}\bar{z}^{i}}{d\bar{\tau}^{2}}=mG^{i}(\bar{\mathbf{z}})+\bar{F}^{i}\ .
\]

For the CP condition the situation is different, because this is now
the case where the field $u^{\alpha}=u_{{\rm lab}}^{\alpha}$ (relative
to which the centroid is computed) is not parallel transported along
$z^{\alpha}(\tau)$, $\nabla_{\mathbf{U}}u_{{\rm lab}}^{\alpha}\ne0$;
therefore there is hidden momentum (cf. Eqs.~(\ref{eq:HiddenInertial})
and (\ref{eq:MomentumCP})): 
\begin{equation}
P_{{\rm hidI}}^{\alpha}=-(h^{U})_{\ \sigma}^{\alpha}S_{\ \beta}^{\sigma}G^{\beta}=-(h^{U})_{\ \sigma}^{\alpha}\epsilon_{\ \beta\gamma\delta}^{\sigma}G^{\beta}S^{\gamma}u_{{\rm lab}}^{\delta}\quad\Rightarrow\quad\vec{P}_{{\rm hidI}}\simeq\vec{S}\times\vec{G}\ .\label{eq:PhidCP}
\end{equation}
The spin-curvature force takes also a different form with this condition,
\begin{equation}
F^{i}=-\epsilon_{jkl}v^{k}G^{l,i}S^{j}+\epsilon_{\ lj}^{i}G_{\ ,k}^{l}v^{k}S^{j}\label{eq:aGEM_CP}
\end{equation}
(notice the relative factor of 2 comparing the first terms of (\ref{eq:aGEM})
and (\ref{eq:aGEM_CP})). The latter difference however is compensated
by the difference between $\vec{G}(\mathbf{z})$ and $\vec{G}(\bar{\mathbf{z}})$,
as explained in Sec.~\ref{sub:The-dependence-of-the-force}: $G^{i}(z)\simeq G^{i}(\bar{z})-G^{i,j}\Delta x_{j}$,
with $\Delta\vec{x}=\vec{S}\times\vec{v}/m$, cf. Eqs.~(\ref{eq:zz'});
that is, $G^{i}(z)\simeq G^{i}(\bar{z})-\epsilon_{jkl}S^{j}v^{k}G^{l,i}/m$,
and therefore $mG^{i}(\mathbf{z})+F^{i}=mG^{i}(\bar{\mathbf{z}})+\bar{F}^{i}$.
The coordinate acceleration is thus given by 
\begin{align}
m\frac{d^{2}z^{i}}{d\tau^{2}} & =mG^{i}(\mathbf{z})+F^{i}-\frac{DP_{{\rm hid}}^{i}}{d\tau}=m\frac{d^{2}\bar{z}^{i}}{d\bar{\tau}^{2}}-\frac{DP_{{\rm hid}}^{i}}{d\tau}\ .\label{eq:coordAczz'2}
\end{align}
That is, the coordinate acceleration of the CP worldline $z^{\alpha}(\tau)$
differs from that of the worldline $\bar{z}^{\alpha}(\bar{\tau})$
of the TD/FMP conditions only by the hidden momentum term involved
in the former. From (\ref{eq:PhidCP}) we have 
\begin{equation}
\frac{DP_{{\rm hid}}^{i}}{d\tau}\simeq P_{{\rm hid},i}^{\alpha}v^{i}=\epsilon_{\ jl}^{i}S^{j}G_{\ ,k}^{l}v^{k}=\frac{m_{{\rm S}}}{r^{3}}\left(\vec{v}\times\vec{S}+3\frac{(\vec{v}\cdot\vec{r})\vec{S}\times\vec{r}}{r^{2}}\right)\label{eq:DPhidCP}
\end{equation}
where we used the fact that, to this accuracy, $D\vec{S}/d\tau\approx0$.

\emph{Nearly circular motion. }--- As in the electromagnetic case,
we assume $\vec{S}\in x_{O}y$, and so, for a nearly circular orbit,
$(\vec{v}\times\vec{r})\cdot\vec{S}\simeq0$, $\vec{v}\cdot\vec{r}\simeq0$;
therefore, the second term of (\ref{eq:maGravMP}) and the last term
of (\ref{eq:maGravMP}) and (\ref{eq:DPhidCP}) vanish. We have thus
for the FMP and TD conditions 
\[
m\frac{d^{2}\bar{z}}{d\bar{\tau}^{2}}=mG^{z}(\bar{\mathbf{z}})-\frac{3m_{{\rm S}}}{r^{3}}(\vec{v}\times\vec{S})^{z}\ ,
\]
and for the CP condition 
\[
m\frac{d^{2}z}{d\tau^{2}}=m\frac{d^{2}\bar{z}}{d\bar{\tau}^{2}}-\frac{m_{{\rm S}}}{r^{3}}(\vec{v}\times\vec{S})^{z}\ .
\]
Both coordinate accelerations oscillate along the orbit, due to the
terms $\vec{v}\times\vec{S}$ (since $\vec{S}$ is approximately constant),
leading to a bobbing motion depicted in Figs. \ref{fig:Examples}c)-d).
Hence, by contrast with the electromagnetic system, in this case a
bobbing is present regardless of the spin condition (or the presence
of hidden momentum); it is just larger for the CP condition, because
the contribution for the bobbing from the hidden momentum adds to
the bobbing caused by the spin-curvature force (they have the same
form, only different factors).

\emph{Nearly radial motion.} --- For a particle in radial motion in
Schwarzschild spacetime, the spin-curvature force under the FMP/TD
conditions is \emph{exactly} zero, $\bar{F}^{\alpha}=\nabla_{\bar{\mathbf{U}}}P^{\alpha}=0$,
as shown in \cite{CostaNatarioZilhao:2012} (in the weak field and
slow motion regime, one can check that from Eq.~(\ref{eq:maGravMP})
above, by noting that the second term is zero, and the first and third
terms cancel out when $\vec{r}\parallel\vec{v}$). The hidden momentum
is also exactly zero for the TD and the non-helical FMP solutions,
so $P^{\alpha}=m\bar{U}^{\alpha}$, and thus $D\bar{U}^{\alpha}/d\bar{\tau}=0$.
When dropped from rest, the particle moves along a geodesic towards
the source. Take the motion to be along the $x$ axis, so that $P^{y}=P^{z}=0$,
$P^{x}=P$, and take $\vec{S}$ along $z$.

For the worldline $z^{\alpha}(\tau)$ given by the CP condition, there
will be a non-vanishing spin-curvature force, cf. Eq.~(\ref{eq:DF});
which, as shown above and in Sec.~\ref{sub:The-dependence-of-the-force},
just compensates for the difference in the Newtonian field $\vec{G}$
on the two worldlines, so that the coordinate acceleration differs
only due to the hidden momentum terms, cf. Eqs.~(\ref{eq:CoordAccelzz'})
and (\ref{eq:coordAczz'2}). Since the momentum $P^{\alpha}$ is the
same regardless of the spin condition, the hidden momentum (\ref{eq:PhidCP})
that arises with this spin condition causes the centroid $z^{\alpha}$
to deflect in the $y$ direction as it approaches the source, in order
to keep 
\[
P^{y}=mU^{y}+(\vec{S}\times\vec{G})^{y}=0\ ,
\]
just like the situation in the electromagnetic system for the FMP
and the TD condition. This is depicted in Fig.~\ref{fig:Examples}c).
Hence, the situation is \emph{opposite} to the electromagnetic analogue:
for the FMP and TD conditions we have no hidden momentum, and $\bar{z}^{\alpha}$
has straight line radial motion; for the CP condition there is hidden
momentum and a centroid that deflects from radial motion.

Finally, if one takes solutions $z'^{\alpha}(\tau')$ of the OKS condition,
other than the one that (to this accuracy) coincides with the centroid
$z^{\alpha}$ of the FMP and TD conditions, we have, from (\ref{eq:CoordAccelzz'}),
$d^{2}z'^{i}/d\tau'^{2}=d^{2}\bar{z}^{i}/d\bar{\tau}^{2}$, and thus
$z'^{\alpha}(\tau')$ are curves that run approximately parallel to
the trajectories of the TD/FMP (non-helical) conditions.

\subsection{Hidden momentum arising from the ``torque'' tensor $\tau^{\alpha\beta}$}

In this section we briefly discuss the hidden momentum (\ref{eq:Phidtau})
that is related to the torque tensor $\tau^{\alpha\beta}$. It is
useful to split

\begin{equation}
\tau^{\alpha\beta}=\tau_{{\rm DEM}}^{\alpha\beta}+\tau_{{\rm QEM}}^{\alpha\beta}+\tau_{{\rm QG}}^{\alpha\beta}+...\label{eq:tau}
\end{equation}
where \cite{Dixon:1964} 
\begin{equation}
\tau_{{\rm DEM}}^{\alpha\beta}=2\mu^{\theta[\beta}F_{\ \ \theta}^{\alpha]}+2d^{[\alpha}F_{\ \ \gamma}^{\beta]}U^{\gamma}\label{eq:tauDEM}
\end{equation}
is the electromagnetic dipole torque, $\tau_{{\rm QEM}}^{\alpha\beta}$
and $\tau_{{\rm QG}}^{\alpha\beta}$ are, respectively, the quadrupole
electromagnetic and gravitational torques (the lowest order torque
in the gravitational case), see \cite{CostaNatarioZilhao:2012} for
the explicit expressions. All these torques (plus the higher order
ones) will contribute to the momentum via Eqs.~(\ref{eq:HiddenMomentum})-(\ref{eq:Phidtau}).
A hidden momentum $P_{{\rm hid}\tau}$ is originated whenever $\tau^{\alpha\beta}$
has a component along the vector field $u^{\alpha}$, cf. Eq.~(\ref{eq:Phidtau}),
and it may be cast into two parts: a part which is pure gauge like
the inertial hidden momentum $P_{{\rm hidI}}^{\alpha}$ (comes from
the choice of the reference worldline $z^{\alpha}(\tau)$; may be
made to vanish by suitable choices), and another part, which arises
in some physical systems, that is not gauge. Let us discuss these
two subtypes of hidden momentum separately.

\subsubsection{The pure gauge hidden momentum that arises from $\tau_{\alpha\beta}$\label{sub:gauge-hidden-torque}}

This contribution is easier to understand if we think about a simple
example. Consider a spinning particle in flat spacetime as depicted
in Fig.~\ref{fig:CMShift}, with no forces ($DP^{\alpha}/d\tau=0$),
but now under a torque. Consider moreover $\tau^{\alpha\beta}$ to
be spatial and orthogonal%
\footnote{e.g., the torque on an electric dipole in an uniform electromagnetic
field, when $z^{\alpha}$ is the common centroid given by the TD or
the (non-helical) FMP condition.%
} to $P^{\alpha}$: $\tau^{\alpha\beta}P_{\beta}=0$. In this case,
just like for a torque-free particle, the centroid $x_{{\rm CM}}^{\alpha}(P)$
(``$x_{{\rm CM}}$'' in Fig.~\ref{fig:CMShift}), given by the
condition $S^{\alpha\beta}P_{\beta}=0$, is \emph{at rest} in the
$P^{i}=0$ frame (note that $P^{\alpha}\parallel U^{\alpha}$ for
the reference worldline $z^{\alpha}=x_{{\rm CM}}^{\alpha}(P)$, as
follows from Eq.~(\ref{eq:Momentum}) with $u^{\alpha}=P^{\alpha}/M$).
Since $x_{{\rm CM}}^{\alpha}(P)$ is unaffected by the torque, it
remains at rest at the body's geometrical center, regardless of the
fact that the spin of the particle is varying. Now consider another
\emph{inertial} observer (4-velocity $\bar{u}^{\alpha}$) moving with
respect to the $P^{i}=0$ frame with \emph{constant} velocity $\vec{v}$
(so that $D\bar{u}^{\alpha}/d\tau=0$, ensuring that no inertial hidden
momentum comes into play); not only the centroid $x_{{\rm CM}}^{\alpha}(\bar{u})$
it measures is shifted to the right relative to $x_{{\rm CM}}^{\alpha}(P)$,
as depicted in Fig.~\ref{fig:CMShift}, as \emph{the shift} (\ref{eq:Shiftuu'Gen})-(\ref{eq:CMShift})
\emph{also varies}, since $S_{\star}^{\alpha}$ varies due to the
torque: 
\begin{equation}
\frac{D\Delta x^{\alpha}}{d\tau}=-\frac{1}{m(\bar{u})}\frac{DS_{\star}^{\alpha\beta}}{d\tau}\bar{u}_{\beta}=-\frac{1}{m(\bar{u})}\tau^{\alpha\beta}\bar{u}_{\beta}\label{eq:DDxTorque}
\end{equation}
(i.e., the body's rotation velocity varies, causing $\Delta x^{\alpha}$
to vary). This means that the centroid $\bar{z}^{\alpha}=x_{{\rm CM}}^{\alpha}(\bar{u})$
will be moving in the $P^{i}=0$ frame; i.e., its 4-velocity $\bar{U}^{\alpha}=d\bar{z}^{\alpha}/d\bar{\tau}$
will have a component orthogonal to $P^{\alpha}$, which reads (in
this special case that $DP^{\alpha}/d\tau=0$, so that Eqs.~(\ref{eq:ShifUperp})
hold) $\bar{U}_{\perp}^{\alpha}=D\Delta x^{\alpha}/d\bar{\tau}$.

In the general case when there are forces acting on the particle,
however, as already mentioned in Sec.~\ref{sub:Center-of-mass-Shift_and_P_U},
one should not think of $\bar{U}_{\perp}^{\alpha}$ as the velocity
of the centroid $\bar{z}^{\alpha}$ relative to $z^{\alpha}=x_{{\rm CM}}^{\alpha}(P)$,
because the latter is \emph{not} at rest in the $P^{i}=0$ frame.
The general argument should be given instead as: \emph{the position
of the centroid} $\bar{z}^{\alpha}=x_{{\rm CM}}^{\alpha}(\bar{u})$
\emph{as measured by a given observer} $\bar{u}^{\alpha}$ \emph{depends
on} \emph{the body's angular momentum}; \emph{when the latter varies
due to the action of a torque,} $x_{{\rm CM}}^{\alpha}(\bar{u})$
\emph{moves accordingly;} $P^{\alpha}$, \emph{however, is unaffected,
leading to} $P^{\alpha}\nparallel U^{\alpha}$. The general (with
$D\bar{u}^{\alpha}/d\tau=0$) expression for $\bar{U}_{\perp}^{\alpha}$
formalizing this statement follows from Eqs.~(\ref{eq:Momentum})
and (\ref{eq:Udecomp})%
\footnote{To obtain (\ref{eq:DDxTorque}) from (\ref{eq:UperpTorque}) in the
special case above, one uses $d\tau=\gamma(\bar{U},P)d\bar{\tau}$
to write $\bar{U}_{\perp}^{\alpha}=\gamma(\bar{U},P)D\Delta x^{\alpha}/d\tau$,
computes $D\bar{S}^{\alpha\beta}/d\bar{\tau}$ from (\ref{eq:SbarS})
using (\ref{eq:DSabdtHidden}) to obtain $\bar{\tau}^{\alpha\beta}=\gamma(\bar{U},P)\tau^{\alpha\beta}$,
and finally uses the assumption above $\tau^{\alpha\beta}P_{\beta}=0\Rightarrow(h^{P})_{\ \sigma}^{\alpha}\bar{\tau}^{\sigma\beta}=\bar{\tau}^{\alpha\beta}$. %
}: 
\begin{equation}
\bar{U}_{\perp}^{\alpha}=-\frac{1}{m(\bar{u})}(h^{P})_{\ \sigma}^{\alpha}\bar{\tau}^{\sigma\beta}\bar{u}_{\beta}\ .\label{eq:UperpTorque}
\end{equation}
Finally, if $U_{\perp}^{\alpha}\ne0$, then $P_{{\rm hid}}^{\alpha}\ne0$
--- i.e., when the centroid moves in the $P^{i}=0$ frame, the momentum
$P^{i}$ is not zero in the centroid frame (the $\bar{U}^{i}=0$ frame);
thus there is hidden momentum, the two effects being reciprocal (and
mere consequences of the fact that $P^{\alpha}\nparallel\bar{U}^{\alpha}$),
cf. Eqs.~(\ref{eq:UperpPhid}).

\subsubsection{``Dynamical'' hidden momentum\label{sub:Dynamical-hidden-momentum}}

In general, the momentum of a multipole particle subject to electromagnetic
and gravitational fields is not parallel to its 4-velocity regardless
of the spin condition; that happens when $\tau^{\alpha\beta}$ is
not a spatial tensor (i.e., when $\tau^{\alpha\beta}u_{\beta}\ne0$
for all unit timelike vectors $u^{\beta}$), and is related to a type
of hidden momentum which occurs in some physical systems and is \emph{not}
gauge. Following \cite{Grallaetal:2010}, we dub this part of $P_{{\rm hid}\tau}^{\alpha}$
the ``dynamical'' hidden momentum. To dipole order, it arises in
magnetic dipoles; let us then consider the case when $\tau^{\alpha\beta}=2\mu^{\theta[\beta}F_{\ \ \theta}^{\alpha]}$
in Eqs.~(\ref{eq:tau})-(\ref{eq:tauDEM}). Take the magnetic dipole
moment to be proportional to the spin, $\mu^{\alpha\beta}=\sigma S^{\alpha\beta}$;
we have from (\ref{eq:Phidtau}), for an arbitrary spin condition
$S^{\alpha\beta}u_{\beta}=0$, 
\[
P_{{\rm hid}\tau}^{\alpha}=-\frac{1}{\gamma(u,U)}(h^{U})_{\ \sigma}^{\alpha}\mu^{\sigma\beta}(E^{u})_{\beta}\equiv P_{{\rm hidEM}}^{\alpha}\ ,
\]
where $(E^{u})^{\alpha}=F_{\ \beta}^{\alpha}u^{\beta}$. If $(E^{u})^{\alpha}\nparallel\mu^{\alpha}$,
$P_{{\rm hidEM}}^{\alpha}$ can never be zero, because then $\mu^{\beta\sigma}(E^{u})_{\beta}\ne0$
and is a space-like vector, thus cannot be parallel to any $U^{\alpha}$.
For the FMP condition ($u^{\alpha}=U^{\alpha})$, $P_{{\rm hidEM}}^{\alpha}$
takes the suggestive form

\begin{equation}
P_{{\rm hidEM}}^{\alpha}=-\mu^{\alpha\beta}E_{\beta}=\epsilon_{\ \beta\gamma\delta}^{\alpha}\mu^{\beta}E^{\gamma}U^{\delta}\ ,\label{eq:PhidEMMP}
\end{equation}
where $E^{\alpha}=F_{\ \beta}^{\alpha}U^{\beta}$ is the electric
field \emph{as measured in the centroid frame. }In such frame, and
in vector notation, $\vec{P}_{{\rm hidEM}}=\vec{\mu}\times\vec{E}$,
which the most usual form in the literature for the hidden momentum
that a magnetic dipole acquires under an external electromagnetic
field (e.g. \cite{Vaidman:1990,HnizdoFluid:1997,ColemanVanVleck:1968,GriffithsAMJPhys:2009}).
It equals \emph{minus} the electromagnetic field momentum $\vec{P}_{\times}$
generated by a magnetic dipole when placed in the external electromagnetic
field, which, in the particle's frame, reads (see e.g.~\cite{Vaidman:1990,GriffithsAMJPhys:2009,CostaNatarioZilhao:2012})
$\vec{P}_{\times}=\int\vec{E}\times\vec{B}_{{\rm dipole}}=-\vec{\mu}\times\vec{E}$.
It should however be noted that $\vec{P}_{{\rm hidEM}}$ is \emph{not}
field momentum; it is \emph{purely mechanical in nature}, which can
be understood through simple models, see e.g. \cite{Vaidman:1990,GriffithsAMJPhys:2009}
(in particular Fig.~9 of \cite{GriffithsAMJPhys:2009}). Such momentum
plays an important role in the conservation laws. Consider, for example,
a magnetic dipole \emph{at rest} in an external electric field; since
no force is exerted on the particle, the setup is \emph{stationary};
it follows from the conservation equations $(T_{{\rm tot}})_{\ \ ;\beta}^{\alpha\beta}=0$
that the total spatial momentum $\vec{P}_{{\rm tot}}\equiv\vec{P}_{{\rm matter}}+\vec{P}_{{\rm EM}}$
(i.e., the matter momentum plus the field momentum) must vanish. The
momentum of the electromagnetic field, $\vec{P}_{{\rm EM}}=\vec{P}_{\times}$,
however, is not zero; it is the momentum $\vec{P}_{{\rm matter}}=\vec{P}_{{\rm hidEM}}=-\vec{P}_{{\rm EM}}$,
hidden in the dipole, that cancels out $\vec{P}_{{\rm EM}}$, ensuring
$\vec{P}_{{\rm tot}}=0$, as required by the conservation laws.

$P_{{\rm hidEM}}^{\alpha}$ also leads to exotic motions, quite analogous
to the ones coming from the inertial hidden momentum studied in Sec.
\ref{sub:Comparison-of-the}, as one would expect from the formal
analogy between (\ref{eq:PhidEMMP}) and the the inertial hidden momentum
under this spin condition, $P_{{\rm hidI}}=-\epsilon_{\ \beta\gamma\delta}^{\alpha}S^{\beta}a^{\gamma}U^{\delta}$,
cf. Eq.~(\ref{eq:momentumhelical}). Indeed, if in the application
in Fig. \ref{fig:Examples}a)-b) we considered particles with dipole
moment $\mu^{\alpha}=\sigma S^{\alpha}\ne0$, there would be a bobbing
(in addition to the one caused by $P_{{\rm hidI}}^{\alpha}$) for
a particle orbiting the source, and, in the case of a particle in
\emph{initially} radial motion, there would be a sideways dipole force
on it, but due to $P_{{\rm hidEM}}^{\alpha}$ the particle's sideways
acceleration would actually be \emph{opposite} to the force. This
effect is discussed in detail in \cite{CostaNatarioZilhao:2012}.
However, a crucial difference exists between these effects and the
effects discussed in the previous sections: the hidden momentum in
Eq.~(\ref{eq:PhidEMMP}) is \emph{not} gauge, nor the motions generated
by it are (in general) made to vanish by any choice of center of mass.

\section{Conclusion}

In this paper we have discussed and compared in detail the different
spin supplementary conditions in the literature, with special attention
being given to the lesser-studied (but potentially useful) Corinaldesi-Papapetrou
(CP) and Ohashi-Kyrian-Semerák (OKS) spin conditions. One of the main
points is that the different solutions allowed by the different spin
conditions are equivalent descriptions of the motion of a given body.
We have shown this equivalence to pole-dipole order, explaining the
change of the spin-curvature force under the different conditions
--- which is seen to be precisely what ensures the consistency of
the different solutions, as it has the magnitude needed to prevent
the worldlines from deviating due to tidal effects of a curved spacetime.
This builds up on the work in \cite{Costaetal:2012} (dealing with
free particles in flat spacetime) and backs the claims in \cite{KyrianSemerak:2007}
about the equivalence of all spin conditions in a curved spacetime.

We clarified the origin of the non-parallelism between $U^{\alpha}$
and $P^{\alpha}$, which can be cast as the particle possessing a
``hidden momentum'', a concept introduced in General Relativity
in \cite{Grallaetal:2010}, and further developed herein. It consists
of two main parts: an ``inertial'' part $P_{{\rm hidI}}^{\alpha}$
that arises solely from the spin condition (i.e., from the choice
of the observers relative to which the center of mass is measured),
which we therefore cast as gauge, and another term $P_{{\rm hid\tau}}^{\alpha}$
arising from the torque tensor $\tau_{\alpha\beta}$, which generically
sub-divides into a part that again is gauge (arising from the motion
of the centroid measured by \emph{some} observers that is induced
when $S^{\alpha\beta}$ varies due to $\tau_{\alpha\beta}$), and
a ``dynamical'' part, which is not gauge. The latter, to dipole
order, consists of a form of hidden momentum that arises in electromagnetic
systems, and was previously known from treatments in classical electrodynamics.

The differences between the various spin conditions were discussed
and illustrated with suitable examples; in particular the reciprocity
(first noted in \cite{Grallaetal:2010}) that exists when one compares
spinning particles under an electromagnetic field in flat spacetime
to spinning particles in a gravitational field: in the first case,
when one picks the centroid as measured in the Laboratory frame (corresponding
to the CP/OKS conditions), there is no inertial hidden momentum, $P_{{\rm hidI}}^{\alpha}=0$,
and thus (if $P_{{\rm hid\tau}}^{\alpha}=0$) the momentum velocity
relation is simply $P^{\alpha}=mU^{\alpha}$; and when one computes
the center of mass in the comoving frame (FMP/TD conditions), $P^{\alpha}$
is no longer parallel to $U^{\alpha}$, leading to exotic motions
(like bobbings). In the gravitational case the situation is reversed:
when one chooses the TD or the (non-helical) FMP conditions, $P^{\alpha}$
is approximately parallel to $U^{\alpha}$; and it is when one chooses
the Laboratory centroid (CP condition) that hidden momentum arises.

All the spin conditions studied present interesting features. The
CP condition yields a natural description, as it amounts to compute
the centroid in the same frame where the motion is observed (the ``Laboratory''
frame, which is given in advance); it leads however to considerable
superfluous motions in gravitational systems. The TD condition defines
always an unique center of mass, which is the central worldline of
the worldtube of centroids (can thus be thought of as describing the
``bulk\textquotedblright{} motion of such worldtube). The FMP condition
yields the most natural transport law for the spin vector, and also
gives rise to exact gravito-electromagnetic analogies (see \cite{CostaNatarioZilhao:2012});
however it is not always easy to single out the non-helical solution
from the (infinite) helical solutions allowed by this condition (the
latter should be avoided, as they are but unnecessarily complicated
descriptions of the motion, as discussed in Sec.~\ref{sub:The-Mathisson-Pirani-condition.}),
and no general prescription for that is known. As for the OKS condition,
it always gauges away the inertial hidden momentum and its induced
motions, ensuring the simplest equations for the centroid motion;
in the absence of torques, one has $F^{\alpha}=ma^{\alpha}$, i.e.,
these are Newtonian-like (or ``dynamical\textquotedblright{}) centroids,
which accelerate only if there is a force.

It is however crucial to notice that in spite of the equivalence of
the descriptions, and the fact that the trajectories of the different
spin conditions are contained within the (convex hull of the) body's
worldtube, their differences, and the superfluous motions induced
by some of them are not negligible (even in weak field, slow motion
approximations), and should not be overlooked. As it is also important
to distinguish these motions from the physical effects. For, as we
have exemplified in Sec.~\ref{sub:Comparison-of-the}, the pure gauge
contribution to the centroid acceleration with the CP condition is
of the same order of magnitude as the one from the spin-curvature
force itself; and it can actually be much larger, as is the case of
the \emph{acceleration} of the outer helical solutions of the FMP
condition, which can be made arbitrarily large.

\appendix

\section{Momentum and angular momentum in curved spacetime\label{sec:Momentum-and-angular in curved spacetime}}

In rectangular coordinates in flat spacetime, the momenta $P^{\alpha}$
and $S^{\alpha\beta}$ of an extended body, as measured by some observer
of 4-velocity $u^{\alpha}$, are well defined by the integrals 
\[
P^{\alpha}=\int_{\Sigma(z,u)}T^{\alpha\beta}d\Sigma_{\beta};\qquad S^{\alpha\beta}=2\int_{\Sigma(z,u)}r^{[\alpha}T^{\beta]\gamma}d\Sigma_{\gamma},
\]
where $\Sigma(z,u)$ is the hyperplane orthogonal to $u^{\alpha}$
(the rest space of $u^{\alpha}$), and $r^{\alpha}=x^{\alpha}-z^{\alpha}$
is the vector connecting the reference worldline $z^{\alpha}$ to
the point of integration $x^{\alpha}$. In curved spacetime the situation
is different, as these integrals amount to summing tensors defined
at different points; different generalizations of the flat spacetime
notions have been proposed in the literature (e.g. \cite{Dixon:1964,Madore:1969,Dixon:1970}),
none of them seeming a priori more natural than the others. Herein
we discuss the mathematical meaning of the definitions used in this
work, and how they relate to the schemes by Dixon \cite{Dixon:1964,Dixon:1970}.

All schemes agree on generalizing $\Sigma(z,u)$ by the \emph{geodesic}
hypersurface orthogonal to $u^{\alpha}$, and on replacing $r^{\alpha}$
by the vector $\mathbf{X}\in\mathcal{T}_{z}$ tangent to the geodesic
connecting $z^{\alpha}$ and $x^{\alpha}$, and whose length equals
that of the geodesic. That is, $\mathbf{X}=\Phi(x)$, where $\Phi\equiv\exp_{z}^{-1}$
is the \emph{inverse} exponential map, mapping points in the spacetime
manifold to vectors in the tangent space $\mathcal{T}_{z}$, $\Phi:\mathcal{M\rightarrow}\mathcal{T}_{z}$.
Where the schemes differ is in the way the vector $\mathcal{A}^{\alpha}\equiv T^{\alpha\beta}d\Sigma_{\beta}$
is integrated. We adhere to the scheme proposed in \cite{Madore:1969}:
using the natural map for tensors induced by $\exp_{z}$ to pull back
the energy-momentum tensor and the volume element to $\mathcal{T}_{z}$,
and integrate therein, which is then a well defined tensor operation.
Let $\bm{\Omega}^{\hat{\alpha}}$ denote an orthonormal co-frame on
$\mathcal{T}_{z}$; the moments can then be written in the manifestly
covariant form 
\begin{eqnarray}
\mathbf{P}(\bm{\Omega}^{\hat{\alpha}}) & = & \int_{\Sigma(z,u)}\mathbf{T}(\Phi^{*}\bm{\Omega}^{\hat{\alpha}},d\bm{\Sigma})\ ;\label{eq:PCov}\\
\mathbf{S}(\bm{\Omega}^{\hat{\alpha}},\bm{\Omega}^{\hat{\beta}}) & = & 2\int_{\Sigma(z,u)}\mathbf{X}(\bm{\Omega}^{[\hat{\alpha}})\mathbf{T}(\Phi^{*}\bm{\Omega}^{\hat{\beta}]},d\bm{\Sigma})\ .\label{SCov}
\end{eqnarray}
Note that since $\mathbf{T}(\Phi^{*}\bm{\Omega}^{\hat{\alpha}},d\bm{\Sigma})=(\exp_{z}^{*}\mathbf{T})(\bm{\Omega}^{\hat{\alpha}},\exp_{z}^{*}d\bm{\Sigma})$,
one is indeed pulling back the integrands from $\mathcal{M}$ to $\mathcal{T}_{z}$.
Note also that Eqs. (\ref{eq:PCov})-(\ref{SCov}) are \emph{equivalent}
to (\ref{eq:Pgeneral})-(\ref{eq:Sab}), i.e., they just amount to
perform the integration in a system of Riemann normal coordinates
$\{x^{\hat{\alpha}}\}$ centered at $z^{\alpha}$ (the coordinates
naturally adapted to the exponential map). This is because such system
is constructed from geodesics radiating out of $z^{\alpha}$; thus
the components of $\mathbf{X}$, in global Lorentz coordinates in
$\mathcal{T}_{z}$, are equal to the coordinates $x^{\hat{\alpha}}$
on $\mathcal{M}$; also the basis 1-forms of such system are the pullbacks
of $\bm{\Omega}^{\hat{\alpha}}$ to $\mathcal{M}$, $dx^{\hat{\alpha}}=\Phi^{*}\bm{\Omega}^{\hat{\alpha}}$;
and, taking it comoving with $u^{\alpha}$ (i.e., at $z$, $\partial_{\hat{0}}=\mathbf{u}$),
$\Sigma(z,u)$ coincides with the spatial hypersurface $x^{\hat{0}}=0$.

Let us now compare these definitions with other schemes in the literature.
In \cite{Dixon:1964}, $P^{\alpha}$ and $S^{\alpha\beta}$ are defined
as 
\begin{equation}
P_{{\rm Dix}}^{\kappa}=\int_{\Sigma(z,u)}\bar{g}_{\alpha}^{\ \kappa}T^{\alpha\beta}d\Sigma_{\beta};\qquad S_{{\rm Dix}}^{\kappa\lambda}=-2\int_{\Sigma(z,u)}\sigma^{[\kappa}\bar{g}_{\alpha}^{\ \lambda]}T^{\alpha\beta}d\Sigma_{\beta},\label{eq:P_SDixon64}
\end{equation}
where $\sigma^{\kappa}(x,z)=-(\Phi(x))^{\kappa}=-X^{\kappa}$, cf.
\cite{Dixon:1974}. These definitions thus differ from (\ref{eq:PCov})-(\ref{SCov})
only in the way the vector $\mathcal{A}^{\alpha}\equiv T^{\alpha\beta}d\Sigma_{\beta}$
is integrated: $\bar{g}_{\alpha}^{\ \kappa}$ is a bitensor which
parallel transports $\mathcal{A}^{\alpha}$ at $x^{\alpha}$ to $z^{\kappa}$
along the geodesic connecting the two points, so that the integral
is performed over vectors $\mathcal{A}^{\kappa}|_{z}=\bar{g}_{\alpha}^{\ \kappa}\mathcal{A}^{\alpha}|_{x}$
defined at $z^{\kappa}$ (in \cite{Dixon:1970,Dixon:1974} different
propagators, $K_{\alpha}^{\ \kappa}$, $H_{\alpha}^{\ \kappa}$ in
the notation therein, are employed; the two schemes are not equivalent
though, as noted in \cite{Dixon:1970}). Writing $\bar{g}_{\hat{\beta}}^{\ \hat{\alpha}}\mathcal{A}^{\hat{\beta}}|_{x}=\mathcal{A}^{\hat{\alpha}}|_{x}+\Delta\mathcal{A}^{\hat{\alpha}}$,
with $\Delta\mathcal{A}^{\hat{\alpha}}=-\int_{x}^{z}\Gamma_{\hat{\beta}\hat{\gamma}}^{\hat{\alpha}}(x')\mathcal{A}^{\hat{\beta}}dx'^{\hat{\gamma}}$,
expanding the integrand in Taylor series around $z^{\alpha}$, and
noting that, in the normal coordinates $\{x^{\hat{\alpha}}\}$ (see
e.g. \cite{MTW}), we have $\Gamma_{\hat{\beta}\hat{\gamma}}^{\hat{\alpha}}(z)=0$
and $\|\Gamma_{\hat{\beta}\hat{\gamma},\hat{\delta}}^{\hat{\alpha}}(z)\|\sim\|\mathbf{R}\|$,
where $\|\mathbf{R}\|\equiv\sqrt{|R_{\alpha\beta\gamma\delta}R^{\alpha\beta\gamma\delta}|}$
denotes the magnitude of the curvature, we have $\Delta\mathcal{A}^{\alpha}=\mathcal{O}(\|\bm{\mathcal{A}}\|\|\mathbf{R}\|x^{2})$.
Therefore $P_{{\rm Dix}}^{\hat{\alpha}}=P^{\hat{\alpha}}+\mathcal{O}(\lambda\|\mathbf{P}\|)$,
where 
\begin{equation}
\lambda=\|\mathbf{R}\|a^{2}\ ,\label{eq:Lambda}
\end{equation}
and $a$ is the largest dimension of the body. Thus, when $\lambda\ll1$,
i.e., when the curvature is not too strong compared to the scale of
the size of the body%
\footnote{\label{fn:Assumption}For example, in the case of the Schwarzschild
spacetime, $\|\mathbf{R}\|\sim m_{{\rm S}}/r^{3}$, $\lambda=(m_{{\rm S}}/r)(a^{2}/r^{2})$;
since $m_{{\rm S}}/r<1$ for any point outside the horizon, $\lambda\ll1$
is guaranteed just by taking the size of the body much smaller than
its the distance to the source, $r^{2}\gg a^{2}$.%
}, $P_{{\rm Dix}}^{\hat{\alpha}}\simeq P^{\hat{\alpha}}$. The two
schemes are actually indistinguishable in a pole-dipole approximation,
where only terms to linear order in $x$ are kept in the integrals
defining the moments; the resulting equations of motion are the same
(compare Eqs.~(43), (49) of \cite{Madore:1969} with Eqs.~(6.31)-(6.32)
of \cite{Dixon:1964}, or Eqs.~(7.1)-(7.2) of \cite{Dixon:1970}),
both schemes leading to the well known Mathisson-Papapetrou equations
(the latter derived using less sophisticated formalisms). These conclusions
are natural, for the metric in Riemann normal coordinates is (e.g.
\cite{MTW}) of the form $g_{\hat{\alpha}\hat{\beta}}=\eta_{\hat{\alpha}\hat{\beta}}+\mathcal{O}(\|\mathbf{R}\|x^{2})$;
hence the assumption $\lambda\ll1$ amounts to say that, \emph{for
the computation} of $P^{\alpha}$ and $S^{\alpha\beta}$, one may,
to a good approximation, take the spacetime as nearly flat throughout
the body.

\subsection{The dependence of the particle's momenta on $\Sigma$\label{sec:AppendixHypersurfaces}}

The momenta (\ref{eq:Pgeneral})-(\ref{eq:Sab}) depend, in general,
on the spacelike hypersurface $\Sigma(z,u)\equiv\Sigma(z(\tau),u)$
on which the integration is performed, see e.g. \cite{Dixon:1964,Madore:1969,Dixon:1970}.
This is so even in flat spacetime; when forces and torques act on
the body, it is clear that $P^{\alpha}(z,u)$, $S^{\alpha\beta}(z,u)$
depend on $z^{\alpha}(\tau)$, and also on the argument $u^{\alpha}$
of $\Sigma$. Curvature brings additional complications, as $u^{\alpha}$
is no longer a ``free vector'', and $\Sigma$ itself is in principle
point dependent. Herein we shall show that, in the absence of electromagnetic
field ($F^{\alpha\beta}=0$), and under the assumption $\lambda\ll1$
made above, for hypersurfaces $\Sigma(z,u)$ \emph{through a point}
$z^{\alpha}$ \emph{within the body}'s convex hull, the $u^{\alpha}$
dependence of the momentum and angular momentum is negligible.

Denote by $\bm{\xi}=dx^{\hat{\alpha}}$ a particular basis 1-form
of the Riemann normal coordinate system $\{x^{\hat{\alpha}}\}$; $P^{\bm{\xi}}\equiv P^{\alpha}\xi_{\alpha}$
is thus the $\bm{\xi}$ component of $P^{\alpha}$. From definition
(\ref{eq:Pgeneral}), and since $\xi_{\alpha}$ has constant components,
we may write the $\bm{\xi}$ component of the momentum as the integral
of a vector $A^{\alpha}\equiv T^{\alpha\beta}\xi_{\beta}$ on a 3-surface,
\[
P^{\bm{\xi}}(z,u)=\xi_{\hat{\alpha}}\int_{\Sigma(z,u)}T^{\hat{\alpha}\hat{\beta}}d\Sigma_{\hat{\beta}}=\int_{\Sigma(z,u)}T^{\hat{\alpha}\hat{\beta}}\xi_{\hat{\alpha}}d\Sigma_{\hat{\beta}}=\int_{\Sigma(z,u)}A^{\beta}d\Sigma_{\beta}\ .
\]

\begin{figure}
\includegraphics[width=1\textwidth]{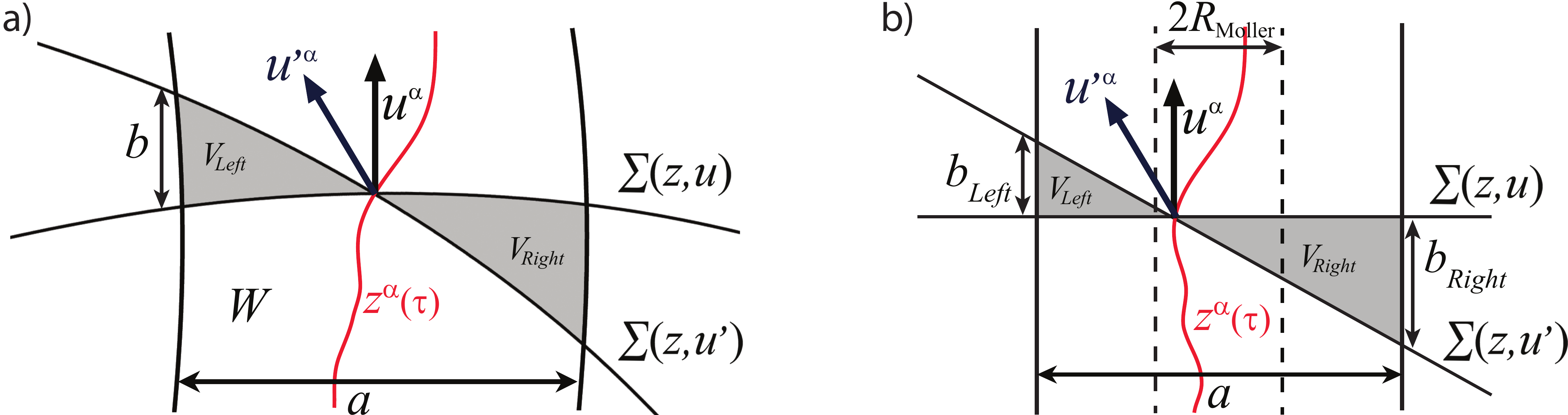}
\caption{\label{fig:Hypersurfaces}Shadowed regions $V_{Left}$ and $V_{Right}$
are the 4-volumes delimited by the hypersurfaces $\Sigma(z,u')$,
$\Sigma(z,u)$, and the boundary of the body's worldtube, of convex
hull $W$. $u^{\alpha}$ is chosen parallel to $P^{\alpha}$. a) Curved
spacetime; b) flat spacetime.}
\end{figure}

Take $u^{\alpha}=P^{\alpha}/M$, and consider another vector $u'^{\alpha}$
at the same point $z^{\alpha}$; the $\bm{\xi}$ component of the
difference between the momenta computed in the hypersurfaces $\Sigma(z,u')$
and $\Sigma(z,u)$, $\Delta P^{\bm{\xi}}\equiv P^{\bm{\xi}}(z,u')-P^{\bm{\xi}}(z,u)$
is, from an application of the Gauss theorem (see Fig. \ref{fig:Hypersurfaces}),

\[
\Delta P^{\bm{\xi}}=\int_{\Sigma(z,u)}A^{\beta}d\Sigma_{\beta}-\int_{\Sigma(z,u')}A^{\beta}d\Sigma_{\beta}=\int_{V_{Left}}A_{\ ;\beta}^{\beta}dV-\int_{V_{Right}}A_{\ ;\beta}^{\beta}dV\ .
\]
Here $V_{Left}$ and $V_{Right}$ denote the shadowed regions of Fig.~\ref{fig:Hypersurfaces}
(where $A^{\alpha}\ne0$), i.e., the ``left'' and ``right'' 4-volumes
delimited by $\Sigma(z,u')$, $\Sigma(z,u)$ and the boundary of the
body's worldtube. Now, using the conservation law $T_{\ \ ;\beta}^{\alpha\beta}=0$,
one notes that

\[
A_{\ ;\beta}^{\beta}=T_{\ \ ;\beta}^{\alpha\beta}\xi_{\alpha}+T^{\alpha\beta}\xi_{\alpha;\beta}=T^{\alpha\beta}\xi_{\alpha;\beta}\ ;
\]
thus 
\[
\Delta P^{\bm{\xi}}=\int_{V_{Left}}T^{\alpha\beta}\xi_{\alpha;\beta}dV-\int_{V_{Right}}T^{\alpha\beta}\xi_{\alpha;\beta}dV\ .
\]
Since $\bm{\xi}$ is a basis 1-form, $\xi_{\hat{\alpha},\hat{\beta}}=0$,
and 
\[
\xi_{\hat{\alpha};\hat{\beta}}=-\Gamma_{\hat{\alpha}\hat{\beta}}^{\hat{\gamma}}\xi_{\hat{\gamma}}=\mathcal{O}(\|\mathbf{R}\|x)\ ;
\]
therefore 
\[
|\Delta P^{\bm{\xi}}|\lesssim\|\mathbf{R}\|\int_{V}T^{\hat{0}\hat{0}}|x|dV=\|\mathbf{R}\|V\left\langle T^{\hat{0}\hat{0}}|x|\right\rangle \ ,
\]
where $V\equiv V_{Left}+V_{Right}$, $\left\langle \,\cdot\,\right\rangle $
denotes the average on the shadowed region of Fig.~\ref{fig:Hypersurfaces}a,
and we noted that $T^{\hat{0}\hat{0}}$ is the largest component of
$T^{\alpha\beta}$ and always positive. Since $b<av(u',u)$ (see Fig.
\ref{fig:Hypersurfaces}), with $v^{\alpha}(u',u)$ defined by Eq.
(\ref{eq:U_u-1}), and $v(u',u)<1$, then $\left\langle |x|\right\rangle <a$;
moreover (assuming $\partial_{\hat{0}}=\mathbf{u}$ at $z$), $V\left\langle T^{\hat{0}\hat{0}}\right\rangle <Mav(u',u)$;
hence we get 
\begin{equation}
|\Delta P^{\bm{\xi}}|\lesssim M\lambda v(u',u)=\|\mathbf{P}\|\lambda v(u',u)\ ,\label{eq:GravDeltaP}
\end{equation}
showing that $\Delta P^{\alpha}$ is negligible%
\footnote{\label{FootDeltaP}The inequality (\ref{eq:GravDeltaP}) means not
only that the components of $\Delta\mathbf{P}$ in the system $\{x^{\hat{\alpha}}\}$
(where $P^{\hat{i}}=0$) are much smaller than $M$, but also much
smaller than the typical spatial momentum in other frames. For instance,
in normal coordinates $\{x^{\alpha'}\}$ comoving with $u'^{\alpha}$,
one has $|P^{i'}|\sim\gamma(u',u)v(u',u)M$ and $|\Delta P^{i'}|\lesssim|\Delta P^{\bm{\xi}}|\gamma(u',u)$,
thus $|\Delta P^{i'}|\ll|P^{i'}|$ when $\lambda\ll1$.%
} compared to $P^{\alpha}$ under the restriction above on the strength
of the gravitational field, $\lambda\ll1$ (the same under which the
different multipole schemes become equivalent, and one can take local
Lorentz coordinates as nearly rectangular throughout the extension
of the body; see also footnote \ref{fn:Assumption}). In the application
in Sec. \ref{sub:Comparison-of-the} --- Schwarzschild spacetime,
far field limit --- we can write 
\[
|\Delta P^{\bm{\xi}}|\lesssim\frac{Mm_{{\rm S}}}{r^{3}}a^{2}v(u',u)\simeq\|P_{{\rm hidI}}\|\frac{a}{R_{{\rm Moller}}}\frac{a}{r}v(u',u)
\]
where $P_{{\rm hidI}}^{\alpha}$ is the inertial hidden momentum \emph{of
the CP condition}, Eq.~(\ref{eq:PhidCP}) ($P_{{\rm hidI}}^{\alpha}$
is zero or negligible for the other solutions). Thus $\Delta P^{\alpha}$
is negligible compared to $P_{{\rm hidI}}^{\alpha}$ under the condition
$\frac{R_{{\rm Moller}}}{a}\gg\frac{a}{r}v(u',u)$, which is reasonable
in a problem where the particle's spin is worth taking into account
(e.g., in the problem of nearly circular motion in Sec. \ref{sub:Comparison-of-the}
this amounts to taking $\omega_{{\rm body}}\gg\omega_{{\rm orbit}}$,
where $\omega_{{\rm body}}$ and $\omega_{{\rm orbit}}$ are the body's
rotation and orbital angular velocities).

Through an analogous procedure, one can show that the dependence of
$S^{\alpha\beta}$ on $u^{\alpha}$ is negligible in this regime.
Let $\mathscr{J}^{\hat{\alpha}\hat{\beta}\hat{\gamma}}\equiv2x^{[\hat{\alpha}}T^{\hat{\beta}]\hat{\gamma}}$,
so that $S^{\hat{\alpha}\hat{\beta}}=\int_{\Sigma(z,u)}\mathscr{J}^{\hat{\alpha}\hat{\beta}\hat{\gamma}}d\Sigma_{\hat{\gamma}}$;
and consider the two basis \emph{spatial} 1-forms $\bm{\xi}$ and
$\bm{\eta}$. Constructing the vector $\mathscr{J}^{\gamma}\equiv\mathscr{J}^{\alpha\beta\gamma}\xi_{\alpha}\eta_{\beta}$,
we can write the $\bm{\xi}\otimes\bm{\eta}$ component of $S^{\alpha\beta}$
as $S^{\bm{\xi\eta}}(z,u)=\int_{\Sigma(z,u)}\mathscr{J}^{\beta}d\Sigma_{\beta}$.
By the Gauss theorem, 
\begin{align*}
\Delta S^{\bm{\xi\eta}}= & \int_{V_{Left}}\mathscr{J}_{\ ;\beta}^{\beta}dV-\int_{V_{Right}}\mathscr{J}_{\ ;\beta}^{\beta}dV\ \sim\ \|\mathbf{R}\|\int_{V}x^{2}|\vec{J}|dV=\|\mathbf{R}\|V\langle|\vec{J}|x^{2}\rangle\\
 & <\|\mathbf{R}\|a^{2}V\langle|\vec{J}|\rangle=\lambda V\langle|\vec{J}|\rangle\lesssim\lambda a^{4}v(u',u)\langle|\vec{J}|\rangle\ ,
\end{align*}
where $J^{\hat{i}}=T^{\hat{0}\hat{i}}$. In the second relation again
we used $\Gamma_{\hat{\alpha}\hat{\beta}}^{\hat{\gamma}}=\mathcal{O}(\|\mathbf{R}\|x)$.
Since $S=\mathcal{O}(a^{4}\langle|\vec{J}|\rangle)$, cf. Eq. (\ref{eq:MinSize}),
we see that indeed when $\lambda\ll1$, $\|\Delta S^{\bm{\xi\eta}}\|\ll S$.

\subsubsection{The case with electromagnetic field\label{sub:The-case-withEM}}

When $F^{\alpha\beta}\ne0$, the conservation law is $T_{\ \ ;\beta}^{\alpha\beta}=F^{\alpha\beta}j_{\beta}$
(denoting by $T^{\alpha\beta}$ \emph{the particle's} energy momentum
tensor). Consider for simplicity flat spacetime, and let $\bm{\xi}$
be a basis 1-form of a \emph{global} Lorentz system; then $T_{\ \ ;\beta}^{\alpha\beta}\xi_{\alpha}=(T^{\alpha\beta}\xi_{\alpha})_{;\beta}\equiv A_{\ ;\beta}^{\beta}=F^{\alpha\beta}j_{\beta}\xi_{\alpha}$.
Note that $f^{\alpha}\equiv F^{\alpha\beta}j_{\beta}$ is the Lorentz
force density. It follows (see Fig. \ref{fig:Hypersurfaces}b)) 
\begin{align*}
\Delta P^{\bm{\xi}}= & \int_{V_{Left}}A_{\ ;\beta}^{\beta}dV-\int_{V_{Right}}A_{\ ;\beta}^{\beta}dV=\xi_{\alpha}\left(V_{Left}\left\langle f^{\alpha}\right\rangle _{Left}-V_{Right}\left\langle f^{\alpha}\right\rangle _{Right}\right)\ .
\end{align*}
We have $V_{Left}=V_{Left}^{(3)}b_{Left}/2$, $V_{Right}=V_{Right}^{(3)}b_{Right}/2$
(where $V^{(3)}$ denote 3-volumes orthogonal to $u^{\alpha}$). Herein
we allow $z^{\alpha}$ to be any point within the worldtube of centroids;
it follows that 
\begin{align*}
b_{Left}\ge v(u,u')\left(\frac{a}{2}-R_{{\rm Moller}}\right); & \qquad b_{Right}\le v(u,u')\left(\frac{a}{2}+R_{{\rm Moller}}\right)\\
V_{Left}^{(3)}\sim a^{2}\left(\frac{a}{2}-R_{{\rm Moller}}\right); & \qquad V_{Right}^{(3)}\sim a^{2}\left(\frac{a}{2}+R_{{\rm Moller}}\right)\ .
\end{align*}
Let $\left\langle f^{\alpha}\right\rangle _{Left}=\left\langle f^{\alpha}\right\rangle _{Right}+\Delta f^{\alpha}$,
with $\|\Delta f^{\alpha}\|\lesssim\|\nabla_{\beta}f^{\alpha}\|a$;
we obtain 
\begin{equation}
|\Delta P^{\bm{\xi}}|\lesssim\|F_{{\rm L}}^{\alpha}\|R_{{\rm Moller}}v(u',u)+\|\nabla_{j}F_{{\rm L}}^{\alpha}\|v(u',u)a^{2}\ .\label{eq:DeltaPEM}
\end{equation}
Hence $\Delta P^{\bm{\xi}}$ has, as upper bound, the sum of two terms:
the impulse of the Lorentz force $F_{{\rm L}}^{\alpha}$ in the time
interval $R_{{\rm Moller}}v(u',u)$ (as measured in the $u^{i}=0$
frame) between the two points where the hyperplane $\Sigma(z,u')$
crosses the worldtube of centroids, plus a term analogous to the gravitational
one (\ref{eq:GravDeltaP}). For the field of a Coulomb charge, discussed
in Sec. \ref{sub:Electromagnetic-system}, they%
{} read 
\begin{align*}
\|F_{{\rm L}}^{\alpha}\|R_{{\rm Moller}}v(u',u) & =|E_{{\rm p}}|v(u',u)\frac{R_{{\rm Moller}}}{r}\sim\|P_{{\rm hidI}}\|v(u',u)\\
\|\nabla_{j}F_{{\rm L}}^{\alpha}\|v(u',u)a^{2} & =|E_{{\rm p}}|v(u',u)\frac{a^{2}}{r^{2}}\sim\|P_{{\rm hidI}}\|\frac{a}{R_{{\rm Moller}}}\frac{a}{r}v(u',u)
\end{align*}
where $E_{{\rm p}}=qQ/r$ is the electric potential energy, and $P_{{\rm hidI}}^{\alpha}$
is the inertial hidden momentum \emph{of the TD/FMP (non-helical)
solutions}, Eq. (\ref{eq:PhidI2}) (for the CP/OKS conditions, $P_{{\rm hidI}}^{\alpha}=0$).
Assuming $|E_{{\rm p}}|<M$, if $R_{{\rm Moller}}/r\ll1$ and $a^{2}/r^{2}\ll1$
(as is the case in the far-field regime), then $|\Delta P^{\bm{\xi}}|\ll M=\|\mathbf{P}\|$,
and $\Delta P^{\alpha}$ is negligible compared to $P^{\alpha}$ by
arguments analogous to the ones given in footnote \ref{FootDeltaP}.
It is also negligible compared to $\|P_{{\rm hidI}}\|$ under the
following conditions: i) $v(u',u)\ll1$ so that the first term of
(\ref{eq:DeltaPEM}) can be neglected (this is guaranteed by the slow
motion assumption in Sec. \ref{sub:Comparison-of-the}); ii) that
$\frac{R_{{\rm Moller}}}{a}\gg\frac{a}{r}v(u',u)$, a condition analogous
to the one we obtained gravitational case above, which is reasonable
whenever the particle's spin is worth taking into account.

Note that the argument above can equally be used to show that $P^{\alpha}$
does not depend on the spin condition. Start with the TD centroid:
$z^{\alpha}=x_{{\rm CM}}^{\alpha}(u)$, with $u^{\alpha}=P^{\alpha}/M$;
the centroids $x_{{\rm CM}}^{\alpha}(u')$ of other spin conditions
are reached by $x_{{\rm CM}}^{\alpha}(u')=x_{{\rm CM}}^{\alpha}(u)+\Delta x^{\alpha}$,
with $\Delta x^{\alpha}\in\Sigma(u,z)$, cf. Eq. (\ref{eq:Shiftuu'Gen}).
Since the argument above applies to any spacelike hyperplane $\Sigma(u',z')$
through any arbitrary centroid $z'^{\alpha}$ on $\Sigma(u,z)$, it
effectively means that, to the accuracy at hand, $P^{\alpha}$ does
not depend on the particular centroid chosen.

\subsection*{Acknowledgments}

We thank the participants of the 524 WE-Heraeus-Seminar for the enlightening
discussions that helped shape this work. We thank O. Semerák also
for very useful correspondence, and Rui Quaresma (quaresma.rui@gmail.com)
for his assistance in the illustrations. L.F.C. is funded by FCT through
grant SFRH/BDP/85664/2012.

\bibliographystyle{unsrt}
\bibliography{CostaNatario_eom_proceedings_2013}

\begin{thebibliography}{10}

\bibitem{FrenkelZphys:1926}
J.~Frenkel.
\newblock {Die Elektrodynamik des rotierenden Elektrons}.
\newblock {\em Z. Phys.}, 37:243, 1926.

\bibitem{FrenkelNature:1926}
J.~Frenkel.
\newblock {Spinning Electrons}.
\newblock {\em Nature}, 117:514, 1926.

\bibitem{BhabhaCorben:1941}
H.~J. {Bhabha} and H.~C. {Corben}.
\newblock {General classical theory of spinning particles in a Maxwell field}.
\newblock {\em Proc. R. Soc. London A}, 178:273, 1940.

\bibitem{Corben:1961}
H.~C. Corben.
\newblock {Spin in Classical and Quantum Theory}.
\newblock {\em Phys. Review}, 121:1833, 1961.

\bibitem{Dixon:1967}
W.~G. Dixon.
\newblock {Description of Extended Bodies by Multipole Moments in Special
  Relativity}.
\newblock {\em J. Math. Phys.}, 8:1591, 1967.

\bibitem{MollerDublin:1949}
C.~Moller.
\newblock {On the definition of the centre of gravity in an arbitrary closed
  system in the theory of relativity}.
\newblock {\em Commun. Dublin Inst. Advanced Studies A}, 5:3, 1949.

\bibitem{Moller:AIH1949}
C.~Moller.
\newblock {Sur la dynamique des systemes ayant un moment angulaire interne}.
\newblock {\em Ann. Inst. Henri Poincar\'e}, 11:251, 1949.

\bibitem{Mathisson:1937}
M.~Mathisson.
\newblock {Neue Mechanik materieller Systeme}.
\newblock {\em Acta Phys. Pol.}, 6:163, 1937.

\bibitem{CorinaldesiPapapetrou:1951}
E.~{Corinaldesi} and A.~{Papapetrou}.
\newblock {Spinning test-particles in general relativity. II}.
\newblock {\em Proc. Roy. Soc. Lond. A.}, 209:259--268, 1951.

\bibitem{Tulczyjew:1959}
W.~Tulczyjew.
\newblock {Motion of multipole particles in General Relativity theory}.
\newblock {\em Acta Phys. Pol.}, 18:393, 1959.

\bibitem{Dixon:1964}
W.~G. Dixon.
\newblock {A covariant multipole formalism for extended test bodies in General
  Relativity}.
\newblock {\em Il Nuovo Cimento}, 34:317, 1964.

\bibitem{Ohashi:2003}
A.~Ohashi.
\newblock {Multipole particle in relativity}.
\newblock {\em Phys. Rev. D}, 68:044009, 2003.

\bibitem{KyrianSemerak:2007}
K.~{Kyrian} and O.~{Semer\'ak}.
\newblock {Spinning test particles in a Kerr field - II}.
\newblock {\em Mon. Not. R. Soc.}, 382:1922, 2007.

\bibitem{Semerak:1999}
O.~Semer\'ak.
\newblock {Spinning test particles in a Kerr field - I}.
\newblock {\em Mon. Not. R. Soc.}, 308:863, 1999.

\bibitem{MathissonZitterbewegung:1937}
M.~Mathisson.
\newblock {Das zitternde Elektron und seine Dynamik}.
\newblock {\em Acta Phys. Pol.}, 6:218, 1937.

\bibitem{Costaetal:2012}
L.~F. {Costa}, C.~{Herdeiro}, J.~{Nat\'ario}, and M.~{Zilh\~ao}.
\newblock {Mathisson's helical motions for a spinning particle: are they
  unphysical?}
\newblock {\em Phys. Rev. D}, 85:024001, 2012.

\bibitem{Grallaetal:2010}
S.~{Gralla}, A.~{Harte}, and R.~{Wald}.
\newblock {Bobbing and Kicks in Electromagnetism and Gravity}.
\newblock {\em Phys. Rev. D}, 81:104012, 2010.

\bibitem{ShockleyJames}
W.~{Shockley} and R.~P. {James}.
\newblock {``Try simplest cases" discovery of ``hidden momentum" forces on
  ``magnetic currents"}.
\newblock {\em Phys. Rev. Lett.}, 18:876, 1967.

\bibitem{Vaidman:1990}
L.~Vaidman.
\newblock {Torque and force on a magnetic dipole}.
\newblock {\em Am. J. Phys.}, 58:978, 1990.

\bibitem{HnizdoFluid:1997}
V.~Hnizdo.
\newblock {Hidden momentum and the electromagnetic mass of a charge and current
  carrying body}.
\newblock {\em Am. J. Phys.}, 65:92, 1997.

\bibitem{ColemanVanVleck:1968}
S.{ Coleman} and J.~H. {Van Vleck}.
\newblock {Origin of ``Hidden Momentum Forces" on Magnets}.
\newblock {\em Phys. Review}, 171:1370, 1968.

\bibitem{GriffithsAMJPhys:2009}
D.~{Babson}, S.~P. {Reynolds}, R.~{Bjorquist}, and D.~J. {Griffiths}.
\newblock {Hidden momentum, field momentum, and electromagnetic impulse}.
\newblock {\em Am. J. Phys.}, 77:826, 2009.

\bibitem{CostaNatarioZilhao:2012}
L.~F. {Costa}, J.~{Nat\'ario}, and M.~{Zilh\~ao}.
\newblock {Spacetime dynamics of spinning particles --- exact
  gravito-electromagnetic analogies}.
\newblock {\em arXiv:1207.0470}, 2012.

\bibitem{Madore:1969}
J.~Madore.
\newblock {The Equations of Motion of an Extended Body in General Relativity}.
\newblock {\em Ann. Inst. Henri Poincar\'e}, 11:221, 1969.

\bibitem{Dixon:1970}
W.~G. Dixon.
\newblock {Dynamics of Extended Bodies in General Relativity. I. Momentum and
  Angular Momentum}.
\newblock {\em Proc. Roy. Soc. Lond. A.}, 314:499, 1970.

\bibitem{MTW}
C.~W. {Misner}, Kip~S. {Thorne}, and J.~A. {Wheeler}.
\newblock {\em {Gravitation}}.
\newblock W. H Freeman and Company, San Francisco, 1973.

\bibitem{JantzenCariniBini:1992}
R.~T. {Jantzen}, P.~{Carini}, and D.~{Bini}.
\newblock {The Many Faces of Gravitoelectromagnetism}.
\newblock {\em Ann. Phys.}, 215:1, 1992.

\bibitem{Synge:1956}
J.~L. Synge.
\newblock {\em {Relativity: the special theory}}.
\newblock North-Holland Pub. Co., Amsterdam, 1956.

\bibitem{CostaNatario:2012}
L.~F. {Costa} and J.~{Nat\'ario}.
\newblock {Gravito-electromagnetic analogies}.
\newblock {\em Gen. Rel. Grav.}, 46:1792, 2014.

\bibitem{GrallaHerrmann:2013}
S.~{Gralla} and F.~{Herrmann}.
\newblock {Hidden momentum and black hole kicks}.
\newblock {\em Class. Quant. Grav.}, 30:205009, 2013.

\bibitem{Bolos:2007}
V.~Bol\'os.
\newblock {Intrinsic definitions of ``relative velocity" in general
  relativity}.
\newblock {\em Commun. Math. Phys.}, 273:217, 2007.

\bibitem{Brewin:2009}
L.~Brewin.
\newblock {Riemann normal coordinate expansions using Cadabra}.
\newblock {\em Class. Quant. Grav.}, 26:175017, 2009.

\bibitem{Beiglbock:1967}
W.~Beiglbock.
\newblock {The Center-of-Mass in Einstein's Theory of Gravitation}.
\newblock {\em Commun. Math. Phys.}, 5:106--130, 1967.

\bibitem{Schattner:1979}
R.~Schattner.
\newblock {The Uniqueness of the Center of Mass in General Relativity}.
\newblock {\em Gen. Rel. Grav.}, 10:395--399, 1979.

\bibitem{Weyssenhoff:1946}
J.~Weyssenhoff.
\newblock {Relativistic Dynamics of Spin-Fluids and Spin-Particles}.
\newblock {\em Nature}, 157:766, 1946.

\bibitem{WeyssenhoffRaabe:1947}
J.~{Weyssenhoff} and A.~{Raabe}.
\newblock {Relativistic dynamics of spin-fluids and spin particles}.
\newblock {\em Acta Phys. Pol.}, 9:7, 1947.

\bibitem{Dixon:1965}
W.~G. Dixon.
\newblock {On a Classical Theory of Charged Particles with Spin and the
  Classical Limit of the Dirac Equation}.
\newblock {\em Il Nuovo Cimento}, 38:1616, 1965.

\bibitem{Plyatsko:2008}
R.~{Plyatsko} and O.~{Stephanyshin}.
\newblock {Mathisson Equations: Non-Oscillatory Solutions in a Schwarzschild
  Field}.
\newblock {\em Acta Phys.Polon. B}, 39:23, 2008.

\bibitem{Plyatsko:2011}
R.~{Plyatsko}, O.~{Stephanyshin}, and M.~{Fenyk}.
\newblock {Mathisson-Papapetrou-Dixon equations in the Schwarzschild and Kerr
  backgrounds}.
\newblock {\em Class. Quantum Grav.}, 28:195025, 2011.

\bibitem{KudryashovaObukhov:2010}
N.~{Kudryashova} and Yu.~N. {Obukhov}.
\newblock {On the dynamics of classical particles with spin}.
\newblock {\em Phys. Lett. A}, 374:3801, 2010.

\bibitem{ObukhovPuetzfeld:2011}
Yu.N. {Obukhov} and D.~{Puetzfeld}.
\newblock {Dynamics of test bodies with spin in de Sitter spacetime}.
\newblock {\em Phys. Rev. D}, 83:044024, 2011.

\bibitem{Dixon:1973}
W.~G. Dixon.
\newblock {The definition of multipole moments for extended test bodies}.
\newblock {\em Gen. Rel. Grav.}, 4:199, 1973.

\bibitem{BiniStrains:2006}
D.~{Bini}, F.~de~{Felice}, and A.~{Geralico}.
\newblock {Strains in general relativity}.
\newblock {\em Class. Quant. Grav.}, 23:7603, 2006.

\bibitem{Dixon:1974}
W.~G. Dixon.
\newblock {Dynamics of Extended Bodies in General Relativity. III. Equations of
  Motion}.
\newblock {\em Phil. Trans. R. Soc. Lond. A}, 277:59, 1974.

\end{thebibliography}

\end{document}